\documentclass[twocolumn,english,groupedaddress, superscriptaddress,10pt,floatfix]{revtex4-1}

\usepackage[T1]{fontenc}
\usepackage[utf8]{inputenc}

\usepackage{babel}
\usepackage{amsmath,amsthm, amssymb}
\usepackage{graphicx}
\usepackage{url}
\usepackage{appendix}

\begin{document}

\title{Open data base analysis of scaling and spatio-temporal properties of power grid frequencies}

\author{Leonardo Rydin Gorj\~ao}
\affiliation{Forschungszentrum J\"ulich, Institute for Energy and Climate Research - Systems Analysis and Technology Evaluation (IEK-STE), Germany}
\affiliation{Institute for Theoretical Physics, University of Cologne, Germany}

\author{Richard Jumar}
\affiliation{Karlsruhe Institute of Technology, Institute for Automation and Applied Informatics, Germany}

\author{Heiko Maass}
\affiliation{Karlsruhe Institute of Technology, Institute for Automation and Applied Informatics, Germany}

\author{Veit Hagenmeyer}
\affiliation{Karlsruhe Institute of Technology, Institute for Automation and Applied Informatics, Germany}

\author{G.Cigdem Yalcin}
\affiliation{Department of Physics, Istanbul University, 34134, Vezneciler, Turkey}

\author{Johannes Kruse}
\affiliation{Forschungszentrum J\"ulich, Institute for Energy and Climate Research - Systems Analysis and Technology Evaluation (IEK-STE), Germany}
\affiliation{Institute for Theoretical Physics, University of Cologne, Germany}

\author{Marc Timme}
\affiliation{Chair for Network Dynamics, Center for Advancing Electronics Dresden (cfaed) and Institute for Theoretical Physics, Technical University of Dresden, Germany}

\author{Christian Beck}
\affiliation{School of Mathematical Sciences, Queen Mary University of London, United Kingdom}

\author{Dirk Witthaut}
\affiliation{Forschungszentrum J\"ulich, Institute for Energy and Climate Research - Systems Analysis and Technology Evaluation (IEK-STE), Germany}
\affiliation{Institute for Theoretical Physics, University of Cologne, Germany}

\author{Benjamin Sch\"afer}
\email{b.schaefer@qmul.ac.uk}
\affiliation{School of Mathematical Sciences, Queen Mary University of London, United Kingdom}

\begin{abstract}
The electrical energy system has attracted much attention from an increasingly diverse research community.
Many theoretical predictions have been made, from scaling laws of fluctuations to propagation velocities of disturbances.
However, to validate any theory, empirical data from large-scale power systems are necessary but are rarely shared openly. 
Here, we analyse an open data base of measurements of electric power grid frequencies across 17 locations in 12 synchronous areas on three continents.
The power grid frequency is of particular interest, as it indicates the balance of supply and demand and carries information on deterministic, stochastic, and control influences.
We perform a broad analysis of the recorded data, compare different synchronous areas and validate a previously conjectured scaling law.
Furthermore, we show how fluctuations change from local independent oscillations to a homogeneous bulk behaviour.
Overall, the presented open data base and analyses constitute a step towards more shared, collaborative energy research.
\end{abstract}

\maketitle
\section*{Introduction}
The energy system, and in particular the electricity system, is undergoing rapid changes due to the introduction of renewable energy sources to mitigate climate change \cite{sawin2018renewables}.
To cope with these changes, new policies and technologies are proposed \cite{meadowcroft2009politics,Markard2018} and a range of business models are implemented in various energy systems across the world \cite{rodriguez2014business}. 
New concepts, such as smart grids \cite{Fang2012}, flexumers \cite{barwaldt2018energy}, or prosumers \cite{parag2016electricity} are developed and tested in pilot regions.
Still, studies rarely systematically compare different approaches, data, or regions, in part because freely available research data are lacking.

The frequency of the electricity grids is a key quantity to monitor as it follows the dynamics of consumption and generation: A surplus of generation, e.g. due to an abundance of wind feed-in, directly translates into an increased frequency.
Vice versa, a shortage of power, e.g. due to a sudden increase in demand, leads to a dropping frequency.
Many control actions monitor and stabilise the power grid frequency when necessary, so that it remains close to its reference value of $50$ or $60$~Hz \cite{Kundur1994}.
Implementing renewable energy generators introduces additional fluctuations since wind or photo-voltaic generation may vary rapidly on various time scales \cite{Anvari2016,wolff2019,wohland2019significant} and reduces the overall inertia available in the grid \cite{hartmann2019effects}.
These fluctuations pose new research questions on how to design and stabilise fully renewable power systems in the future.

Analysis and modelling of the power grid frequency and its statistics and complex dynamics have become increasingly popular in the interdisciplinary community, attracting also much attention from mathematicians and physicists.
Studies have investigated for example different dynamical models \cite{Filatrella2008,schmietendorf2014self,nishikawa2015}, compared centralised vs. decentralised topologies \cite{Rohden2012,Menck2013,rodrigues2016kuramoto}, investigated the effect of fluctuations on the grid's stability \cite{schafer2017escape,hindes2019network}, or how fluctuations propagate \cite{zhang2019fluctuation,HaehnePropagation2019}.
Further research proposed real-time pricing schemes \cite{Schaefer2015}, optimised the placement of (virtual) inertia \cite{poolla2017optimal,pagnier2019inertia}, or investigated cascading failures in power grids \cite{Simonsen2008,yang2017small,schafer2019dynamical,nesti2018emergent}.
However, these theoretical findings or predictions are rarely connected with real data of multiple existing power grids.

In addition to the need raised by theoretical models from the physics and mathematics community, there is also a great need for open data bases and analysis from an engineering perspective.
While there exist data bases of frequency time series, such as GridEye/FNET \cite{chai_wide-area_2016} or GridRadar \cite{Gridradar}, these data bases are not open, which limits their value for the research community. 
In particular, different scientists with access to selected, individual types of data only, from grid frequencies to electricity prices, demand, and consumption dynamics, cannot combine their data with these data bases, thereby hindering to study more complex questions, such as the impact of prices dynamics or demand control on system stability.

Hence, open empirical data are necessary to validate theoretical predictions, adjust models, and apply new data analysis methods. Furthermore, a direct comparison of different existing power grids would be very helpful when designing future systems that include high shares of wind energy, as they are already implemented in the Nordic grid, or by moving towards liberal markets, such as the one in Continental Europe.
Proposals of creating small autonomous cells, i.e., dividing large synchronous areas into \emph{microgrids} \cite{Lasseter2004} should be evaluated by comparing synchronous power grids of different size to estimate fluctuation and stability risks.
In addition, cascading failures, spreading of perturbations, and other analyses of spatial properties of the power system may be evaluated by recording and analysing the frequency at multiple measurement sites.

In this article, we present an analysis of an open data base for power grid frequency measurements \cite{Jumar2020Data} recorded with an Electrical Data Recorder (EDR) across multiple synchronous areas \cite{maass2013first,maass2015data}.
Details on how the recordings were made are described in \cite{Jumar2020Data}, while we focus on an initial analysis and interpretation of the recordings, which are publicly available \cite{database2020}.
First, we discuss the statistical properties of the various synchronous areas and observe a trend of decreasing fluctuation amplitudes for larger power systems. 
Next, we provide a detailed analysis of a synchronised wide-area measurement carried out in Continental Europe.
We perform a detailed analysis showing that short time fluctuations are independent, while long time scale trends are highly correlated throughout the network.
We extract the precise time scales on which the power grid frequency transitions from localised to bulk dynamics.
Finally, we extract inter-area oscillations emerging in the Continental European (CE) area. 
Overall, by establishing this data base and performing a first analysis, we demonstrate the value of a data-driven analysis in an interdisciplinary context.

\section*{Results}
\subsection*{Data overview}
\begin{figure*}[ht]
~ ~ \quad ~ ~\! \includegraphics[width=0.917\linewidth]{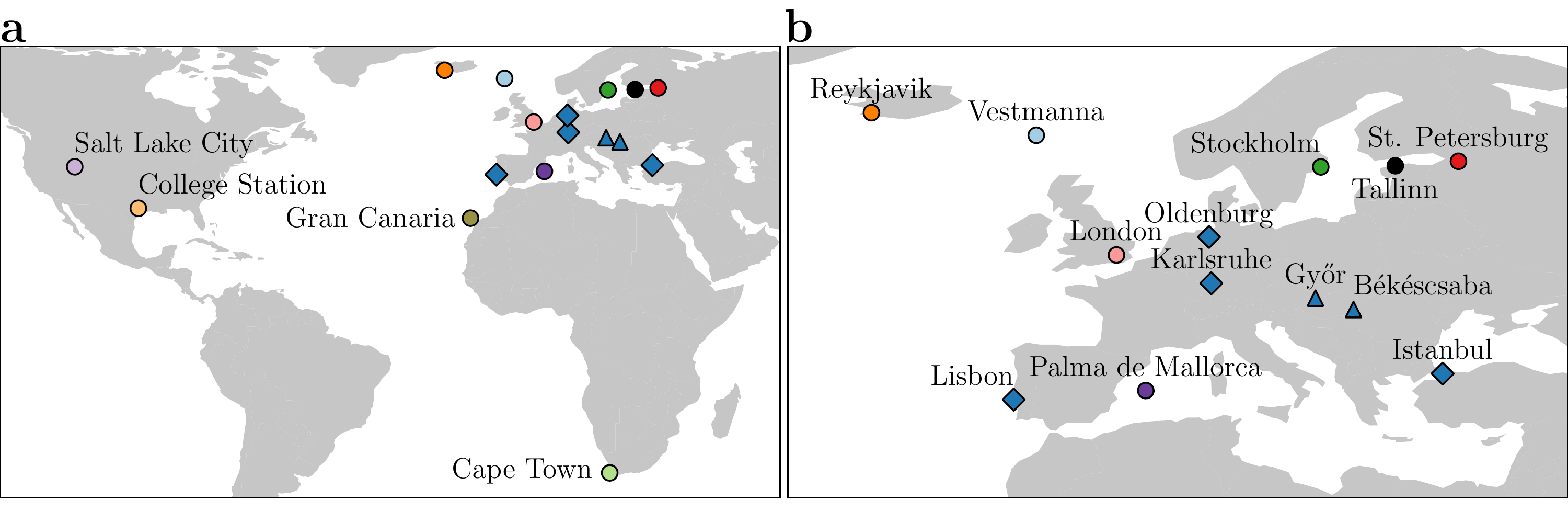}
\includegraphics[width=1\linewidth]{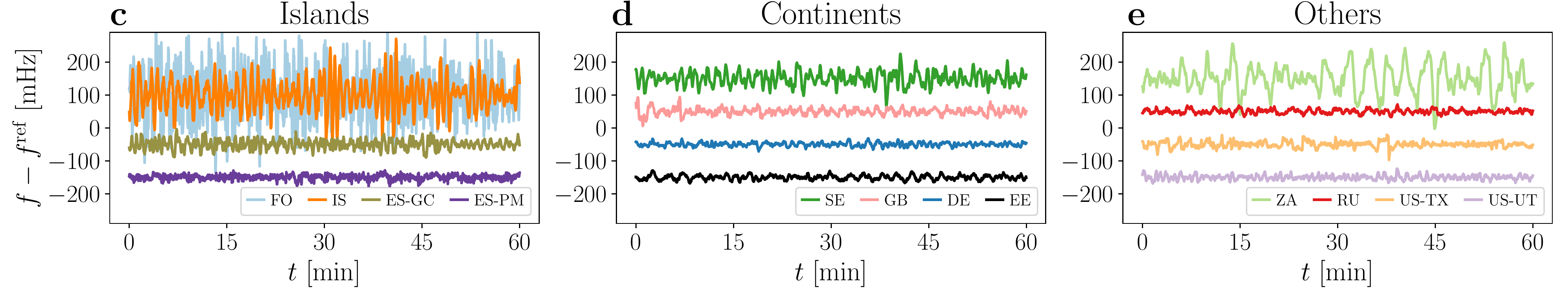}
\caption{Overview of available frequency data.
a: Different locations in Europe, Africa, and Northern America at which frequency measurements were taken.
Australia and large parts of Asia or not displayed as there were no measurements recorded.
b: Zoom of the European region (excluding Gran Canaria) with all locations labelled.
Circles indicate measurement sites where single measurements for several days were taken, diamonds mark the four locations where we performed synchronised measurements and triangles mark sites for which we received additional data.
c-e: Frequency trajectories display very different characteristics.
We plot one hour extracts of the deviations from the reference frequency of $f^\text{ref}=50$ Hz (or $60$ Hz for the US power grids), which are offset from the zero mean to improve readability.
Panels c-e and following plots abbreviate the measurement sites using the ISO 3166 code for each country and each location is assigned a colour code, as in the maps in panels a and b. For more details on the data acquisition and measurement locations see Supplementary Note 1 and \cite{Jumar2020Data}. 
Maps were created using Python 3 and geoplots.}
\label{fig:Map}
\end{figure*}

\begin{figure*}[ht]
\includegraphics[width=1\linewidth]{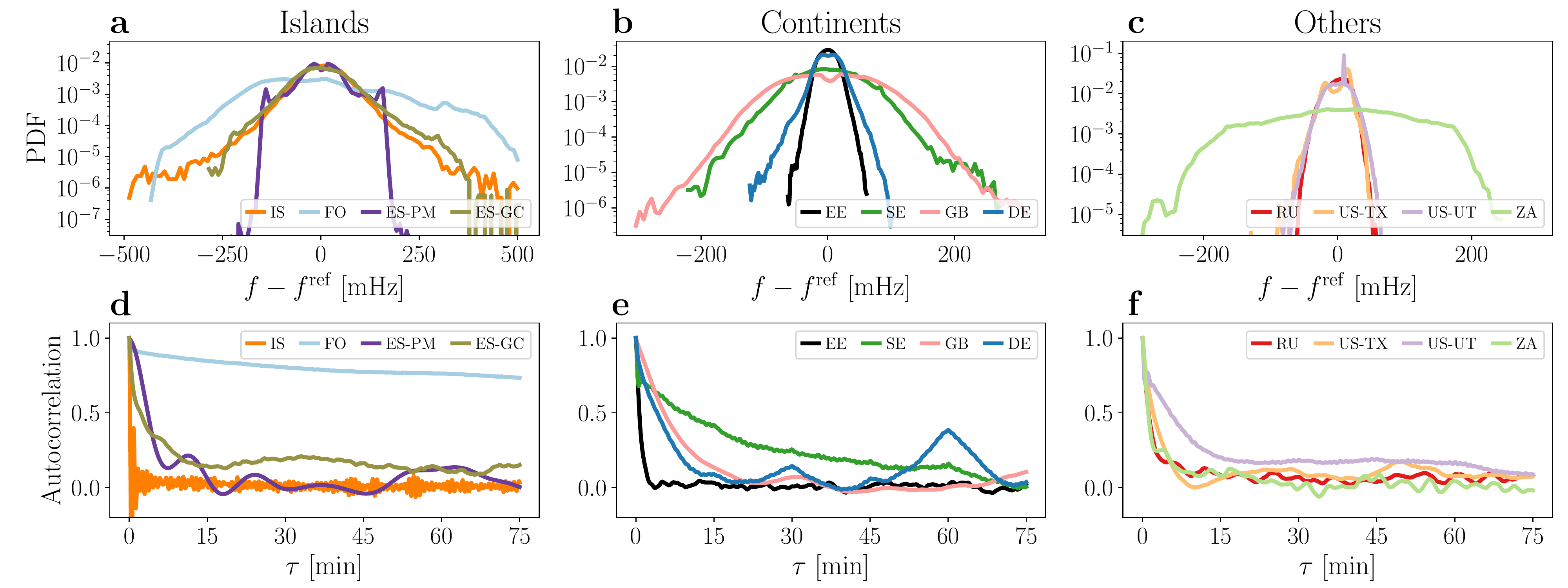}
\caption{Heterogeneity in power grid statistics.
Both histograms and autocorrelation functions display very distinct features between the different synchronous areas.
a-c: Histograms of the different synchronous areas provide insight on heavy tails but also the different scales of the fluctuations.
We visualise the empirical probability distributions of the various areas by histograms on a logarithmic scale.
d-f: The complex autocorrelation decay reveals distinct time scales in the different grids.
We compute the autocorrelation of each area for a time lag of up to 75 mins.}
\label{fig:Histograms_Autocorrelation}
\end{figure*}

We recorded power grid frequency time series using a GPS synchronizsed frequency acquisition device called Electrical Data Recorder (EDR) \cite{maass2013first,maass2015data}, providing similar data as a Phasor Measurement Unit (PMU) would.
Recordings were taken at local power plugs, which have been shown to give similar measurement results as monitoring the transmission grid with GPS time stamps  \cite{kakimoto2006monitoring}, see also \cite{Jumar2020Data} for details on the data acquisition and a description of the open data base.
In addition, we received a one week measurement from the Hungarian TSO for the two cities Békéscsaba and Győr.
We marked the locations of the measurement locations on a geographic map in Fig.~\ref{fig:Map} a-b.
Still, many more synchronous areas in the Americas, Asia, Africa, and Australia should be covered in the future. 

To gain a first impression of the frequency dynamics, we visualise frequency trajectories in different synchronous areas and note quite distinct behaviour, see Fig.~\ref{fig:Map}, c-e.
We refer to each measurement by the country or state in which it was recorded, see also Supplementary Note 1.
We group the measurements into (European) continental areas, (European) islands and other (non-European) regions, which are also mostly continental.
Most islands, such as Gran Canaria (ES-GC), Faroe Islands (FO), and Iceland (IS), but also South Africa (ZA), display large deviations from the reference frequency, while the continental areas, such as the Baltic (EE) and Continental European areas (DE), as well as the measurements taken in the United States (US-UT and US-TX) and Russia (RU), stay close to the reference frequency.
There are still more differences within each group: For example, the dynamics in Gran Canaria (ES-GC) and South Africa (ZA) are much more regular then the very erratic jumps of the frequency over time observable in the Faroe Islands (FO) and Iceland (IS).
Finally, we do not observe any qualitative difference between $50$ and $60$~Hz areas (right), when adjusting for the different reference frequency.
Note that some of the synchronous areas considered here are indeed coupled via high-voltage direct current (HVDC) lines but still possess independent synchronous behavior. Specifically, the British (GB), Continental (DE), Baltic (EE) and Nordic (SE) European areas as well as Mallorca are connected in this way. The HVDC connection of Mallorca towards Continental Europe might be the reason it displays overall smaller deviations than FO or IS, which cannot access another large synchronous area for balance.

Let us quantify the different statistics in a more systematic way by investigating distributions (histograms) and autocorrelation functions of the various areas.
The distributions contain important information of how likely deviations from the reference frequency are, how large typical deviations are (width of the distribution) and whether fluctuations are Gaussian (histogram displays an inverted parabola in log-scale) and whether they are skewed (asymmetric distribution).
Analysing the distributions (histograms) of the individual synchronous areas (Fig.~\ref{fig:Histograms_Autocorrelation} a-c), we note that the islands tend to exhibit broader and more heavy-tailed distributions than the larger continental areas.
Still, there are considerable differences within each group.
For example, we observe a larger standard deviation and thereby broader distribution in the Nordic (SE) and British (GB) areas compared to Continental Europe (DE), which is in agreement with earlier studies \cite{Schaefer2017a,gorjao_data-driven_2020}.
Some distributions, such as those for Russia (RU) or the Baltic grid (EE), do show approximately Gaussian characteristics while for several other areas, such as Gran Canaria (ES-GC) and Iceland (IS) they exhibit a high kurtosis ($\kappa^\text{Iceland}\approx 7$, as compared to $\kappa=3$ for a Gaussian), i.e., heavy tails and thereby a high probability for large frequency deviations.
We provide a more detailed analysis of the first statistical moments, i.e., standard deviation $\sigma$, skewness $\beta$ and kurtosis $\kappa$ in Supplementary Note 1.

Complementary to the aggregated statistics observable in histograms, the autocorrelation function contains information on intrinsic time scales of the observed stochastic process, see Fig.~\ref{fig:Histograms_Autocorrelation} d-e.
For simple stochastic processes such as Ornstein--Uhlenbeck processes, we would expect an exponential decay $\exp(-\gamma \tau)$ of the autocorrelation with some damping constant $\gamma$ \cite{Gardiner1985}.
While most synchronous areas do show an approximately exponential decay, the decay constants vary widely.
For example, the autocorrelation of the Icelandic data (IS) rapidly drops to zero, while the autocorrelation of the Nordic grid (SE) has an initial sharp drop, followed by a very slow decay.
Other grids, such as the Faroe Islands (FO) or the Western interconnection (US-UT) do show a slow decay, indicating long-lasting correlations, induced e.g. via correlated noise.
Finally, regular power dispatch actions every $15$ minutes are clearly observable in the Continental European (DE), British (GB) and also the Mallorcan (ES-PM) grids, consistent with earlier findings \cite{Schaefer2017a,anvari_stochastic_2020,gorjao_data-driven_2020}.
 
Concluding, we see that histograms are a good indicator of how heavy-tailed the frequency distributions are, while the autocorrelation function reveals information on regular patterns and long term correlations.
These correlations are likely connected to market activity or regulatory action, demand and generation mixture, and other aspects specific to each synchronous area.
Instead of going deep into individual comparisons let us search for general applicable scaling laws instead.

\subsection*{Scaling of individual grids}

\begin{figure}[ht]
\includegraphics[width=1\linewidth]{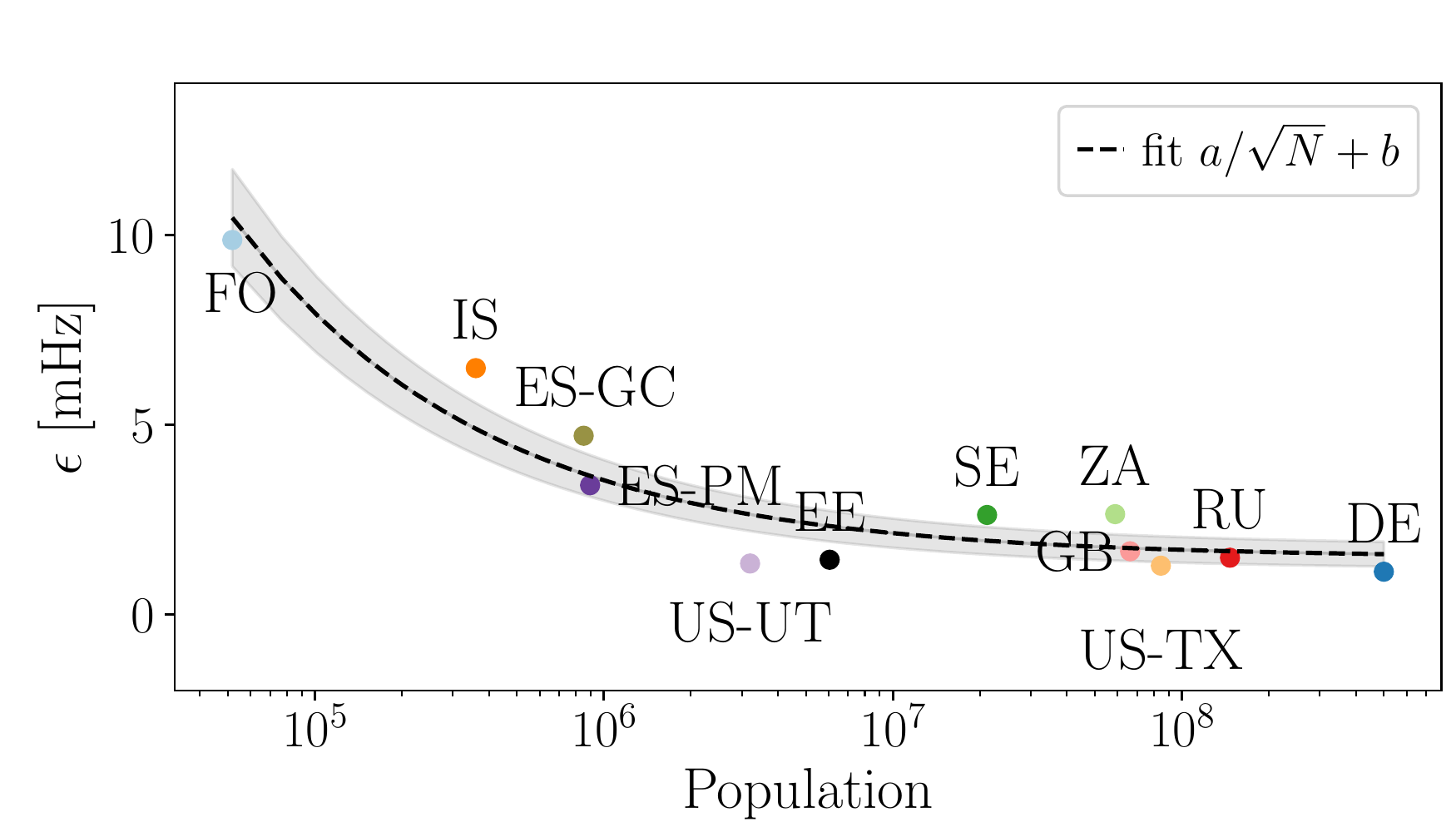}
\caption{The noise tends to decrease with an increasing size of the synchronous area until it saturates.
We plot the extracted noise amplitude $\epsilon$ compared to the logarithm of the population in a given synchronous area.
The population size serves as a proxy for the total generation and consumption of that area, as data on the size of the power grids is not commonly available.
The shaded area is the standard deviation of the $\epsilon$ estimation.}
\label{fig:Scaling}
\end{figure}

For the first time, we have the opportunity to analyse numerous synchronous areas of different size, ranging from Continental Europe with a yearly power generation of about $3000$~TWh \cite{ENTSOEFactsheet2018} and a population of hundreds of millions to the Faroe Islands with a population of only tens of thousands.
These various areas allow us to test a previously conjectured scaling law \cite{Schaefer2017a} of fluctuation amplitudes given as $\epsilon \sim 1/\sqrt{N}$, i.e., the aggregated noise amplitude $\epsilon$ in a synchronous area should decrease like the square root of the effective size of the area.

To derive this scaling relation, we formulate a Stochastic Differential Equation (SDE) of the aggregated frequency dynamics.
A basic model, also known as the aggregated swing equation \cite{Machowski2011, Ulbig2014}, is given as 
\begin{equation}\label{eq:Freq_bulk_dynamics}
    M\frac{\text{d}}{\text{d}t}\bar{\omega}\left(t\right)=-M\gamma\bar{\omega}\left(t\right)+ \Delta P(t),
\end{equation}
with bulk angular velocity $\bar\omega$, total inertia of a region $M$, power imbalance $\Delta P(t)$, and effective damping to inertia ratio $\gamma$, which also comprises primary control.
The bulk angular velocity is the scaled deviation of the frequency from the reference: $\bar\omega=2\pi\! \left(f-f^\text{ref} \right)$ and $\Delta P(t)$ effectively represents noise acting on the system with mean $\left<\Delta P (t)\right> = 0$, as generation and load are balanced on average.
A simple scaling law for the frequency variability can be derived if the short-term power fluctuations at each grid node are assumed to be Gaussian.
If the grid has $N$ nodes with identical noise amplitudes, the standard deviation of the power imbalance scales as
\begin{equation}
    \sigma_{\Delta P} \sim \sqrt{N}.
\end{equation}
At the same time, the total inertia typically scales linearly with the size of the grid, i.e., $M \sim N$.
As a result, the amplitude of the total noise acting on the angular velocity dynamics scales as 
\begin{equation}\label{eq:Proposed_Noise_Scaling}
    \epsilon \sim \frac{1}{M}\sigma_{\Delta P}   \sim \frac{1}{\sqrt{N}}.    
\end{equation} 
A more detailed derivation is provided in Supplementary Note 2 and discussed in \cite{Schaefer2017a,gorjao_data-driven_2020}. And a technical discussion of extracting the aggregated noise amplitude is presented in  \cite{Gorjao2019}. We note that the scaling law has to be modified if the noise at the nodes is not Gaussian \cite{Schaefer2017a}.

To verify the proposed scaling law in eq.~\eqref{eq:Proposed_Noise_Scaling}, we approximate the number of nodes $N$ by the population of an area, since generation data are not available for all synchronous areas and population and generation tend to be approximately proportional \cite{ENTSOEFactsheet2018}.
We utilise the population size as a proxy for size of the grid $N$.
Indeed, we note that the aggregated noise amplitude $\epsilon$ does approximately decay with the inverse square root of the population size, as predicted, see Fig.~\ref{fig:Scaling}. At a certain size, the noise saturates.
The deviations from the prediction, such as by South Africa (ZA) and Iceland (IS) are likely caused by different local control mechanisms, or non-Gaussian noise distributions, which we focus on in the next section. 
Interestingly, while Faroe Islands (FO) and Mallorca (ES-PM) do display non-Gaussian probability density functions, they follow the proposed scaling law. 
Why this is the case and how a fully non-Gaussian scaling law could capture this even better remain open questions for future work.
Still, we observe a decay of the noise, approximately following the prediction over four orders of magnitude. 

\subsection*{Increment analysis}

\begin{figure*}[ht]
\includegraphics[width=1\linewidth]{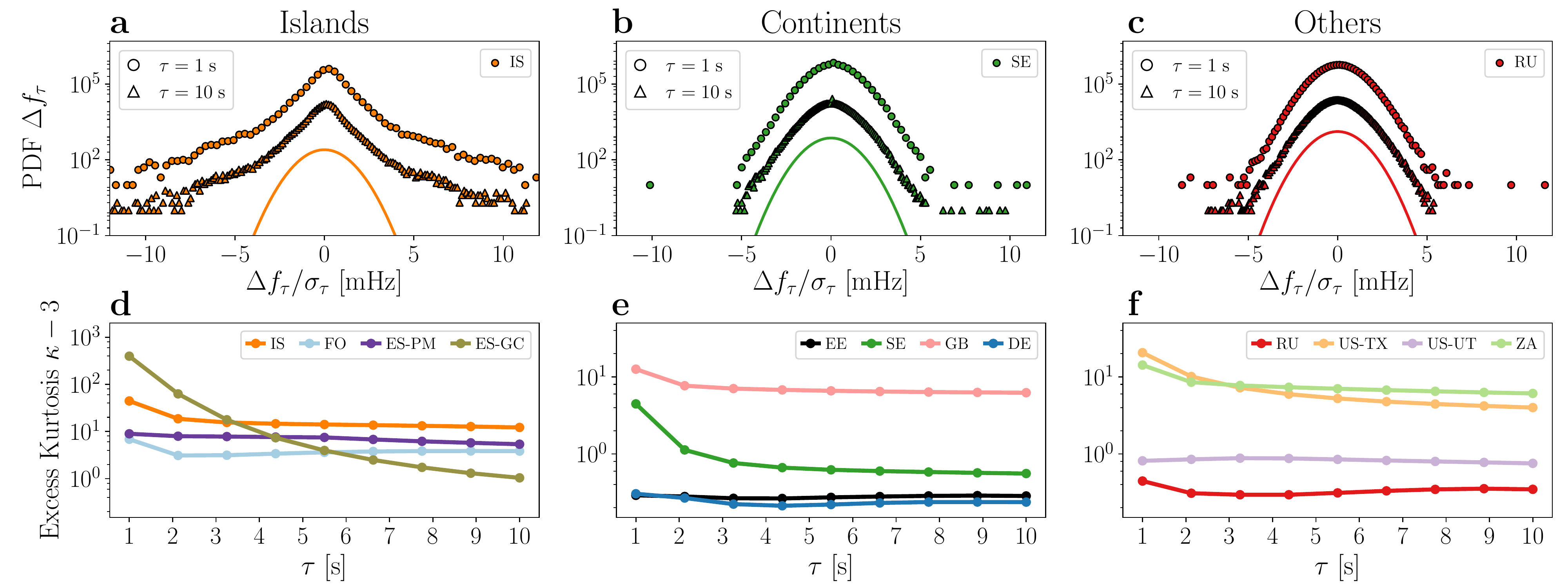}
\caption{Increment analysis reveals non-Gaussian characteristics, dominantly in islands. 
Top:(a-c): We display histograms of the increments $\Delta f_\tau$ for the lag values $\tau=1,10$ seconds for selected areas. The curves are shifted for visibility and compared to a Gaussian distribution as reference.
a: Iceland (IS) displays clear deviations from Gaussianity, even for larger increments $\tau$.
b: The Nordic area (SE) displays a non-Gaussian distribution for $\tau=1~s$, but approaches a Gaussian distribution for larger delays $\tau$.
c: Russia (RU) has a Gaussian increment distribution for all lags $\tau$.
Bottom (d-e): We plot the excess kurtosis $\kappa-3$ for the different examined power grid frequency recordings on a log scale.
We observe a non-vanishing intermittency in Gran Canaria (ES-GC), Iceland (IS), Faroe Islands (FO), Mallorca (ES-PM), Britain (GB), Texas (US-TX), and South Africa (ZA).
In contrast, the increments' distribution of the Baltic (EE), Continental Europe (DE), Nordic (SE), Russia (RU) synchronous areas, and the Western Interconnection (US-UT) approach a Gaussian distribution.
See also Fig.~\ref{fig:SynchronousMeasurements_Trajectory} for an illustration how increments are computed from a trajectory.
}
\label{fig:Increment_Analysis}
\end{figure*}

In the previous section, we approximated the noise acting on each synchronous area as Gaussian to derive an approximate scaling law.
In the following, we want to go beyond this simplification and investigate the rich short-time statistics present in each synchronous area. We will see in particular how non-Gaussian distributions clearly emerge on the time scale of a few seconds.

This short time scale is investigated via increments $\Delta f_\tau$.
The increment of a frequency time series is computed as the difference of two values of the frequency with a time lag $\tau$
\begin{equation}
    \Delta f_\tau = f(t+\tau) - f(t).
\end{equation}
An analysis of $\Delta f_\tau$ provides information on how the time series changes from one time lag $\tau$ to the next.
On a short time scale of $\tau \approx 1$ second the increments can be used as a proxy for the noise $\epsilon$ acting on the system, see also Supplementary Note 2.

The increments for a Wiener process, an often used reference stochastic process, are  Gaussian regardless of the lag $\tau$ \cite{Gardiner1985}.
However, for many real world time series, ranging from heart beats \cite{peng1993long} and turbulence to solar and wind generation \cite{Anvari2016}, we observe non-Gaussian distributions for small lags $\tau$.
For many such processes with non-Gaussian increments, the probability distribution functions (PDFs) of the increments tend to approach Gaussian distributions for larger increments \cite{Anvari2016}.
We observe a similar behaviour for the frequency statistics, see Fig.~\ref{fig:Increment_Analysis}.
The Nordic area (SE) displays deviations from Gaussianity for small lags $\tau$ but approximates a Gaussian distribution for larger $\tau$.
The Russian area (RU) even starts out with an almost Gaussian increment distributions. Contrary, the Icelandic area (IS) shows clear deviations from a Gaussian distribution for all lags $\tau$ investigated here.
Still, for larger lags the pronounced tails flatten and the increment distribution slowly approaches a Gaussian distribution. The non-Gaussian increments on a short time scale point to non-Gaussian driving forces, e.g. in terms of generation or demand fluctuations acting on the power grid.

To investigate the deviations of the frequency increments from Gaussian properties, we utilise the excess kurtosis $\kappa-3$ of the distribution.
Since the kurtosis $\kappa$, the normalised fourth moment of a distribution, is  $\kappa^\text{Gauss}=3$ for a Gaussian distribution, a positive excess kurtosis points to heavy tails of the distribution.

Computing the excess kurtosis $\kappa-3$ for all our data sets, we observe variable degrees of deviation across the various synchronous areas (Fig.~\ref{fig:Increment_Analysis}).
In some areas, the intermittent behaviour of the increments $\Delta f_\tau$ is subdued and the overall distribution approaches a Gaussian distribution (in EE, DE, SE, RUS, and US-UT), i.e., the excess kurtosis $\kappa-3$ becomes very small ($\lesssim 10^0$).
In contrast, all islands as well as GB, US-TX, and ZA display large and non-vanishing intermittent behaviour, with a large excess kurtosis ($\sim 10^1...10^2$).
Iceland (IS), as well as Gran Canaria (ES-GC) show impressive deviations from Gaussianity, which require detailed modelling in the future. 

\begin{figure*}[ht]
\includegraphics[width=\linewidth]{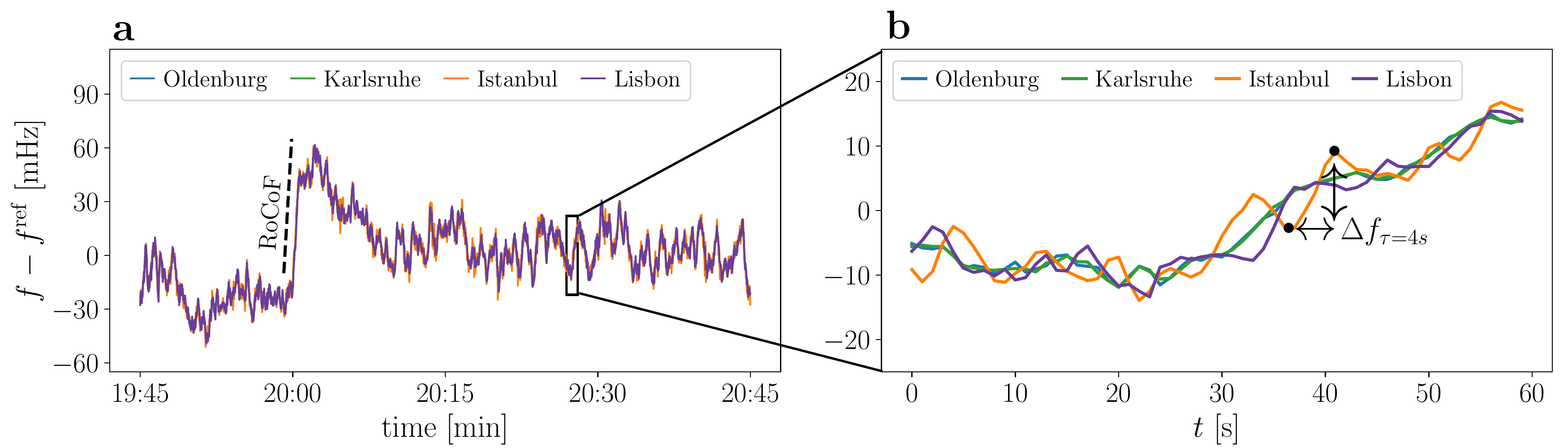}
\caption{Synchronised measurements within the Continental European (CE) synchronous area differ on the short time scale.
We show a $1$ hour frequency trajectory recorded at four different sites in the CE area: Oldenburg, Karlsruhe, Lisbon, and Istanbul.
We further illustrate the RoCoF (rate of change of frequency) as the slope of the frequency every hour and the increment statistics $\Delta f_\tau$ as the frequency difference between two points with time lag $\tau$.
For clarity, we do not include the two Hungarian measurement sites here, which produce similar results.}
\label{fig:SynchronousMeasurements_Trajectory}
\end{figure*}

We summarise that smaller regions tend to display more intermittency in their increments than larger regions, again consistent with findings on the scaling of the aggregated noise amplitude $\epsilon$ (Fig.~\ref{fig:Scaling}).
Furthermore, we observe that increment distributions tend to approach Gaussian distributions for larger increments, as expected \cite{Anvari2016},  but with distinct time horizons that depend on the grid area.
For most of the islands the excess kurtosis remains high even for lags of ten seconds.
Contrary, in most areas of continental size the excess kurtosis is very small already for lags larger than one second.
Very interesting is also the following observation: Non-Gaussian distributions in the aggregated frequency statistics (Fig.~\ref{fig:Histograms_Autocorrelation}) are not necessarily linked with non-Gaussian increments.
For example in Continental Europe (DE) we observe Gaussian increments but a non-Gaussian aggregated distribution. The deviation from Gaussianity in the aggregated distribution, e.g. in terms of frequent extreme events, is likely explained by the external drivers, such as market activities \cite{schafer2018isolating}.
Finally, this analysis presented here extends previous increment analyses \cite{Haehne2018,HaehnePropagation2019}, which only considered increments of less than a second ($\tau<1~s$), while we observe relevant non-Gaussian behaviour for larger increments ($\tau \geq 1~s$). 
We further analyse the differences between aggregated kurtosis and increment kurtosis in Supplementary Note 1 and discuss Castaing's model \cite{Castaing1990} and superstatistics \cite{Beck2003} as more theoretical approaches towards increment analysis in Supplementary Note 3.

\subsection*{Correlated dynamics within one area}

\begin{figure*}[ht]
\includegraphics[width=0.45\linewidth]{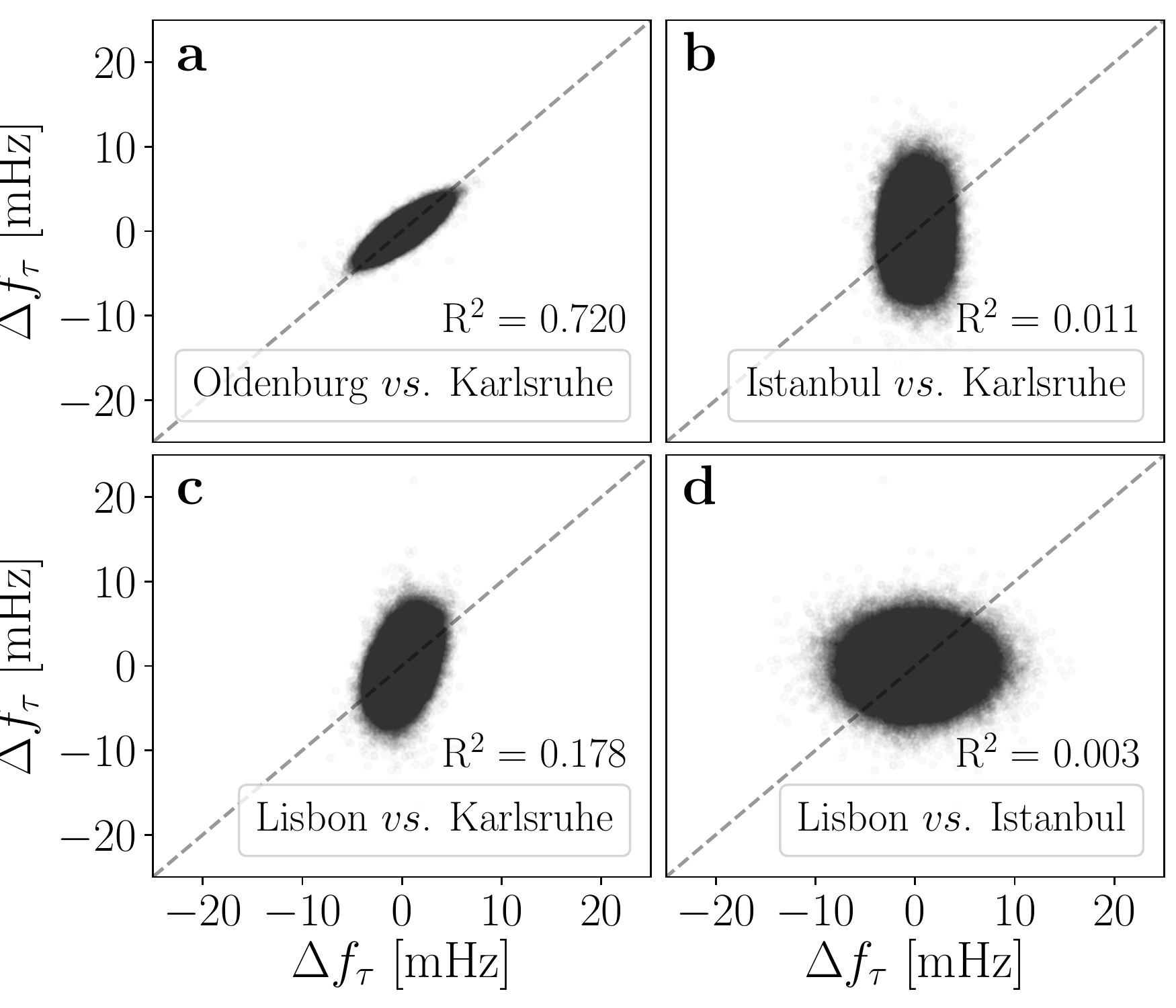}
\includegraphics[width=0.45\linewidth]{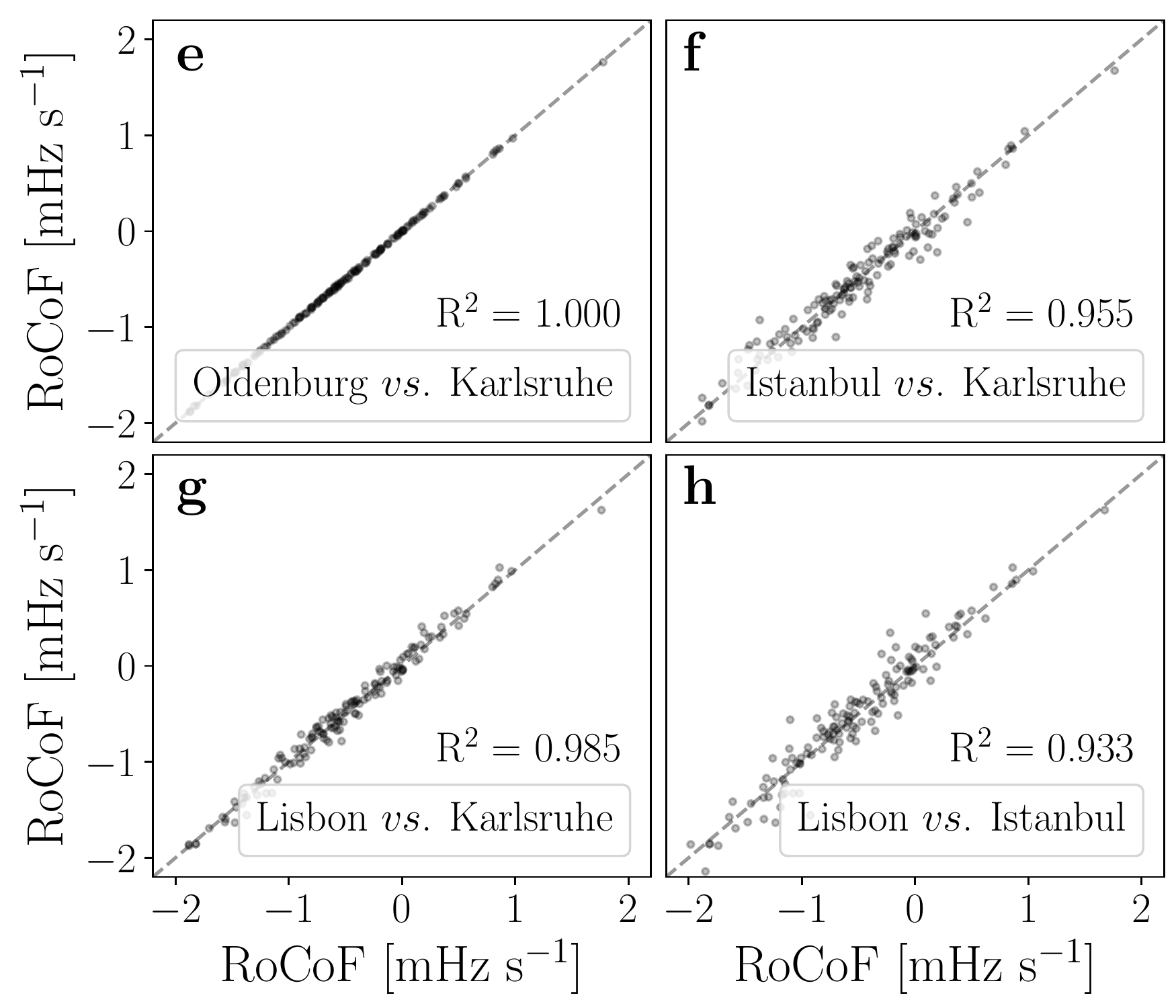}
\caption{From localised fluctuations to bulk behaviour. From left to right, we move our focus from short time scales (increments) to long time scales (RoCoF). 
\textbf{a-d:} Short time increments are mostly independent.
We compute the increment statistics $\Delta f_\tau=f(t+\tau)-f(t)$ for the increment time $\tau=1$\!~s at the four sites in Continental Europe.
The squared correlation coefficient $R^2$ is rounded to 4 digits.
See also Supplementary Note 4 for larger lags $\tau$ and more details.
\textbf{e-h:} Correlations at the long time scale.
We record the estimated rate of change of frequency (RoCoF) $\text{d}f/\text{d}t$ every $60$ minutes for all four grid locations. 
In the scatter plots, each point represents the RoCoF or increment value $\Delta f_\tau$ computed at two different locations at the same time $t$.
\label{fig:SynchronousMeasurements_ROCOF_Spectrum_Increments}
}
\end{figure*}

\begin{figure}[ht]
\includegraphics[width=1\linewidth]{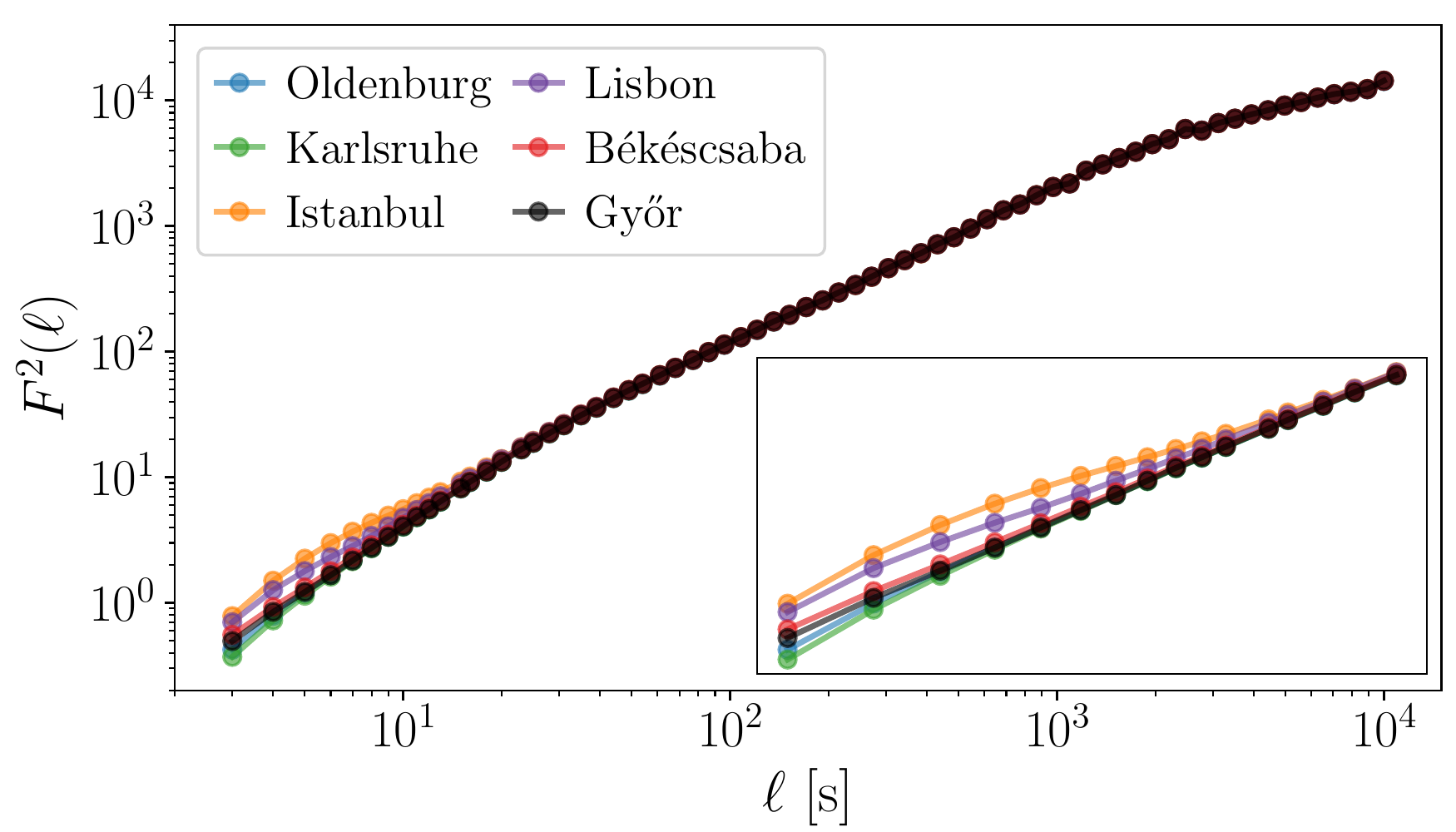}
\caption{Detrended Fluctuation Analysis (DFA) connects short and long time scales.
We perform a DFA~\cite{MFDFA2020}, with order $m=1$, in accordance with Ref.~\cite{meyer2020identifying}, and plot the fluctuation function $F^2 (\ell)$ as a function of the time window length $l$.
The inset magnifies the values for $\ell \in\{ 10^0\dots 2 \cdot 10^1 \} $.
The lines connect data points to each other to guide the eye.
\label{fig:SynchronousMeasurements_DFA}}
\end{figure}

Moving away from comparing individual synchronous areas, we use GPS-synchronised measurements at multiple locations within the same synchronous area,
the Continental European (CE) area, marked as diamonds and triangles in Fig.~\ref{fig:Map}.
These measurements reveal that the frequency at different locations is almost identical on long time scales but differs on shorter time scales, see Fig.~\ref{fig:SynchronousMeasurements_Trajectory}.
While the trajectories of the two German locations, Oldenburg and Karlsruhe, are almost identical, there are visible oscillations between the frequency values recorded in Central Europe (Karlsruhe) compared to the values recorded in the peripheries (Istanbul and Lisbon). 

Let us quantify this by analysing the time series at the time scale of $1$ second and hours, see Fig.~\ref{fig:SynchronousMeasurements_ROCOF_Spectrum_Increments}. Increments $\Delta f_\tau$, as also introduced above, reveal the short-term variability of a time series. In addition, we measure the long-term correlations on a time scale of hours by determining the rate of change of frequency (RoCoF). The RoCoF is the temporal derivative of the frequency and thereby very similar to increments. However, here it has a very different meaning as we evaluate it only at every full hour and take into account several data points, see \cite{gorjao_data-driven_2020} and Methods. Thereby, the RoCoF mirrors the hourly power dispatch \cite{Weissbach2009} and gives a good indication of long-term dynamics and deterministic external forcing.
In the next section, we will also investigate the intermediate time scale of several seconds and inter-area oscillations.

Short time scale dynamics, as determined by frequency increments $\Delta f_\tau$, are almost independent on the time scale of $\tau = 1$ second, see Fig.~\ref{fig:SynchronousMeasurements_ROCOF_Spectrum_Increments} a-d. We generate scatter plots of the increment value $\Delta f_\tau(t)$ at the same time $t$ at two different locations. 
If the increments are always identical, all points should lie on a straight line with slope 1. If the increments are completely uncorrelated, we would expect a circle or an ellipse aligned with one axis.
Indeed, the increments taken at the same time for Oldenburg and Karlsruhe are highly correlated and almost always identical, i.e., the points in a scatter plots follow a narrow tilted ellipse (Fig.~\ref{fig:SynchronousMeasurements_ROCOF_Spectrum_Increments} a).
Moving geographically further away from Karlsruhe, the increments of Istanbul (Fig.~\ref{fig:SynchronousMeasurements_ROCOF_Spectrum_Increments} b)  are completely uncorrelated with those recorded in Karlsruhe, i.e., large frequency jumps in Istanbul may take place at the same time as small jumps happen in Karlsruhe.
A similar picture of uncorrelated increments emerges when comparing Lisbon and Istanbul  (Fig.~\ref{fig:SynchronousMeasurements_ROCOF_Spectrum_Increments} d), while Lisbon vs. Karlsruhe displays some small correlation (Fig.~\ref{fig:SynchronousMeasurements_ROCOF_Spectrum_Increments} c). 
At the two peripheral locations, Lisbon and Istanbul, the increment distributions are much wider, i.e., larger jumps on a short time scale are much more common in Istanbul and Lisbon than they are in Karlsruhe. For larger lags $\tau>1~s$, the increments between all pairs become more correlated, see Supplementary Note 4.

Let us move to longer time scales. At the $60$ minute time stamps, power is dispatched in the CE grid to match the current demand, leading to a sudden surge in the frequency \cite{Weissbach2009, gorjao_data-driven_2020, anvari_stochastic_2020}.
Interestingly, the frequency dynamics at the different grid sites are very similar, i.e., the deterministic event of the power dispatch is seen unambiguously everywhere in the synchronous area, almost regardless of distance, see Fig.~\ref{fig:SynchronousMeasurements_ROCOF_Spectrum_Increments}.
All locations closely follow the same trajectory on the $1$ hour time scale.
This is reflected in highly correlated RoCoF values, with a particularly good match between Oldenburg and Karlsruhe and a linear regression coefficient of at least $R^2 \geq 0.93$ for all pairs (Fig.~\ref{fig:SynchronousMeasurements_ROCOF_Spectrum_Increments} a-d). 

We combine these different time scales in a single detrended fluctuation analysis (DFA), where we also integrate the two Hungarian locations, see Fig.~\ref{fig:SynchronousMeasurements_DFA}.
At short time scales, the DFA results differ for the four locations, while starting at the time scale of $t \sim  10^1$ seconds, the four curves coincide.
For the time scale of $1$ second, all locations are subject to different fluctuations, with Istanbul and Lisbon displaying the largest values of the fluctuation function. This is coherent with results of the increment analysis, where  Istanbul and Lisbon have the broadest increment distributions (Fig.~\ref{fig:SynchronousMeasurements_ROCOF_Spectrum_Increments} a-d).
Moving to longer time scales of tens or hundreds of seconds, we observe a coincidence of the fluctuation function. This coincidence, i.e., identical behaviour for large time scales is in good agreement with the highly correlated RoCoF results (Fig.~\ref{fig:SynchronousMeasurements_ROCOF_Spectrum_Increments} e-h). We may also interpret this change from short term and localised dynamics to long term and bulk behaviour as a change from stochastic to deterministic dynamics, i.e., the random fluctuations are localised and take place on a short time scale, while the deterministic dispatch actions and overall trends penetrate the whole grid on a long time scale. See also Methods and Supplementary Note 5 for details on the DFA methodology.

\subsection*{Spatio-temporal dynamics}

\begin{figure*}[ht]
\includegraphics[width=0.49\linewidth]{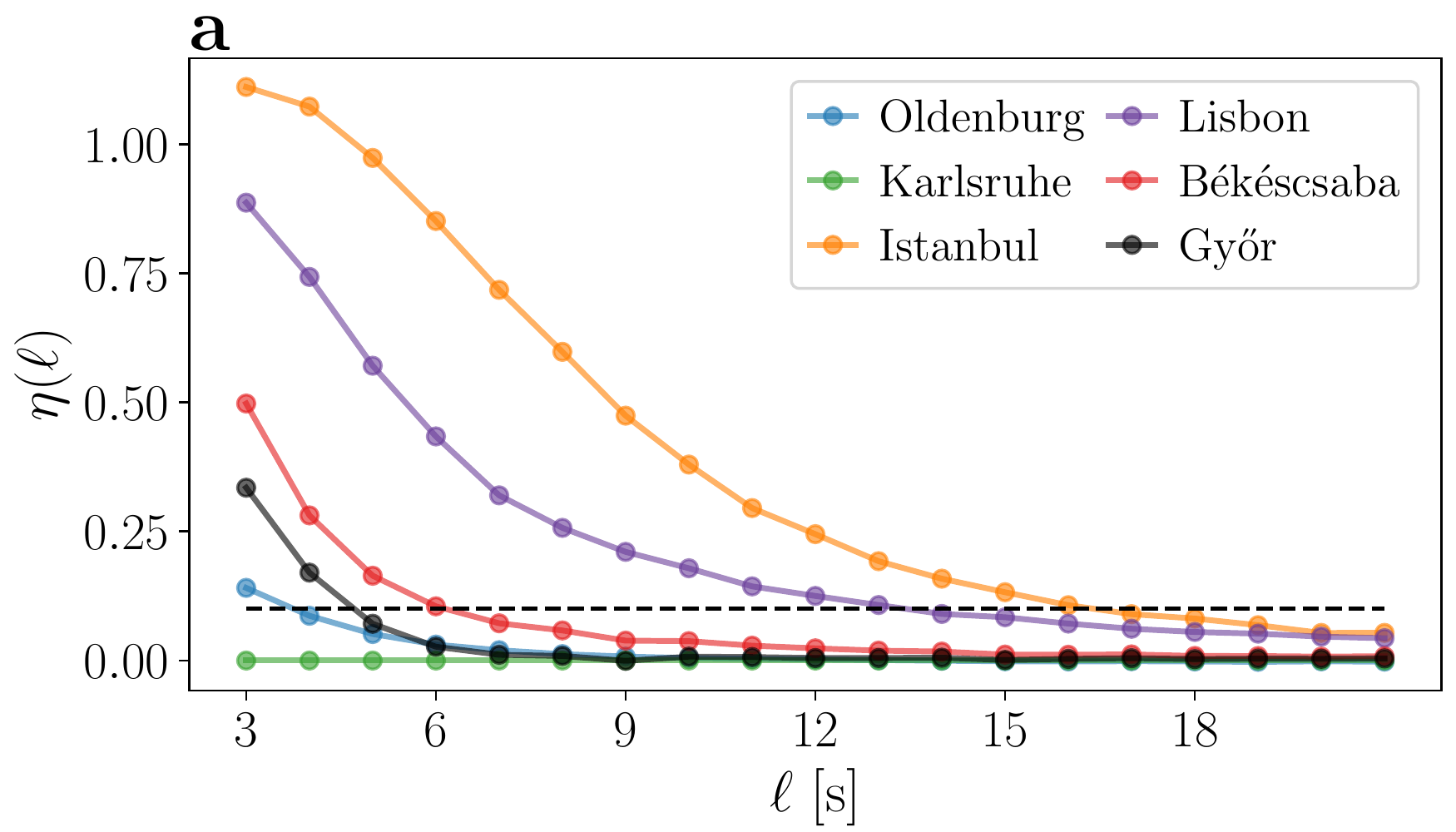} \includegraphics[width=0.49\linewidth]{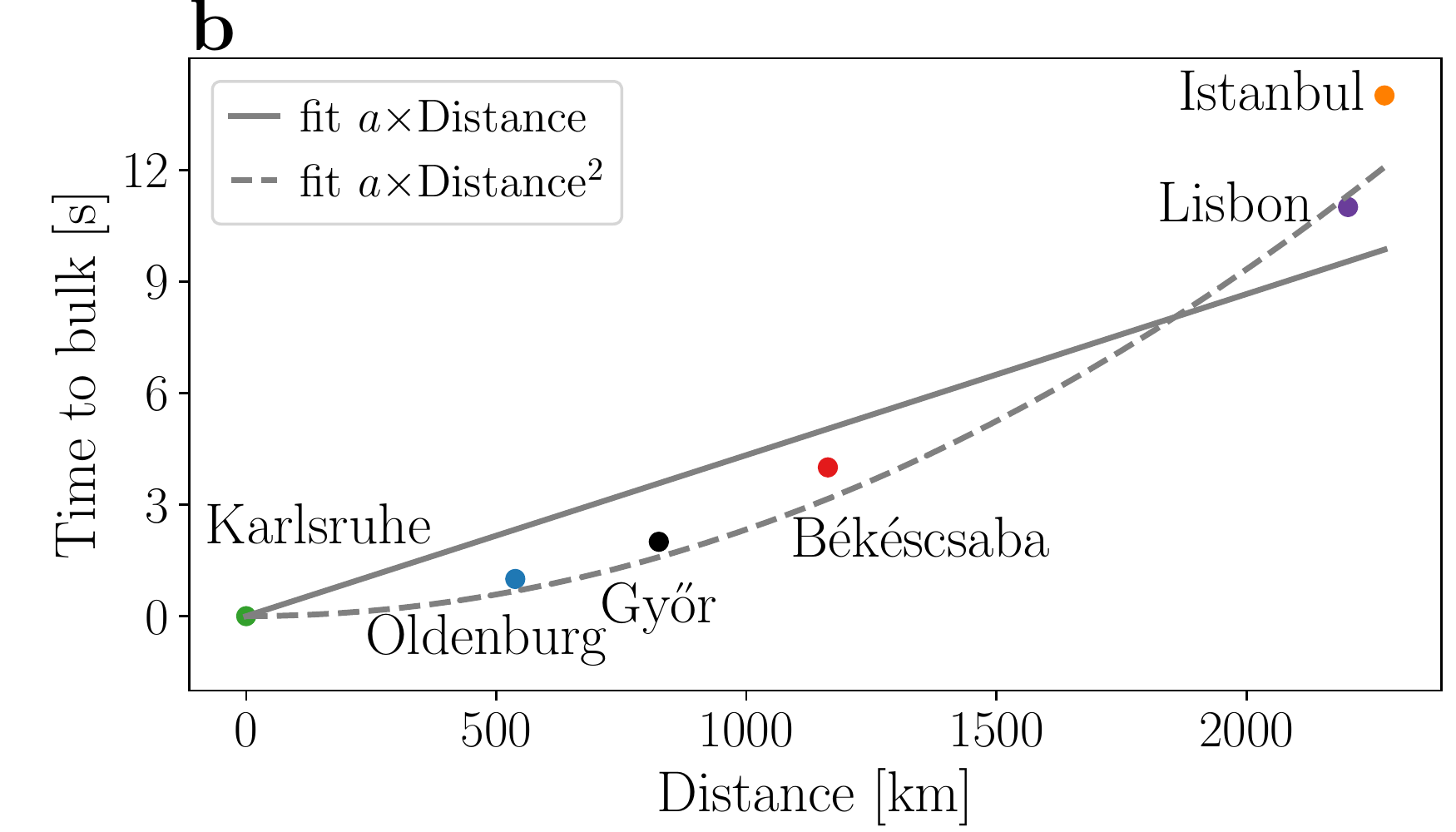}
\caption{The time-to-bulk increases with distance. a: The relative DFA function $\eta(\ell)$ with Karlsruhe as reference; we determine the time-to-bulk as the time when this value reaches $0.1$ (dashed line), see also eq.~\eqref{eq:relative_DFA}.
b: We plot the so extracted time-to-bulk vs. the distance from Karlsruhe and provide a simple fit for the first five points (i.e. excluding Istanbul as it clearly behaves differently).
We obtain a value for the linear fit $a=5.2\times10^{-3}$\!~s\! km$^{-1}$, see Methods for details on the distance.
}
\label{fig:Time_to_bulk}
\end{figure*}

Next, let us investigate the spatio-temporal aspect of the synchronised measurements.
We connect the transition from local fluctuations towards bulk behaviour with the geographical distance of the measurement points, complementing earlier analysis based on voltage angles \cite{cresap1981emergence,chompoobutrgool2012identification}.
We determine the typical time-to-bulk, i.e., the time necessary so that the dynamics at a given node approximates the bulk behaviour.
To this end, we choose Karlsruhe, Germany, as our reference, which is very central within the Continental European synchronous area.
The choice of the reference does not qualitatively change the results. For each of the remaining five locations, we compute the relative DFA function
\begin{equation} \label{eq:relative_DFA}
    \eta(\ell) = \frac{{F^2}^\text{location}(\ell)-{F^2}^\text{Karlsruhe}(\ell)}{{F^2}^\text{Karlsruhe}(\ell)}
\end{equation} 
with respect to Karlsruhe and ask, when does this difference drop below $0.1$ (or $10\%$), i.e., when are the fluctuation at $\text{location}$ almost indistinguishable from the ones in Karlsruhe?

The further apart two locations are, the later they reach the bulk behaviour, i.e., the larger their time-to-bulk, see Fig.~\ref{fig:Time_to_bulk}. 
This observation can be intuitively understood: Two sites in close geographical vicinity are typically tightly coupled and can be synchronised by their neighbours, whereas sites far away have to stabilise on their own.
Our time-to-bulk analysis quantifies this intuition. 
We consider both a linear and a quadratic fit. A linear dependence is expected if the bulk behaviour is realised by coupling via the shortest available path. In contrast, if the propagation is following a diffusive pattern via multiple independent paths, we would expect a quadratic dependence of the time with respect to the distance.
Indeed, the quadratic fit, following diffusive coupling, is a much better fit than a linear one, as indicated by a lower 
Root-Mean-Squared-Error $0.5$, compared to $1.2$ seconds in the linear case.
Using the newly obtained fits, we find that a location only $100$~km from Karlsruhe will have to independently stabilise fluctuations on the scale of $0.5$ to $1$ second and will then closely synchronise with the dynamics in Karlsruhe (our bulk reference).
Contrary, a site $1000$~km away has to stabilise already for about $3$ to $5$ seconds before it is fully integrated in the bulk. 
This gives additional guidance for the control within large synchronous areas, in particular for remote and weakly coupled sites.
Clearly, these first estimates demonstrate that further research is necessary to validate and adjust spatio-temporal models of the power grid \cite{zhang2019fluctuation}.

\subsection*{Principal Component Analysis}
\begin{figure*}[t]
\includegraphics[width=1\linewidth]{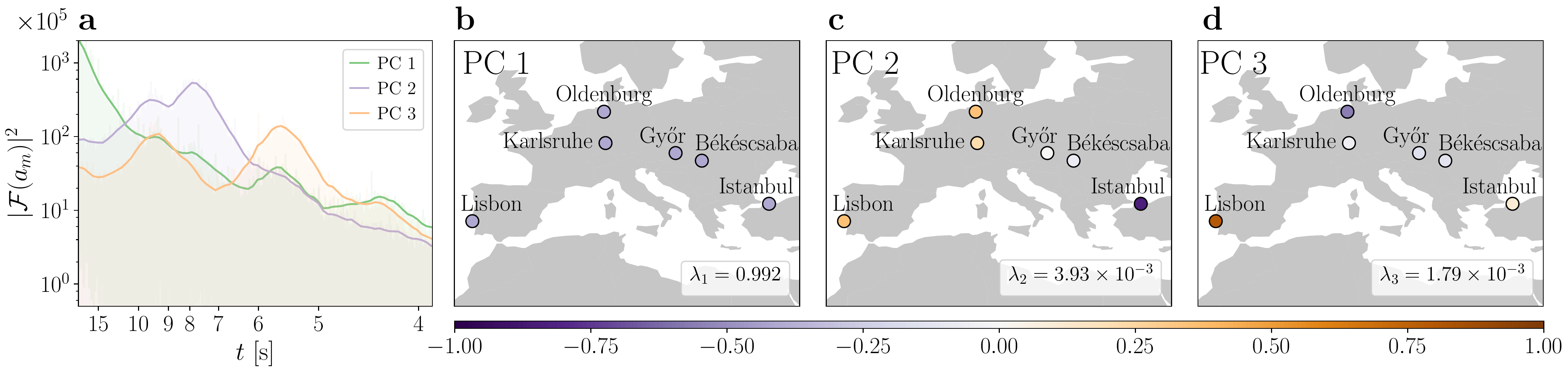}
\caption{Principal component analysis (PCA) of frequency recordings reveals inter-areas oscillations. 
\textbf{a}: In Continental Europe, the squared Fourier amplitudes $\left|\mathcal{F}\left(a_m(t)\right)\right|^2$ of the three dominant principal components (PCs) exhibits typical period lengths of $t\approx 7$s and $t\approx4.5$s.
\textbf{b}: The first spatial mode (PC1) corresponds to the bulk behaviour of the frequency and explains already $\lambda_1=99.2\%$ of the total variance.
Contrary, the second (PC2) and third (PC3) mode reveal asynchronous inter-area modes (c-d).
We refer to Supplementary Note 6 for details on the method and the results.} 
\label{fig:PCA_modes}
\end{figure*}

So far, we have focused on when and how the localised fluctuations transition into a bulk behaviour.
During this transition, on the intermediate time scale of about $5$ seconds, we observe another phenomenon: 'Inter-area oscillations', i.e., oscillations between sites in different geographical areas far apart but still within one synchronous area. 
Different methods are available to extract spatial inter-area modes, ranging from Empirical Mode Decomposition \cite{messina_extraction_2007} to nonlinear Koopman modes \cite{susuki_nonlinear_2011}.
Here, we use a principal component analysis (PCA)~\cite{bishop_pattern_2007}, which was already introduced to power systems when analysing inter-area modes and identifying coherent regions \cite{anaparthi_coherency_2005}.
A PCA separates the aggregated dynamics observed in the full system into ordered principal components, which we interpret as oscillation modes.
Ideally, we can explain most of the observed dynamics of the full system by interpreting a few dominant modes.
Each of these modes contains information of which geographical sites are involved in the modes dynamics, similar to an eigenvector.
Typical behaviour includes a translational dynamics of all sites (the eigenvector with entries 1 everywhere) or distinct oscillations between individual sites (an eigenvector with entry 1 at one site and -1 at another site). 

Indeed, applying a PCA to the synchronised measurements in CE, we can capture almost the entire dynamics with just three modes, see Fig.~\ref{fig:PCA_modes}. 
In Fig.~\ref{fig:PCA_modes} a, we provide the squared Fourier amplitudes of each mode and in Fig.~\ref{fig:PCA_modes} b-d we visualise the first three modes geographically.
These three modes already explain the largest shares $\lambda_m$ of the total variance, see Supplementary Note 6 for the remaining modes and more details.
The first mode (PC1) explains $\lambda_1\approx99.2\%$ of the variance and represents the synchronous bulk behaviour of the frequency.
The second (PC2) and third (PC3) mode correspond to asynchronous inter-area modes.
They contribute much less to the total variance due to their small amplitude (cf. Fig.~\ref{fig:SynchronousMeasurements_Trajectory}).
In PC2 (Fig.~\ref{fig:PCA_modes} c), Western Europe forms a coherent region that is in phase opposition to Istanbul (East-West dipole), while in PC3 (Fig.~\ref{fig:PCA_modes} d), Lisbon and Istanbul swing in opposition to Oldenburg (North-South dipole).
Similar results were found in an earlier theoretical study of the CE area, which also revealed global inter-area modes with dipole structures \cite{grebe_low_2010}.

The temporal dynamics of the spatial modes exhibit typical frequencies of inter-area oscillations.
Fig.~\ref{fig:PCA_modes} a shows the squared Fourier amplitudes $\left|\mathcal{F}\left(a_m(t)\right)\right|^2$ of the spatial modes.
The components PC2 and PC3 have their largest peaks at $t\approx7$\!~s and $t\approx 4.5$\!~s, which are the periods of these inter-area modes.
These periods correspond well to the typical periods of inter-area oscillations, which are reported to be $1.25$--$8$\!~s \cite{Klein1991}.
On larger time scales $t>12$\!~s, the amplitudes $\left|\mathcal{F}\left(a_m(t)\right)\right|^2$ of the inter-area modes drop below the values of PC1.
Thus, the frequency dynamics is dominated by the its bulk behaviour again, which is consistent with the estimated time-to-bulk of $12$-$15$s (Fig.~\ref{fig:Time_to_bulk}).

\section*{Discussion}
In this article, we have presented a detailed analysis of a recently published open data base of power grid frequency measurements \cite{Jumar2020Data}.
We have compared various independent synchronous areas, from small regions, such as the Faroe Islands and Mallorca, to large synchronous areas, like the Western Interconnection in North America and the Continental European grid, spanning areas with only tens of thousand customers to those with hundreds of millions.
Especially the smaller areas tend to show a larger volatility in terms of aggregated noise but also increment intermittency, such as Iceland and Gran Canaria.
We have complemented this analysis of independent grids by GPS-synchronised measurements within the Continental European power grid, revealing high correlations of the frequency at long time scales but mostly independent dynamics on fluctuation-dominated short time scales.
Compared to other studies applying synchronised, wide-area measurements, such as  FNET/Grideye in the US \cite{chai_wide-area_2016} or evaluations from Iceland \cite{tuttelberg2018estimation}, the data we analysed here is freely available for further research \cite{Jumar2020Data}.

The comparison of different synchronous areas gives us a solid foundation to test previously conjectured scaling laws of fluctuations in power grids with their size \cite{Schaefer2017a}, helps us to develop synthetic models \cite{gorjao_data-driven_2020} or predict the frequency \cite{Kruse2020Predict} of small grids, such as microgrids.
Furthermore, aggregating standardised measurements from different areas, we can compare countries with high shares of renewables (high wind penetration in Iceland) with areas with almost no renewable generation (Mallorca) to learn how they influence the frequency dynamics and thereby the power grid stability.
Similarly, this comparison also gives insights on how different market structures impact the frequency statistics and stability of a power grid.

Our results on the spatial dependencies in the Continental European (CE) synchronous area are also highly relevant for the operation of power grids and other research in the field.
The observations that the long term behaviour is almost identical throughout the synchronous area but short time fluctuations differ, are in agreement with earlier theoretical findings \cite{zhang2019fluctuation}.
Based on the DFA results (Figs.~\ref{fig:SynchronousMeasurements_DFA} and \ref{fig:Time_to_bulk}), we provide a quantitative estimate that at least for the CE area already at time scales of about ten seconds, we observe an almost uniform bulk behaviour, even for locations thousands of kilometres apart.
This bulk behaviour emerges much faster when locations are closer to one another. 

In the regime of resonant behaviour \cite{zhang2019fluctuation}, we observe inter-area oscillations with period lengths of $t=7$\!~s and $t=4.5$\!~s, which we extract using a principal component analysis (PCA).
These time scales agree well with frequencies of inter-area oscillations reported in other studies in Europe \cite{Klein1991, grebe_low_2010, vanfretti2013spectral} but also in the US \cite{cui_inter-area_2017}.  However, we notice that the time scales separating bulk, resonance and local behaviour are different than the authors in a theoretical work \cite{zhang2019fluctuation} assumed.
There, local fluctuations were described for the $0.1$ second time scale and bulk dynamics already started at times between about $2$\!~s and $5$\!~s. This raises the question on how these time scales depend on the size and the dynamics of the power grid under consideration.
Finally, we note that the PCA is a prime example for a model-free and data-driven analysis that still allows interpretations. 

Our observation of frequency increments being independent on time scales of one second is consistent with earlier studies \cite{Haehne2018}.
For Continental Europe, we find that $1$\!~s-increments are correlated at small distances (below $500$~km), but independent at locations far apart.
On time scales of one second and below, we cannot observe global inter-area modes anymore. Instead, we expect local fluctuations that quickly decay with distance to their origin \cite{zhang2019fluctuation, HaehnePropagation2019}, which is consistent with our findings.
The distribution of these short term fluctuation was reported to exhibit a strongly non-Gaussian distribution when subject to intermittent wind power feed-in \cite{Haehne2018}.
In agreement with these results, the non-Gaussian effects vanish on time scales above one second in our recordings from Continental Europe.
However, in other, particularly smaller, synchronous areas we even observe heavy-tailed increment distributions on times scales up to $10$\!~s.
This is likely related to the grid size and control regulations, although a detailed explanation still remains open.

In this paper, we connect the mathematics and physics communities with the engineering community, by providing potent data analysis tools from the theoretical side and then connecting these findings in the practical domain of power grid dynamics without the use of an explicit model.
Both the data analysis and its interpretation could be very useful for the operation of individual grids.
Our insights for the scaling could be used to improve control mechanisms, such as demand side management \cite{tchuisseu2017}, while our spreading insights give further indications about how fast cascading failures will spread throughout the power grid \cite{schafer2019dynamical}.
Several grid operators as well as other researchers have likely recorded power grid frequency time series at many more grid locations than we could provide in this single study.
All such recordings from different sources should be combined to enable more comparisons between the dynamics of synchronous areas of different sizes and under different conditions.
The data base studied here \cite{Jumar2020Data} may offer a valuable starting point for such endeavours.

Because data are still only scarcely available, there remain many open questions: Can we systematically determine a propagation velocity of disturbances through the grid and compare these with theoretical predictions \cite{pagnier2019inertia,zhang2019fluctuation, schroeder2019dynamic}?
Can we identify other time series influencing the power grid frequency dynamics and quantify their correlation such as hydro power plants in the Nordic area or demand of aluminium plants in Iceland?
Can we extract the impact of market activities on the frequency dynamics in all synchronous areas?
From a statistical modelling perspective, it would be interesting to investigate the  scaling of higher moments, i.e. skewness and kurtosis, with time lag and size in more detail.
These questions constitute only a small selection from a multitude that an open data base may help to address from a broad, interdisciplinary perspective, including engineering, mathematics, data science, time series analysis, and many other fields.

\section*{Methods}
\subsection*{Data selection}

We make use of the open data base, described in detail in \cite{Jumar2020Data} to perform all analysis presented in the main text and in Supplementary Notes 1 to 6. This data set contains recordings of twelve independent synchronous regions recorded between 2017 and 2020. While some locations, such as the Faroe Islands only contain a single week of data, other regions, such as Continental Europe have been monitored for several months or years, for more details see \cite{Jumar2020Data}.
However, due to some technical difficulties, e.g. loss of GPS signal or unplugging the device, some measurements are not a number, i.e., "NaN" and are tagged as not reliable in the data base.
These entries have been deleted to compute the histograms and statistical measures in Supplementary Note 1.
To compute the autocorrelation function, as well as for the analysis of the synchronised measurement in Continental Europe, we selected the longest possible trajectory without any "NaN" entries. 
As a final note: From the available Gran Canaria data, we are using the March 2018 data.

\subsection*{RoCoF computation}
When determining the rate of change of frequency (RoCoF), i.e., the time derivative of the frequency, we follow the same procedure as has been outlined in \cite{gorjao_data-driven_2020}:
We select a short time window centred around the anticipated dispatch jumps at $60$~minutes of about $25$ seconds length, i.e. starting at (X):59:48 and lasting until (X+1):00:12 for all hours X.
Then, we fit this short frequency trajectory with a linear function $f(t)=a + b t$.
We are not interested in the offset $a$ but the value of $b$ gives us the slope of the frequency changes, i.e., the time derivative of the frequency is approximately given as $\frac{\text{d}f}{\text{d}t} \approx b$.

\subsection*{Detrended Fluctuation Analysis (DFA)}
To carry out the detrended fluctuation analysis (DFA), we follow a similar procedure as outlined in \cite{meyer2020identifying}, using the package outlined in ref. \cite{MFDFA2020}.
The main idea is to detrend the data and extract the most dominant time scales by measuring the scaling behaviour of the data from increasing segments of data.
The commonly denoted fluctuation function $F^2(\ell)$, function of the segment size $\ell$ on the time series, accounts for the variance of segmented data of increasing size.
The scaling of the underlying process or processes can thus be extracted.
In \cite{meyer2020identifying} a detailed study of the different time scales in power grid frequencies can be found, largely focusing on scales of about ten seconds and above, whereas we put particular emphasis on the smallest time scales available, of the order of $1$ second. More details are given in Supplementary Note 5.

\subsection*{Time-to-bulk}
To extract the time-to-bulk, seen in Fig.~\ref{fig:Time_to_bulk}, we take the measurements of the Detrended Fluctuation Analysis (DFA) in Fig.~\ref{fig:SynchronousMeasurements_DFA} and utilise Karlsruhe as the reference for comparison.
Having Karlsruhe as a reference, we compare the normalised fluctuations $\eta(\ell)$:
\begin{equation} 
    \eta(\ell) = \frac{{F^2}^\text{location}(\ell)-{F^2}^\text{Karlsruhe}(\ell)}{{F^2}^\text{Karlsruhe}(\ell)},
\end{equation}(eq. \eqref{eq:relative_DFA} in the main text), to extract the excess fluctuation at the different locations.
As there is no standard, we choose a threshold value of $10\%$ for fluctuations at the different recordings to be identical.
Once $\eta(\ell)$ drops below this threshold of $10\%$, the datasets are considered to be identical.
In this manner, we determine the time-to-bulk as the necessary time of a recording to exhibit the same fluctuation behaviour as the reference of Karlsruhe.
The distance measures taken are the geographic distances with respect to Karlsruhe, applying OpenStreet Maps \url{https://www.openstreetmap.org/} and using the routing by Foot(OSRM).
This yields the following distances from Karlsruhe:  Oldenburg: $538$\!~km, Győr: $825$\!~km, Békéscsaba: $1163$\!~km, Lisbon:  $2203$\!~km, Istanbul: $2276$\!~km.
The reason to use route finding by foot is that the power grid is not taking any air plane routes but is limited also to the shortest routes available in the transmission grid.
These distances in the power system might be even longer where transmission line density is low.
Note that our choice of geographical distance does not apply any assumption on the underlying power grid topology.
With fully (yet currently unavailable) information about all operational transmission lines, a shortest path distance on the transmission network would be an alternative \cite{HaehnePropagation2019}.

\subsection*{Data availability}
Frequency recordings are described in detail in \cite{Jumar2020Data}. An open repository containing all recordings can be accessed here: \url{https://osf.io/by5hu/}.
The Hungarian TSO data is available here: \url{https://osf.io/m43tg/}.
All data that support the results presented in the figures of this study are available from the authors upon reasonable request.

\subsection*{Code availability}
Code to produce the presented analysis and figures is available on github: \url{https://github.com/LRydin}

\begin{acknowledgments}
We would like to express our deepest gratitude to everyone who helped create the data base by connecting the EDR at their hotel room, home, or office: 
Damià Gomila, Malte Schröder, Jan Wohland, Filipe Pereira, André Frazão, Kaur Tuttelberg, Jako Kilter, Hauke Hähne, and Bálint Hartmann. 
We gratefully acknowledge support from the Federal Ministry of Education and Research (BMBF grants No 03SF0472 and 03EK3055), the Helmholtz Association (via the joint initiative \textit{Energy System 2050 - A Contribution of the Research Field Energy} and the grant No VH-NG-1025), the associative \textit{Uncertainty Quantification – From Data to Reliable Knowledge (UQ)} with grant No ZT-I-0029, the German Science Foundation (DFG) by a grant toward the Cluster of Excellence \textit{Center for Advancing Electronics Dresden} (cfaed), and the Scientific Research Projects Coordination Unit of Istanbul University, Project No 32990.
This work was performed as part of the Helmholtz School for Data Science in Life, Earth and Energy (HDS-LEE).
This project has received funding from the European Union’s Horizon 2020 research and innovation programme under the Marie Skłodowska-Curie grant agreement No 840825.
\end{acknowledgments}

\subsection*{Author contributions}
L.R.G., R.J., H.M., V.H., G.C.Y., J.K., M.T., C.B., D.W., and B.S. conceived and designed the research. R.J. and H.M. constructed the measurement device and evaluated the experimental data. L.R.G., J.K., and B.S. performed the data analysis and generated the figures. All authors contributed to discussing and interpreting the results and writing the manuscript. 

\subsection*{Competing interests}
The Authors declare no Competing Financial or Non-Financial Interests.

\bibliographystyle{naturemag}
\addcontentsline{toc}{section}{\refname}
\bibliography{References}

\begin{thebibliography}{10}
\expandafter\ifx\csname url\endcsname\relax
  \def\url#1{\texttt{#1}}\fi
\expandafter\ifx\csname urlprefix\endcsname\relax\def\urlprefix{URL }\fi
\providecommand{\bibinfo}[2]{#2}
\providecommand{\eprint}[2][]{\url{#2}}

\bibitem{sawin2018renewables}
\bibinfo{author}{Sawin, J.~L.} \emph{et~al.}
\newblock \bibinfo{title}{Renewables 2018--{G}lobal status report}
  (\bibinfo{year}{2018}).

\bibitem{meadowcroft2009politics}
\bibinfo{author}{Meadowcroft, J.}
\newblock \bibinfo{title}{What about the politics? {S}ustainable development,
  transition management, and long term energy transitions}.
\newblock \emph{\bibinfo{journal}{Policy Sciences}}
  \textbf{\bibinfo{volume}{42}}, \bibinfo{pages}{323} (\bibinfo{year}{2009}).

\bibitem{Markard2018}
\bibinfo{author}{Markard, J.}
\newblock \bibinfo{title}{The next phase of the energy transition and its
  implications for research and policy}.
\newblock \emph{\bibinfo{journal}{Nature Energy}} \textbf{\bibinfo{volume}{3}},
  \bibinfo{pages}{628--633} (\bibinfo{year}{2018}).

\bibitem{rodriguez2014business}
\bibinfo{author}{Rodr{\'\i}guez-Molina, J.},
  \bibinfo{author}{Mart{\'\i}nez-N{\'u}{\~n}ez, M.},
  \bibinfo{author}{Mart{\'\i}nez, J.-F.} \& \bibinfo{author}{P{\'e}rez-Aguiar,
  W.}
\newblock \bibinfo{title}{Business models in the smart grid: Challenges,
  opportunities and proposals for prosumer profitability}.
\newblock \emph{\bibinfo{journal}{Energies}} \textbf{\bibinfo{volume}{7}},
  \bibinfo{pages}{6142--6171} (\bibinfo{year}{2014}).

\bibitem{Fang2012}
\bibinfo{author}{Fang, X.}, \bibinfo{author}{Misra, S.}, \bibinfo{author}{Xue,
  G.} \& \bibinfo{author}{Yang, D.}
\newblock \bibinfo{title}{{Smart Grids - The new and improved Power Grid: A
  Survey}}.
\newblock \emph{\bibinfo{journal}{{Communications Surveys \& Tutorials, IEEE}}}
  \textbf{\bibinfo{volume}{14}}, \bibinfo{pages}{944--980}
  (\bibinfo{year}{2012}).

\bibitem{barwaldt2018energy}
\bibinfo{author}{B{\"a}rwaldt, G.}
\newblock \bibinfo{title}{Energy revolution needs interpreters}.
\newblock \emph{\bibinfo{journal}{ATZelektronik worldwide}}
  \textbf{\bibinfo{volume}{13}}, \bibinfo{pages}{68--68}
  (\bibinfo{year}{2018}).

\bibitem{parag2016electricity}
\bibinfo{author}{Parag, Y.} \& \bibinfo{author}{Sovacool, B.~K.}
\newblock \bibinfo{title}{Electricity market design for the prosumer era}.
\newblock \emph{\bibinfo{journal}{Nature Energy}} \textbf{\bibinfo{volume}{1}},
  \bibinfo{pages}{16032} (\bibinfo{year}{2016}).

\bibitem{Kundur1994}
\bibinfo{author}{Kundur, P.}, \bibinfo{author}{Balu, N.~J.} \&
  \bibinfo{author}{Lauby, M.~G.}
\newblock \emph{\bibinfo{title}{Power system stability and control}},
  vol.~\bibinfo{volume}{7} (\bibinfo{publisher}{McGraw-Hill, New York},
  \bibinfo{year}{1994}).

\bibitem{Anvari2016}
\bibinfo{author}{Anvari, M.} \emph{et~al.}
\newblock \bibinfo{title}{Short term fluctuations of wind and solar power
  systems}.
\newblock \emph{\bibinfo{journal}{New Journal of Physics}}
  \textbf{\bibinfo{volume}{18}}, \bibinfo{pages}{063027}
  (\bibinfo{year}{2016}).

\bibitem{wolff2019}
\bibinfo{author}{Wolff, M.~F.} \emph{et~al.}
\newblock \bibinfo{title}{Heterogeneities in electricity grids strongly enhance
  non-{G}aussian features of frequency fluctuations under stochastic power
  input}.
\newblock \emph{\bibinfo{journal}{Chaos: An Interdisciplinary Journal of
  Nonlinear Science}} \textbf{\bibinfo{volume}{29}}, \bibinfo{pages}{103149}
  (\bibinfo{year}{2019}).

\bibitem{wohland2019significant}
\bibinfo{author}{Wohland, J.}, \bibinfo{author}{Omrani, N.~E.},
  \bibinfo{author}{Keenlyside, N.} \& \bibinfo{author}{Witthaut, D.}
\newblock \bibinfo{title}{{Significant multidecadal variability in German wind
  energy generation}}.
\newblock \emph{\bibinfo{journal}{Wind Energy Science}}
  \textbf{\bibinfo{volume}{4}}, \bibinfo{pages}{515--526}
  (\bibinfo{year}{2019}).

\bibitem{hartmann2019effects}
\bibinfo{author}{Hartmann, B.}, \bibinfo{author}{Vokony, I.} \&
  \bibinfo{author}{T{\'a}czi, I.}
\newblock \bibinfo{title}{Effects of decreasing synchronous inertia on power
  system dynamics--{O}verview of recent experiences and marketisation of
  services}.
\newblock \emph{\bibinfo{journal}{International Transactions on Electrical
  Energy Systems}} \textbf{\bibinfo{volume}{29}}, \bibinfo{pages}{e12128}
  (\bibinfo{year}{2019}).

\bibitem{Filatrella2008}
\bibinfo{author}{Filatrella, G.}, \bibinfo{author}{Nielsen, A.~H.} \&
  \bibinfo{author}{Pedersen, N.~F.}
\newblock \bibinfo{title}{{Analysis of a Power Grid using a Kuramoto-like
  Model}}.
\newblock \emph{\bibinfo{journal}{{The European Physical Journal B}}}
  \textbf{\bibinfo{volume}{61}}, \bibinfo{pages}{485--491}
  (\bibinfo{year}{2008}).

\bibitem{schmietendorf2014self}
\bibinfo{author}{Schmietendorf, K.}, \bibinfo{author}{Peinke, J.},
  \bibinfo{author}{Friedrich, R.} \& \bibinfo{author}{Kamps, O.}
\newblock \bibinfo{title}{Self-organized synchronization and voltage stability
  in networks of synchronous machines}.
\newblock \emph{\bibinfo{journal}{The European Physical Journal Special
  Topics}} \textbf{\bibinfo{volume}{223}}, \bibinfo{pages}{2577--2592}
  (\bibinfo{year}{2014}).

\bibitem{nishikawa2015}
\bibinfo{author}{Nishikawa, T.} \& \bibinfo{author}{Motter, A.~E.}
\newblock \bibinfo{title}{Comparative analysis of existing models for
  power-grid synchronization}.
\newblock \emph{\bibinfo{journal}{New Journal of Physics}}
  \textbf{\bibinfo{volume}{17}}, \bibinfo{pages}{015012}
  (\bibinfo{year}{2015}).

\bibitem{Rohden2012}
\bibinfo{author}{Rohden, M.}, \bibinfo{author}{Sorge, A.},
  \bibinfo{author}{Timme, M.} \& \bibinfo{author}{Witthaut, D.}
\newblock \bibinfo{title}{{Self-organized Synchronization in Decentralized
  Power Grids}}.
\newblock \emph{\bibinfo{journal}{{Physical Review Letters}}}
  \textbf{\bibinfo{volume}{109}}, \bibinfo{pages}{064101}
  (\bibinfo{year}{2012}).

\bibitem{Menck2013}
\bibinfo{author}{Menck, P.~J.}, \bibinfo{author}{Heitzig, J.},
  \bibinfo{author}{Marwan, N.} \& \bibinfo{author}{Kurths, J.}
\newblock \bibinfo{title}{{How Basin Stability Complements the Linear-stability
  Paradigm}}.
\newblock \emph{\bibinfo{journal}{{Nature Physics}}}
  \textbf{\bibinfo{volume}{9}}, \bibinfo{pages}{89--92} (\bibinfo{year}{2013}).

\bibitem{rodrigues2016kuramoto}
\bibinfo{author}{Rodrigues, F.~A.}, \bibinfo{author}{Peron, T. K.~D.},
  \bibinfo{author}{Ji, P.} \& \bibinfo{author}{Kurths, J.}
\newblock \bibinfo{title}{The {K}uramoto model in complex networks}.
\newblock \emph{\bibinfo{journal}{Physics Reports}}
  \textbf{\bibinfo{volume}{610}}, \bibinfo{pages}{1--98}
  (\bibinfo{year}{2016}).

\bibitem{schafer2017escape}
\bibinfo{author}{Sch{\"a}fer, B.} \emph{et~al.}
\newblock \bibinfo{title}{Escape routes, weak links, and desynchronization in
  fluctuation-driven networks}.
\newblock \emph{\bibinfo{journal}{Physical Review E}}
  \textbf{\bibinfo{volume}{95}}, \bibinfo{pages}{060203}
  (\bibinfo{year}{2017}).

\bibitem{hindes2019network}
\bibinfo{author}{Hindes, J.}, \bibinfo{author}{Jacquod, P.} \&
  \bibinfo{author}{Schwartz, I.~B.}
\newblock \bibinfo{title}{Network desynchronization by non-{G}aussian
  fluctuations}.
\newblock \emph{\bibinfo{journal}{Physical Review E}}
  \textbf{\bibinfo{volume}{100}}, \bibinfo{pages}{052314}
  (\bibinfo{year}{2019}).

\bibitem{zhang2019fluctuation}
\bibinfo{author}{Zhang, X.}, \bibinfo{author}{Hallerberg, S.},
  \bibinfo{author}{Matthiae, M.}, \bibinfo{author}{Witthaut, D.} \&
  \bibinfo{author}{Timme, M.}
\newblock \bibinfo{title}{Fluctuation-induced distributed resonances in
  oscillatory networks}.
\newblock \emph{\bibinfo{journal}{Science Advances}}
  \textbf{\bibinfo{volume}{5}}, \bibinfo{pages}{eaav1027}
  (\bibinfo{year}{2019}).

\bibitem{HaehnePropagation2019}
\bibinfo{author}{Hähne, H.}, \bibinfo{author}{Schmietendorf, K.},
  \bibinfo{author}{Tamrakar, S.}, \bibinfo{author}{Peinke, J.} \&
  \bibinfo{author}{Kettemann, S.}
\newblock \bibinfo{title}{Propagation of wind-power-induced fluctuations in
  power grids}.
\newblock \emph{\bibinfo{journal}{Physical Review E}}
  \textbf{\bibinfo{volume}{99}}, \bibinfo{pages}{050301}
  (\bibinfo{year}{2019}).

\bibitem{Schaefer2015}
\bibinfo{author}{Sch{\"a}fer, B.}, \bibinfo{author}{Matthiae, M.},
  \bibinfo{author}{Timme, M.} \& \bibinfo{author}{Witthaut, D.}
\newblock \bibinfo{title}{{Decentral Smart Grid Control}}.
\newblock \emph{\bibinfo{journal}{{New Journal of Physics}}}
  \textbf{\bibinfo{volume}{17}}, \bibinfo{pages}{015002}
  (\bibinfo{year}{2015}).

\bibitem{poolla2017optimal}
\bibinfo{author}{Poolla, B.~K.}, \bibinfo{author}{Bolognani, S.} \&
  \bibinfo{author}{D{\"o}rfler, F.}
\newblock \bibinfo{title}{Optimal placement of virtual inertia in power grids}.
\newblock \emph{\bibinfo{journal}{IEEE Transactions on Automatic Control}}
  \textbf{\bibinfo{volume}{62}}, \bibinfo{pages}{6209--6220}
  (\bibinfo{year}{2017}).

\bibitem{pagnier2019inertia}
\bibinfo{author}{Pagnier, L.} \& \bibinfo{author}{Jacquod, P.}
\newblock \bibinfo{title}{Inertia location and slow network modes determine
  disturbance propagation in large-scale power grids}.
\newblock \emph{\bibinfo{journal}{PLOS ONE}} \textbf{\bibinfo{volume}{14}},
  \bibinfo{pages}{e0213550} (\bibinfo{year}{2019}).

\bibitem{Simonsen2008}
\bibinfo{author}{Simonsen, I.}, \bibinfo{author}{Buzna, L.},
  \bibinfo{author}{Peters, K.}, \bibinfo{author}{Bornholdt, S.} \&
  \bibinfo{author}{Helbing, D.}
\newblock \bibinfo{title}{Transient dynamics increasing network vulnerability
  to cascading failures}.
\newblock \emph{\bibinfo{journal}{Physical Review Letters}}
  \textbf{\bibinfo{volume}{100}}, \bibinfo{pages}{218701}
  (\bibinfo{year}{2008}).

\bibitem{yang2017small}
\bibinfo{author}{Yang, Y.}, \bibinfo{author}{Nishikawa, T.} \&
  \bibinfo{author}{Motter, A.~E.}
\newblock \bibinfo{title}{Small vulnerable sets determine large network
  cascades in power grids}.
\newblock \emph{\bibinfo{journal}{Science}} \textbf{\bibinfo{volume}{358}},
  \bibinfo{pages}{eaan3184} (\bibinfo{year}{2017}).

\bibitem{schafer2019dynamical}
\bibinfo{author}{Sch{\"a}fer, B.} \& \bibinfo{author}{Yalcin, G.~C.}
\newblock \bibinfo{title}{{Dynamical modeling of cascading failures in the
  Turkish power grid}}.
\newblock \emph{\bibinfo{journal}{Chaos: An Interdisciplinary Journal of
  Nonlinear Science}} \textbf{\bibinfo{volume}{29}}, \bibinfo{pages}{093134}
  (\bibinfo{year}{2019}).

\bibitem{nesti2018emergent}
\bibinfo{author}{Nesti, T.}, \bibinfo{author}{Zocca, A.} \&
  \bibinfo{author}{Zwart, B.}
\newblock \bibinfo{title}{Emergent failures and cascades in power grids: a
  statistical physics perspective}.
\newblock \emph{\bibinfo{journal}{Physical Review Letters}}
  \textbf{\bibinfo{volume}{120}}, \bibinfo{pages}{258301}
  (\bibinfo{year}{2018}).

\bibitem{chai_wide-area_2016}
\bibinfo{author}{Chai, J.} \emph{et~al.}
\newblock \bibinfo{title}{Wide-area measurement data analytics using
  {FNET}/{GridEye}: {A} review}.
\newblock In \emph{\bibinfo{booktitle}{2016 {Power} {Systems} {Computation}
  {Conference} ({PSCC})}} (\bibinfo{year}{2016}).

\bibitem{Gridradar}
\bibinfo{author}{{MagnaGen GmbH}}.
\newblock \bibinfo{title}{Gridradar--an independent grid monitoring system}.
\newblock \bibinfo{howpublished}{\url{https://gridradar.net/}}
  (\bibinfo{year}{2020}).

\bibitem{Lasseter2004}
\bibinfo{author}{Lasseter, R.~H.} \& \bibinfo{author}{Paigi, P.}
\newblock \bibinfo{title}{Microgrid: A conceptual solution}.
\newblock In \emph{\bibinfo{booktitle}{{2004 IEEE 35th Annual Power Electronics
  Specialists Conference (PESC)}}}, vol.~\bibinfo{volume}{6},
  \bibinfo{pages}{4285--4290} (\bibinfo{organization}{IEEE},
  \bibinfo{year}{2004}).

\bibitem{Jumar2020Data}
\bibinfo{author}{Jumar, R.}, \bibinfo{author}{Maaß, H.},
  \bibinfo{author}{Schäfer, B.}, \bibinfo{author}{Gorjão, L.~R.} \&
  \bibinfo{author}{Hagenmeyer, V.}
\newblock \bibinfo{title}{Power grid frequency data base}.
\newblock \emph{\bibinfo{journal}{arXiv preprint arXiv:2006.01771}}
  (\bibinfo{year}{2020}).

\bibitem{maass2013first}
\bibinfo{author}{Maa{\ss}, H.} \emph{et~al.}
\newblock \bibinfo{title}{First evaluation results using the new electrical
  data recorder for power grid analysis}.
\newblock \emph{\bibinfo{journal}{IEEE Transactions on Instrumentation and
  Measurement}} \textbf{\bibinfo{volume}{62}}, \bibinfo{pages}{2384--2390}
  (\bibinfo{year}{2013}).

\bibitem{maass2015data}
\bibinfo{author}{Maa{\ss}, H.} \emph{et~al.}
\newblock \bibinfo{title}{Data processing of high-rate low-voltage distribution
  grid recordings for smart grid monitoring and analysis}.
\newblock \emph{\bibinfo{journal}{EURASIP Journal on Advances in Signal
  Processing}} \textbf{\bibinfo{volume}{2015}}, \bibinfo{pages}{14}
  (\bibinfo{year}{2015}).

\bibitem{database2020}
\bibinfo{author}{Jumar, R.}, \bibinfo{author}{Maaß, H.},
  \bibinfo{author}{Schäfer, B.}, \bibinfo{author}{Gorjão, L.~R.} \&
  \bibinfo{author}{Hagenmeyer, V.}
\newblock \bibinfo{title}{Power grid frequency data base}.
\newblock \bibinfo{howpublished}{\url{https://osf.io/by5hu/}}
  (\bibinfo{year}{2020}).

\bibitem{kakimoto2006monitoring}
\bibinfo{author}{Kakimoto, N.}, \bibinfo{author}{Sugumi, M.},
  \bibinfo{author}{Makino, T.} \& \bibinfo{author}{Tomiyama, K.}
\newblock \bibinfo{title}{Monitoring of interarea oscillation mode by
  synchronized phasor measurement}.
\newblock \emph{\bibinfo{journal}{IEEE Transactions on Power Systems}}
  \textbf{\bibinfo{volume}{21}}, \bibinfo{pages}{260--268}
  (\bibinfo{year}{2006}).

\bibitem{Schaefer2017a}
\bibinfo{author}{Sch{\"a}fer, B.}, \bibinfo{author}{Beck, C.},
  \bibinfo{author}{Aihara, K.}, \bibinfo{author}{Witthaut, D.} \&
  \bibinfo{author}{Timme, M.}
\newblock \bibinfo{title}{Non-{G}aussian power grid frequency fluctuations
  characterized by {L{\'e}vy-stable} laws and superstatistics}.
\newblock \emph{\bibinfo{journal}{Nature Energy}} \textbf{\bibinfo{volume}{3}},
  \bibinfo{pages}{119--126} (\bibinfo{year}{2018}).

\bibitem{gorjao_data-driven_2020}
\bibinfo{author}{Rydin~Gorjão, L.} \emph{et~al.}
\newblock \bibinfo{title}{Data-driven model of the power-grid frequency
  dynamics}.
\newblock \emph{\bibinfo{journal}{{IEEE} Access}} \textbf{\bibinfo{volume}{8}},
  \bibinfo{pages}{43082--43097} (\bibinfo{year}{2020}).

\bibitem{Gardiner1985}
\bibinfo{author}{Gardiner, C.~W.}
\newblock \emph{\bibinfo{title}{Handbook of Stochastic Methods: for Physics,
  Chemistry and the Natural Sciences}} (\bibinfo{publisher}{Springer},
  \bibinfo{year}{1985}).

\bibitem{anvari_stochastic_2020}
\bibinfo{author}{Anvari, M.} \emph{et~al.}
\newblock \bibinfo{title}{Stochastic properties of the frequency dynamics in
  real and synthetic power grids}.
\newblock \emph{\bibinfo{journal}{Physical Review Research}}
  \textbf{\bibinfo{volume}{2}}, \bibinfo{pages}{013339} (\bibinfo{year}{2020}).

\bibitem{ENTSOEFactsheet2018}
\bibinfo{author}{{ENTSO-E}}.
\newblock \bibinfo{title}{Statistical factsheet 2018}.
\newblock
  \bibinfo{howpublished}{\url{https://docstore.entsoe.eu/Documents/Publications/Statistics/Factsheet/entsoe_sfs2018_web.pdf}}
  (\bibinfo{year}{2018}).

\bibitem{Machowski2011}
\bibinfo{author}{Machowski, J.}, \bibinfo{author}{Bialek, J.} \&
  \bibinfo{author}{Bumby, J.}
\newblock \emph{\bibinfo{title}{{Power System Dynamics: Stability and
  Control}}} (\bibinfo{publisher}{{John Wiley \& Sons, Chichester}},
  \bibinfo{year}{2011}).

\bibitem{Ulbig2014}
\bibinfo{author}{Ulbig, A.}, \bibinfo{author}{Borsche, T.~S.} \&
  \bibinfo{author}{Andersson, G.}
\newblock \bibinfo{title}{Impact of low rotational inertia on power system
  stability and operation}.
\newblock \emph{\bibinfo{journal}{IFAC Proceedings Volumes}}
  \textbf{\bibinfo{volume}{47}}, \bibinfo{pages}{7290--7297}
  (\bibinfo{year}{2014}).

\bibitem{Gorjao2019}
\bibinfo{author}{Rydin~Gorjão, L.} \& \bibinfo{author}{Meirinhos, F.}
\newblock \bibinfo{title}{kramersmoyal: Kramers--{M}oyal coefficients for
  stochastic processes}.
\newblock \emph{\bibinfo{journal}{Journal of Open Source Software}}
  \textbf{\bibinfo{volume}{4}}, \bibinfo{pages}{1693} (\bibinfo{year}{2019}).

\bibitem{peng1993long}
\bibinfo{author}{Peng, C.-K.} \emph{et~al.}
\newblock \bibinfo{title}{Long-range anticorrelations and non-{G}aussian
  behavior of the heartbeat}.
\newblock \emph{\bibinfo{journal}{Physical Review Letters}}
  \textbf{\bibinfo{volume}{70}}, \bibinfo{pages}{1343} (\bibinfo{year}{1993}).

\bibitem{schafer2018isolating}
\bibinfo{author}{Sch\"afer, B.}, \bibinfo{author}{Timme, M.} \&
  \bibinfo{author}{Witthaut, D.}
\newblock \bibinfo{title}{Isolating the impact of trading on grid frequency
  fluctuations}.
\newblock In \emph{\bibinfo{booktitle}{2018 IEEE PES Innovative Smart Grid
  Technologies Conference Europe (ISGT-Europe)}}, \bibinfo{pages}{1--5}
  (\bibinfo{organization}{IEEE}, \bibinfo{year}{2018}).

\bibitem{Haehne2018}
\bibinfo{author}{Hähne, H.}, \bibinfo{author}{Schottler, J.},
  \bibinfo{author}{Wächter, M.}, \bibinfo{author}{Peinke, J.} \&
  \bibinfo{author}{Kamps, O.}
\newblock \bibinfo{title}{The footprint of atmospheric turbulence in power grid
  frequency measurements}.
\newblock \emph{\bibinfo{journal}{Europhysics Letters}}
  \textbf{\bibinfo{volume}{121}}, \bibinfo{pages}{30001}
  (\bibinfo{year}{2018}).

\bibitem{Castaing1990}
\bibinfo{author}{Castaing, B.}, \bibinfo{author}{Gagne, Y.} \&
  \bibinfo{author}{Hopfinger, E.}
\newblock \bibinfo{title}{Velocity probability density functions of high
  {R}eynolds number turbulence}.
\newblock \emph{\bibinfo{journal}{Physica D: Nonlinear Phenomena}}
  \textbf{\bibinfo{volume}{46}}, \bibinfo{pages}{177--200}
  (\bibinfo{year}{1990}).

\bibitem{Beck2003}
\bibinfo{author}{Beck, C.} \& \bibinfo{author}{Cohen, E. G.~D.}
\newblock \bibinfo{title}{Superstatistics}.
\newblock \emph{\bibinfo{journal}{Physica A}} \textbf{\bibinfo{volume}{322}},
  \bibinfo{pages}{267--275} (\bibinfo{year}{2003}).

\bibitem{MFDFA2020}
\bibinfo{author}{Rydin~Gorjão, L.}
\newblock \bibinfo{title}{{MFDFA: Multifractal Detrended Fluctuation Analysis
  in Python}}.
\newblock \bibinfo{howpublished}{\url{https://zenodo.org/record/3625759}}
  (\bibinfo{year}{2020}).

\bibitem{meyer2020identifying}
\bibinfo{author}{Meyer, P.~G.}, \bibinfo{author}{Anvari, M.} \&
  \bibinfo{author}{Kantz, H.}
\newblock \bibinfo{title}{Identifying characteristic time scales in power grid
  frequency fluctuations with dfa}.
\newblock \emph{\bibinfo{journal}{Chaos: An Interdisciplinary Journal of
  Nonlinear Science}} \textbf{\bibinfo{volume}{30}}, \bibinfo{pages}{013130}
  (\bibinfo{year}{2020}).

\bibitem{Weissbach2009}
\bibinfo{author}{Wei{\ss}bach, T.} \& \bibinfo{author}{Welfonder, E.}
\newblock \bibinfo{title}{{High Frequency Deviations within the European Power
  System--Origins and Proposals for Improvement}}.
\newblock \emph{\bibinfo{journal}{VGB powertech}}
  \textbf{\bibinfo{volume}{89}}, \bibinfo{pages}{26} (\bibinfo{year}{2009}).

\bibitem{cresap1981emergence}
\bibinfo{author}{Cresap, R.} \& \bibinfo{author}{Hauer, J.}
\newblock \bibinfo{title}{Emergence of a new swing mode in the western power
  system}.
\newblock \emph{\bibinfo{journal}{IEEE Transactions on Power Apparatus and
  Systems}} \bibinfo{pages}{2037--2045} (\bibinfo{year}{1981}).

\bibitem{chompoobutrgool2012identification}
\bibinfo{author}{Chompoobutrgool, Y.} \& \bibinfo{author}{Vanfretti, L.}
\newblock \bibinfo{title}{Identification of power system dominant inter-area
  oscillation paths}.
\newblock \emph{\bibinfo{journal}{IEEE Transactions on Power Systems}}
  \textbf{\bibinfo{volume}{28}}, \bibinfo{pages}{2798--2807}
  (\bibinfo{year}{2012}).

\bibitem{messina_extraction_2007}
\bibinfo{author}{Messina, A.~R.} \& \bibinfo{author}{Vittal, V.}
\newblock \bibinfo{title}{Extraction of dynamic patterns from wide-area
  measurements using empirical orthogonal functions}.
\newblock \emph{\bibinfo{journal}{IEEE Transactions on Power Systems}}
  \textbf{\bibinfo{volume}{22}}, \bibinfo{pages}{682--692}
  (\bibinfo{year}{2007}).

\bibitem{susuki_nonlinear_2011}
\bibinfo{author}{Susuki, Y.} \& \bibinfo{author}{Mezic, I.}
\newblock \bibinfo{title}{Nonlinear {K}oopman modes and coherency
  identification of coupled swing dynamics}.
\newblock \emph{\bibinfo{journal}{IEEE Transactions on Power Systems}}
  \textbf{\bibinfo{volume}{26}}, \bibinfo{pages}{1894--1904}
  (\bibinfo{year}{2011}).

\bibitem{bishop_pattern_2007}
\bibinfo{author}{Bishop, C.~M.}
\newblock \emph{\bibinfo{title}{Pattern {Recognition} and {Machine}
  {Learning}}} (\bibinfo{publisher}{Springer}, \bibinfo{address}{New York},
  \bibinfo{year}{2007}), \bibinfo{edition}{1} edn.

\bibitem{anaparthi_coherency_2005}
\bibinfo{author}{Anaparthi, K.}, \bibinfo{author}{Chaudhuri, B.},
  \bibinfo{author}{Thornhill, N.} \& \bibinfo{author}{Pal, B.}
\newblock \bibinfo{title}{Coherency identification in power systems through
  principal component analysis}.
\newblock \emph{\bibinfo{journal}{IEEE Transactions on Power Systems}}
  \textbf{\bibinfo{volume}{20}}, \bibinfo{pages}{1658--1660}
  (\bibinfo{year}{2005}).

\bibitem{grebe_low_2010}
\bibinfo{author}{Grebe, E.}, \bibinfo{author}{Kabouris, J.},
  \bibinfo{author}{Lopez~Barba, S.}, \bibinfo{author}{Sattinger, W.} \&
  \bibinfo{author}{Winter, W.}
\newblock \bibinfo{title}{Low frequency oscillations in the interconnected
  system of {Continental} {Europe}}.
\newblock In \emph{\bibinfo{booktitle}{{IEEE} {PES} {General} {Meeting}}},
  \bibinfo{pages}{1--7} (\bibinfo{publisher}{IEEE},
  \bibinfo{address}{Minneapolis, MN}, \bibinfo{year}{2010}).

\bibitem{Klein1991}
\bibinfo{author}{Klein, M.}, \bibinfo{author}{Rogers, G.~J.} \&
  \bibinfo{author}{Kundur, P.}
\newblock \bibinfo{title}{A fundamental study of inter-area oscillations in
  power systems}.
\newblock \emph{\bibinfo{journal}{IEEE Transactions on Power Systems}}
  \textbf{\bibinfo{volume}{6}}, \bibinfo{pages}{914--921}
  (\bibinfo{year}{1991}).

\bibitem{tuttelberg2018estimation}
\bibinfo{author}{Tuttelberg, K.}, \bibinfo{author}{Kilter, J.},
  \bibinfo{author}{Wilson, D.} \& \bibinfo{author}{Uhlen, K.}
\newblock \bibinfo{title}{Estimation of power system inertia from ambient wide
  area measurements}.
\newblock \emph{\bibinfo{journal}{IEEE Transactions on Power Systems}}
  \textbf{\bibinfo{volume}{33}}, \bibinfo{pages}{7249--7257}
  (\bibinfo{year}{2018}).

\bibitem{Kruse2020Predict}
\bibinfo{author}{Kruse, J.}, \bibinfo{author}{Schäfer, B.} \&
  \bibinfo{author}{Witthaut, D.}
\newblock \bibinfo{title}{Predicting the power grid frequency}.
\newblock \emph{\bibinfo{journal}{arXiv preprint arXiv:2004.09259}}
  (\bibinfo{year}{2020}).

\bibitem{vanfretti2013spectral}
\bibinfo{author}{Vanfretti, L.}, \bibinfo{author}{Bengtsson, S.},
  \bibinfo{author}{Peri{\'c}, V.~S.} \& \bibinfo{author}{Gjerde, J.~O.}
\newblock \bibinfo{title}{Spectral estimation of low-frequency oscillations in
  the {N}ordic grid using ambient synchrophasor data under the presence of
  forced oscillations}.
\newblock In \emph{\bibinfo{booktitle}{2013 IEEE Grenoble Conference}},
  \bibinfo{pages}{1--6} (\bibinfo{organization}{IEEE}, \bibinfo{year}{2013}).

\bibitem{cui_inter-area_2017}
\bibinfo{author}{Cui, Y.} \emph{et~al.}
\newblock \bibinfo{title}{Inter-area oscillation statistical analysis of the
  {U}.{S}. {Eastern} interconnection}.
\newblock \emph{\bibinfo{journal}{The Journal of Engineering}}
  \textbf{\bibinfo{volume}{2017}}, \bibinfo{pages}{595--605}
  (\bibinfo{year}{2017}).

\bibitem{tchuisseu2017}
\bibinfo{author}{Tchuisseu, E.~T.}, \bibinfo{author}{Gomila, D.},
  \bibinfo{author}{Brunner, D.} \& \bibinfo{author}{Colet, P.}
\newblock \bibinfo{title}{Effects of dynamic-demand-control appliances on the
  power grid frequency}.
\newblock \emph{\bibinfo{journal}{Physical Review E}}
  \textbf{\bibinfo{volume}{96}}, \bibinfo{pages}{022302}
  (\bibinfo{year}{2017}).

\bibitem{schroeder2019dynamic}
\bibinfo{author}{Schröder, M.}, \bibinfo{author}{Zhang, X.},
  \bibinfo{author}{Wolter, J.} \& \bibinfo{author}{Timme, M.}
\newblock \bibinfo{title}{Dynamic perturbation spreading in networks}.
\newblock \emph{\bibinfo{journal}{IEEE Transactions on Network Science and
  Engineering}}  (\bibinfo{year}{2019}).

\end{thebibliography}


\begin{thebibliography}{10}
\expandafter\ifx\csname url\endcsname\relax
  \def\url#1{\texttt{#1}}\fi
\expandafter\ifx\csname urlprefix\endcsname\relax\def\urlprefix{URL }\fi
\providecommand{\bibinfo}[2]{#2}
\providecommand{\eprint}[2][]{\url{#2}}

\bibitem{Schaefer2017a}
\bibinfo{author}{Sch{\"a}fer, B.}, \bibinfo{author}{Beck, C.},
  \bibinfo{author}{Aihara, K.}, \bibinfo{author}{Witthaut, D.} \&
  \bibinfo{author}{Timme, M.}
\newblock \bibinfo{title}{Non-{G}aussian power grid frequency fluctuations
  characterized by {L{\'e}vy-stable} laws and superstatistics}.
\newblock \emph{\bibinfo{journal}{Nature Energy}} \textbf{\bibinfo{volume}{3}},
  \bibinfo{pages}{119--126} (\bibinfo{year}{2018}).

\bibitem{ENTSOE2013}
\bibinfo{author}{{ENTSO-E}}.
\newblock \bibinfo{title}{Network code on requirements for grid connection
  applicable to all generators (rfg)}.
\newblock
  \bibinfo{howpublished}{\url{https://www.entsoe.eu/major-projects/network-code-development/requirements-for-generators/}}
  (\bibinfo{year}{2013}).

\bibitem{Gardiner1985}
\bibinfo{author}{Gardiner, C.~W.}
\newblock \emph{\bibinfo{title}{Handbook of Stochastic Methods: for Physics,
  Chemistry and the Natural Sciences}} (\bibinfo{publisher}{Springer},
  \bibinfo{year}{1985}).

\bibitem{gorjao_data-driven_2020}
\bibinfo{author}{Rydin~Gorjão, L.} \emph{et~al.}
\newblock \bibinfo{title}{Data-driven model of the power-grid frequency
  dynamics}.
\newblock \emph{\bibinfo{journal}{{IEEE} Access}} \textbf{\bibinfo{volume}{8}},
  \bibinfo{pages}{43082--43097} (\bibinfo{year}{2020}).

\bibitem{schafer2018isolating}
\bibinfo{author}{Sch\"afer, B.}, \bibinfo{author}{Timme, M.} \&
  \bibinfo{author}{Witthaut, D.}
\newblock \bibinfo{title}{Isolating the impact of trading on grid frequency
  fluctuations}.
\newblock In \emph{\bibinfo{booktitle}{2018 IEEE PES Innovative Smart Grid
  Technologies Conference Europe (ISGT-Europe)}}, \bibinfo{pages}{1--5}
  (\bibinfo{organization}{IEEE}, \bibinfo{year}{2018}).

\bibitem{Haehne2018}
\bibinfo{author}{Hähne, H.}, \bibinfo{author}{Schottler, J.},
  \bibinfo{author}{Wächter, M.}, \bibinfo{author}{Peinke, J.} \&
  \bibinfo{author}{Kamps, O.}
\newblock \bibinfo{title}{The footprint of atmospheric turbulence in power grid
  frequency measurements}.
\newblock \emph{\bibinfo{journal}{Europhysics Letters}}
  \textbf{\bibinfo{volume}{121}}, \bibinfo{pages}{30001}
  (\bibinfo{year}{2018}).

\bibitem{HaehnePropagation2019}
\bibinfo{author}{Hähne, H.}, \bibinfo{author}{Schmietendorf, K.},
  \bibinfo{author}{Tamrakar, S.}, \bibinfo{author}{Peinke, J.} \&
  \bibinfo{author}{Kettemann, S.}
\newblock \bibinfo{title}{Propagation of wind-power-induced fluctuations in
  power grids}.
\newblock \emph{\bibinfo{journal}{Physical Review E}}
  \textbf{\bibinfo{volume}{99}}, \bibinfo{pages}{050301}
  (\bibinfo{year}{2019}).

\bibitem{ENTSOEFactsheet2018}
\bibinfo{author}{{ENTSO-E}}.
\newblock \bibinfo{title}{Statistical factsheet 2018}.
\newblock
  \bibinfo{howpublished}{\url{https://docstore.entsoe.eu/Documents/Publications/Statistics/Factsheet/entsoe_sfs2018_web.pdf}}
  (\bibinfo{year}{2018}).

\bibitem{Weissbach2009}
\bibinfo{author}{Wei{\ss}bach, T.} \& \bibinfo{author}{Welfonder, E.}
\newblock \bibinfo{title}{{High Frequency Deviations within the European Power
  System--Origins and Proposals for Improvement}}.
\newblock \emph{\bibinfo{journal}{VGB powertech}}
  \textbf{\bibinfo{volume}{89}}, \bibinfo{pages}{26} (\bibinfo{year}{2009}).

\bibitem{Machowski2011}
\bibinfo{author}{Machowski, J.}, \bibinfo{author}{Bialek, J.} \&
  \bibinfo{author}{Bumby, J.}
\newblock \emph{\bibinfo{title}{{Power System Dynamics: Stability and
  Control}}} (\bibinfo{publisher}{{John Wiley \& Sons, Chichester}},
  \bibinfo{year}{2011}).

\bibitem{Weixelbraun2009}
\bibinfo{author}{Weixelbraun, M.}, \bibinfo{author}{Renner, H.},
  \bibinfo{author}{Schmaranz, R.} \& \bibinfo{author}{Marketz, M.}
\newblock \bibinfo{title}{Dynamic simulation of a 110-kv-network during grid
  restoration and in islanded operation}.
\newblock In \emph{\bibinfo{booktitle}{{20th International Conference and
  Exhibition on Electricity Distribution-Part 1, 2009}}}, \bibinfo{pages}{1--4}
  (\bibinfo{organization}{IET}, \bibinfo{year}{2009}).

\bibitem{Lamouroux09}
\bibinfo{author}{Lamouroux, D.} \& \bibinfo{author}{Lehnertz, K.}
\newblock \bibinfo{title}{Kernel-based regression of drift and diffusion
  coefficients of stochastic processes}.
\newblock \emph{\bibinfo{journal}{Physics Letters A}}
  \textbf{\bibinfo{volume}{373}}, \bibinfo{pages}{3507--3512}
  (\bibinfo{year}{2009}).

\bibitem{Gorjao2019}
\bibinfo{author}{Rydin~Gorjão, L.} \& \bibinfo{author}{Meirinhos, F.}
\newblock \bibinfo{title}{kramersmoyal: Kramers--{M}oyal coefficients for
  stochastic processes}.
\newblock \emph{\bibinfo{journal}{Journal of Open Source Software}}
  \textbf{\bibinfo{volume}{4}}, \bibinfo{pages}{1693} (\bibinfo{year}{2019}).

\bibitem{Rinn16}
\bibinfo{author}{Rinn, P.}, \bibinfo{author}{Lind, P.~G.},
  \bibinfo{author}{Wächter, M.} \& \bibinfo{author}{Peinke, J.}
\newblock \bibinfo{title}{The {L}angevin approach: An {R} package for modeling
  {M}arkov processes}.
\newblock \emph{\bibinfo{journal}{Journal of Open Research Software}}
  \textbf{\bibinfo{volume}{4}}, \bibinfo{pages}{e34} (\bibinfo{year}{2016}).

\bibitem{Castaing1990}
\bibinfo{author}{Castaing, B.}, \bibinfo{author}{Gagne, Y.} \&
  \bibinfo{author}{Hopfinger, E.}
\newblock \bibinfo{title}{Velocity probability density functions of high
  {R}eynolds number turbulence}.
\newblock \emph{\bibinfo{journal}{Physica D: Nonlinear Phenomena}}
  \textbf{\bibinfo{volume}{46}}, \bibinfo{pages}{177--200}
  (\bibinfo{year}{1990}).

\bibitem{Castaing1994}
\bibinfo{author}{Castaing, B.}
\newblock \bibinfo{title}{Scalar intermittency in the variational theory of
  turbulence}.
\newblock \emph{\bibinfo{journal}{Physica D: Nonlinear Phenomena}}
  \textbf{\bibinfo{volume}{73}}, \bibinfo{pages}{31--37}
  (\bibinfo{year}{1994}).

\bibitem{Castaing1996}
\bibinfo{author}{Castaing, B.}
\newblock \bibinfo{title}{The temperature of turbulent flows}.
\newblock \emph{\bibinfo{journal}{Journal de Physique II}}
  \textbf{\bibinfo{volume}{6}}, \bibinfo{pages}{105--114}
  (\bibinfo{year}{1996}).

\bibitem{Tabar2019}
\bibinfo{author}{Tabar, M. R.~R.}
\newblock \emph{\bibinfo{title}{The {F}riedrich--{P}einke Approach to
  Reconstruction of Dynamical Equation for Time Series: {C}omplexity in View of
  Stochastic Processes}}, \bibinfo{pages}{143--164}
  (\bibinfo{publisher}{Springer International Publishing},
  \bibinfo{address}{Cham}, \bibinfo{year}{2019}).

\bibitem{Beck2003}
\bibinfo{author}{Beck, C.} \& \bibinfo{author}{Cohen, E. G.~D.}
\newblock \bibinfo{title}{Superstatistics}.
\newblock \emph{\bibinfo{journal}{Physica A}} \textbf{\bibinfo{volume}{322}},
  \bibinfo{pages}{267--275} (\bibinfo{year}{2003}).

\bibitem{Beck2005}
\bibinfo{author}{Beck, C.}, \bibinfo{author}{Cohen, E. G.~D.} \&
  \bibinfo{author}{Swinney, H.~L.}
\newblock \bibinfo{title}{From time series to superstatistics}.
\newblock \emph{\bibinfo{journal}{Physical Review E}}
  \textbf{\bibinfo{volume}{72}}, \bibinfo{pages}{056133}
  (\bibinfo{year}{2005}).

\bibitem{Anvari2016}
\bibinfo{author}{Anvari, M.} \emph{et~al.}
\newblock \bibinfo{title}{Short term fluctuations of wind and solar power
  systems}.
\newblock \emph{\bibinfo{journal}{New Journal of Physics}}
  \textbf{\bibinfo{volume}{18}}, \bibinfo{pages}{063027}
  (\bibinfo{year}{2016}).

\bibitem{zhang2019fluctuation}
\bibinfo{author}{Zhang, X.}, \bibinfo{author}{Hallerberg, S.},
  \bibinfo{author}{Matthiae, M.}, \bibinfo{author}{Witthaut, D.} \&
  \bibinfo{author}{Timme, M.}
\newblock \bibinfo{title}{Fluctuation-induced distributed resonances in
  oscillatory networks}.
\newblock \emph{\bibinfo{journal}{Science Advances}}
  \textbf{\bibinfo{volume}{5}}, \bibinfo{pages}{eaav1027}
  (\bibinfo{year}{2019}).

\bibitem{Peng1994}
\bibinfo{author}{Peng, C.-K.} \emph{et~al.}
\newblock \bibinfo{title}{Mosaic organization of {DNA} nucleotides}.
\newblock \emph{\bibinfo{journal}{Physical Review E}}
  \textbf{\bibinfo{volume}{49}}, \bibinfo{pages}{1685--1689}
  (\bibinfo{year}{1994}).

\bibitem{Kantelhardt2002}
\bibinfo{author}{Kantelhardt, J.~W.} \emph{et~al.}
\newblock \bibinfo{title}{Multifractal detrended fluctuation analysis of
  nonstationary time series}.
\newblock \emph{\bibinfo{journal}{Physica A}} \textbf{\bibinfo{volume}{316}},
  \bibinfo{pages}{87--114} (\bibinfo{year}{2002}).

\bibitem{Ihlen2012}
\bibinfo{author}{Ihlen, E.}
\newblock \bibinfo{title}{Introduction to {M}ultifractal {D}etrended
  {F}luctuation {A}nalysis in {M}atlab}.
\newblock \emph{\bibinfo{journal}{Frontiers in Physiology}}
  \textbf{\bibinfo{volume}{3}}, \bibinfo{pages}{141} (\bibinfo{year}{2012}).

\bibitem{MFDFA2020}
\bibinfo{author}{Rydin~Gorjão, L.}
\newblock \bibinfo{title}{{MFDFA: Multifractal Detrended Fluctuation Analysis
  in Python}}.
\newblock \bibinfo{howpublished}{\url{https://zenodo.org/record/3625759}}
  (\bibinfo{year}{2020}).

\bibitem{grebe_low_2010}
\bibinfo{author}{Grebe, E.}, \bibinfo{author}{Kabouris, J.},
  \bibinfo{author}{Lopez~Barba, S.}, \bibinfo{author}{Sattinger, W.} \&
  \bibinfo{author}{Winter, W.}
\newblock \bibinfo{title}{Low frequency oscillations in the interconnected
  system of {Continental} {Europe}}.
\newblock In \emph{\bibinfo{booktitle}{{IEEE} {PES} {General} {Meeting}}},
  \bibinfo{pages}{1--7} (\bibinfo{publisher}{IEEE},
  \bibinfo{address}{Minneapolis, MN}, \bibinfo{year}{2010}).

\bibitem{bishop_pattern_2007}
\bibinfo{author}{Bishop, C.~M.}
\newblock \emph{\bibinfo{title}{Pattern {Recognition} and {Machine}
  {Learning}}} (\bibinfo{publisher}{Springer}, \bibinfo{address}{New York},
  \bibinfo{year}{2007}), \bibinfo{edition}{1} edn.

\bibitem{Klein1991}
\bibinfo{author}{Klein, M.}, \bibinfo{author}{Rogers, G.~J.} \&
  \bibinfo{author}{Kundur, P.}
\newblock \bibinfo{title}{A fundamental study of inter-area oscillations in
  power systems}.
\newblock \emph{\bibinfo{journal}{IEEE Transactions on Power Systems}}
  \textbf{\bibinfo{volume}{6}}, \bibinfo{pages}{914--921}
  (\bibinfo{year}{1991}).

\end{thebibliography}

\end{document}


\title{Supplementary Information accompanying the manuscript \\ Open data base analysis of scaling and  spatio-temporal properties of power grid frequencies}

\author{Leonardo Rydin Gorj\~ao}
\affiliation{Forschungszentrum J\"ulich, Institute for Energy and Climate Research - Systems Analysis and Technology Evaluation (IEK-STE), Germany}
\affiliation{Institute for Theoretical Physics, University of Cologne, Germany}

\author{Richard Jumar}
\affiliation{Karlsruhe Institute of Technology, Institute for Automation and Applied Informatics, Germany}

\author{Heiko Maass}
\affiliation{Karlsruhe Institute of Technology, Institute for Automation and Applied Informatics, Germany}

\author{Veit Hagenmeyer}
\affiliation{Karlsruhe Institute of Technology, Institute for Automation and Applied Informatics, Germany}

\author{G.Cigdem Yalcin}
\affiliation{Department of Physics, Istanbul University, 34134, Vezneciler, Turkey}

\author{Johannes Kruse}
\affiliation{Forschungszentrum J\"ulich, Institute for Energy and Climate Research - Systems Analysis and Technology Evaluation (IEK-STE), Germany}
\affiliation{Institute for Theoretical Physics, University of Cologne, Germany}

\author{Marc Timme}
\affiliation{Chair for Network Dynamics, Center for Advancing Electronics Dresden (cfaed) and Institute for Theoretical Physics, Technical University of Dresden, Germany}

\author{Christian Beck}
\affiliation{School of Mathematical Sciences, Queen Mary University of London, United Kingdom}

\author{Dirk Witthaut}
\affiliation{Forschungszentrum J\"ulich, Institute for Energy and Climate Research - Systems Analysis and Technology Evaluation (IEK-STE), Germany}
\affiliation{Institute for Theoretical Physics, University of Cologne, Germany}

\author{Benjamin Sch\"afer}
\email{b.schaefer@qmul.ac.uk}
\affiliation{School of Mathematical Sciences, Queen Mary University of London, United Kingdom}

\begin{abstract}
 Within this Supplementary Information, we give further details on the analysis presented in the main text. In particular, we introduce the abbreviations used to describe the various measurement sites, report the first four statistical moments of all areas. Furthermore, we provide more details on the PCA analysis, including a visualization of the modes and a comparison with the spectrum.
\end{abstract}

\maketitle

\section*{Supplementary Note 1}
\subsection*{Details on individual measurements: Abbreviations and statistical measures}
\subsubsection*{Abbreviations}

The power grid frequency has been recorded in several synchronous areas across Europe and beyond. We introduce the abbreviations used  when referring to the measurement sites, e.g. in plots, in Supplementary Table~\ref{tab:abbreviations}: We list the town, and country where the measurement was taken and the synchronous area to which this particular device was connected.
For example, the measurement taking place in Karlsruhe, Germany is abbreviated as DE (ISO 3166 for Germany) and is connected to the Continental European synchronous area (CE). When discussing the results, we will often refer to the synchronous area and name the measurement site abbreviation in parenthesis, e.g. Continental Europe (DE) is known to be influenced by market dynamics, see also  \cite{Schaefer2017a}. Unless we are specifically interested in the short time scale, a measurement taken in one location is representative for the whole synchronous area, see also discussion on time-to-bulk in the main text.
For areas that are part of a larger country, we first name the country and then specify the area further, e.g. US-UT stands for the United States of America, State Utah, which in turn is part of the Western Interconnection. 

\begin{table*}[h]
\caption{Abbreviations of measurement locations and their connection to synchronous areas. For each country, we adopt the ISO 3166 code. The population is extracted from Wikipedia and the sources therein in March 2020.}
\begin{centering}
\begin{tabular}{|c|c|c|c|}
\hline 
Abbreviation & Measurement location & Synchronous area & Population \tabularnewline
\hline 
\hline 
\multicolumn{4}{|c|}{\textbf{Islands}}\tabularnewline
\hline 
IS & Reykjavík, Iceland & Iceland & 360,390 \tabularnewline
\hline 
FO & Vestmanna, Faroe Islands & Faroe Islands & 51,783 \tabularnewline
\hline 
ES-GC & Las Palmas de Gran Canaria, Canary Islands, Spain & Gran Canaria & 851,231  \tabularnewline
\hline 
ES-PM & Palma de Mallorca, Balearic Islands, Spain & Mallorca & 896,038 \tabularnewline
\hline 
\multicolumn{4}{|c|}{\textbf{Continental}}\tabularnewline
\hline 
DE & Karlsruhe, Germany & Continental Europe (CE) & 500,000,000 \tabularnewline
\hline 
GB & London, United Kingdom & Great Britain (GB) & 66,224,800
 \tabularnewline
\hline 
EE & Tallin, Estonia & Baltic & 6,042,657
  \tabularnewline
\hline 
SE & Stockholm, Sweden & Nordic & 21,180,931
 \tabularnewline
\hline 
\multicolumn{4}{|c|}{\textbf{Others}}\tabularnewline
\hline 
US-UT & Salt Lake City, Utah, US & Western Interconnection  & 3,205,958 \tabularnewline
\hline 
US-TX & College Station, Texas, US & Texas Interconnection  & 84,600,000 \tabularnewline
\hline 
ZA & Cape Town, South Africa & South Africa & 58,775,022 \tabularnewline
\hline 
RU & St. Petersburg, Russia & Russia & 146,745,098 \tabularnewline
\hline 
\end{tabular}
\par\end{centering}
\label{tab:abbreviations}
\end{table*}

\subsubsection*{Statistical measures}

Let us systematically investigate the statistical properties of the various areas by computing their mean deviation from the reference $\mu$, as well as their standard deviation $\sigma$, skewness $\beta$, and kurtosis $\kappa$ in Supplementary Table~\ref{tab:Statisticalmeasures}.
First, we note that most of the synchronous areas are very close to their reference frequency of $50$ Hz (or $60$ Hz for US areas) on average. Except for Great Britain (GB) and South Africa (ZA), continental areas display a smaller standard deviation than islanded areas. For GB, we can attribute this large deviations to the very different frequency regulation framework (when compared to  Continental Europe), which allows large deviations \cite{ENTSOE2013}. The skewness and kurtosis are more difficult to interpret. Some areas, such as Texas (US-TX) and Faroe Islands show a substantial skewness, while other areas, such as Continental Europe (DE) and Iceland (IS) display a large kurtosis $\kappa>3$, hinting at heavy tails in the distribution.
\begin{table}
\caption{Statistical properties of different measurement sites. We report the mean $\mu$ as the mean deviation  from the reference frequency $f-f^\text{ref}$, standard deviation $\sigma$, skewness $\beta$ and kurtosis $\kappa$ of the data.}
\centering{}%
\begin{tabular}{|c|c|c|c|c|}
\hline 
Area & Mean $\mu$ [mHz] & Std. $\sigma$ [mHz] & Skew. $\beta$ & Kurtosis $\kappa$\tabularnewline
\hline 
\hline 
\multicolumn{5}{|c|}{\textbf{Islands}}\tabularnewline
\hline 
IS & 0.90 & 55.81 & -0.04 & 7.02\tabularnewline
\hline 
FO & -36.65 & 133.15 & 0.27 & 2.39 \tabularnewline
\hline 
ES-GC & -0.60  & 61.63 & 0.38 & 5.00 \tabularnewline
\hline 
ES-PM & 1.32 & 70.13 & 0.06 & 2.62 \tabularnewline
\hline 
\multicolumn{5}{|c|}{\textbf{Continental}}\tabularnewline
\hline 
DE & 0.22 & 18.77 & 0.08 & 3.95 \tabularnewline
\hline 
GB & -0.06 & 62.29 & 0.04 & 2.42 \tabularnewline
\hline 
EE & -0.29 & 12.44 & -0.09 & 2.95 \tabularnewline
\hline 
SE & 0.61 & 48.91 & 0.19 & 4.05 \tabularnewline
\hline 
\multicolumn{5}{|c|}{\textbf{Others}}\tabularnewline
\hline 
US-UT & -0.73 & 18.67 & -0.04 & 2.56 \tabularnewline
\hline 
US-TX & 0.20 & 17.79 & -0.49 & 2.42 \tabularnewline
\hline 
ZA & 1.47 & 93.43 & -0.15 & 2.39 \tabularnewline
\hline 
RU & -1.63 & 16.44 & -0.24 & 2.52 \tabularnewline
\hline 
\end{tabular}
\label{tab:Statisticalmeasures}
\end{table}

\paragraph*{Kurtosis of aggregated data and increment statistics}
Let us have a closer look at the kurtosis values. We utilise the kurtosis and its deviation from the Gaussian value of $\kappa^\text{Gauss}=3$ to quantify whether a given distribution displays heavy tails and thereby deviates from a Gaussian distribution. For the aggregated statistics studied above this is relevant as it tells us how likely we will observe extreme deviations, which could lead to a curtailment of demand or a shutdown of generators. For the increments, this is of particular interest for the statistical modelling since a classical Ornstein--Uhlenbeck Process would lead to Gaussian increments \cite{Gardiner1985}.

A kurtosis of $\kappa>3$ is only clearly observed in the aggregated data of IS, ES-GC, DE, and SE. In all cases, the numerous extreme events are a remarkable observation. For such a highly regulated and controlled system as the power grid to deviate so strongly so often demands an explanation.
We are confident that the heavy tails at the two continental areas DE and SE are related to extreme deviations at the trading intervals \cite{gorjao_data-driven_2020,schafer2018isolating,Schaefer2017a}. The tails in the two islands might also be caused by market activities or arise due to a different  operation by the TSO, as will be clarified in future work.

The role of the kurtosis changes when moving to increments. A large kurtosis of the increments indicates that the effective noise acting on the system does indeed not follow a Gaussian distribution, as expected in many stochastic processes. Naturally, investigating the increments of an empirical trajectory, we will observe a much more volatile behaviour than if we only observe the trajectory itself. Large jumps will be present and on a short time scale we expect extreme tails in real-world data. Some of these large jumps could be caused by renewable generators, others by fast control mechanisms or inverters. Still it is remarkable that kurtosis values of $\kappa~10^2$, are observed for IS and ES-GC in Fig. 4 of the main text. While the general trend of decreasing kurtosis with increasing time lag is in coherence with previous work on increment analysis \cite{Haehne2018,HaehnePropagation2019}, further work is necessary to arrive at a full statistical description of frequency increment.

\clearpage
\section*{Supplementary Note 2}
\subsection*{Scaling of fluctuations}

Let us investigate whether the fluctuation in terms of regular deviations, measured by the standard deviation $\sigma$, or extreme deviations, measured by the kurtosis $\kappa$ scale with the size of the power system. Intuitively, we would expect that the more generation is present in a given area, the smaller the fluctuations we observe are going to be. As a proxy of the total generation, we use the total population of a synchronous area as this information is easily available for all areas and population and total generation are approximately proportional \cite{ENTSOEFactsheet2018}. 
As we can see in Supplementary Fig~\ref{fig:Scaling_kurtosis}, the kurtosis $\kappa$ is not a simple function of the size of a synchronous area.
The occurrence of heavy tails, as measured by the kurtosis $\kappa$, depends on dispatch strategies, market regulations \cite{Weissbach2009} and control requirements, which are standardised within the European Network of Transmission System Operators for Electricity (ENTSO-E) \cite{ENTSOE2013}, leading to very similar values of the kurtosis in most synchronous areas. 

\begin{figure}[ht]
\includegraphics[width=0.5\linewidth]{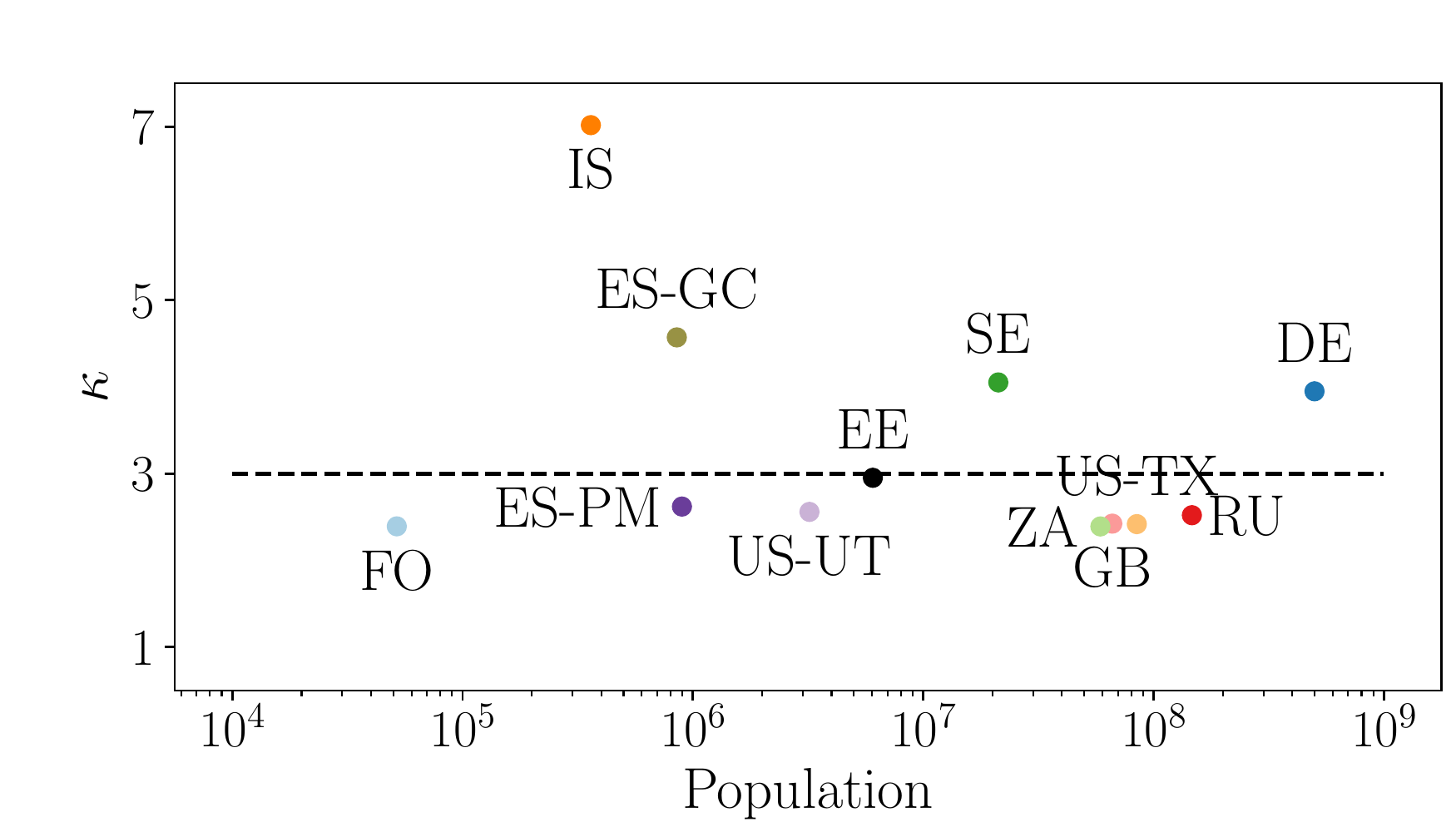}
\caption{Extreme fluctuation occurrence do not decrease with increasing grid size.
We plot the kurtosis as a measure of heavy tails as a function of the population of a synchronous area.}
\label{fig:Scaling_kurtosis}
\end{figure}

In contrast, we have seen in the main text that the aggregated noise amplitude $\epsilon$ decreases approximately as the square root of the size of a synchronous area. Let us review the derivation of this relation in more detail and discuss alternative approaches to observe the scaling. We follow the arguments presented in \cite{Schaefer2017a}: Let us use the standard swing equation \cite{Machowski2011} to describe the synchronous frequency dynamics at each node $i$ as \begin{equation}
   M_{i} \dot{\omega_i}\left(t\right)=-D_i \omega_i \left(t\right)+ \epsilon_i \Gamma_i(t) + P^m_i +P^e_i,
\end{equation}
where $\omega_i=f_i/(2\pi)$ is the nodal angular velocity, $M_{i}$ is the inertia, $D_i$ is the damping, $\epsilon_i \Gamma_i(t)$ is a noise term and $P^m_i$ and $P^e_i$ are the mechanical power generated or consumed and the transmitted electrical power respectively.
Moving to the bulk description, i.e., defining the total inertia $M:=\sum_{i=1}^{N}M_{i}$ and the bulk angular velocity $\bar{\omega}:=\sum_{i=1}^{N}M_{i}\omega_{i}/M$ and assuming a constant damping to inertia ratio $\gamma=D_{i}/M_{i}$ \cite{Weixelbraun2009} as well as balanced electrical and mechanical power on average, i.e., $\sum_{i=1}^{N}P^m_{i}=0,~ \sum_{i=1}^{N}P^e_{i}=0$, we obtain 
\begin{equation}
    \frac{\text{d}}{\text{d}t}\bar{\omega}\left(t\right)=-\gamma\bar{\omega}\left(t\right)+\frac{1}{M}\sum_{i=1}^{N}\epsilon_{i} \Gamma_i \left(t\right).
    \label{eq:Freq_OrnsteinUhlenBeck}
\end{equation}
The equation from the main text is re-obtained by identifying $\Delta P$ as $\sum_{i=1}^{N}\epsilon_{i} \Gamma_i \left(t\right)$.
If we assume that the noise $\Gamma_i(t)$ at each node is approximately Gaussian with zero mean and standard deviation $1$ (the amplitude is included in $\epsilon_i$), we can formulate the Fokker--Planck equation \cite{Gardiner1985} of this Ornstein--Uhlenbeck process \eqref{eq:Freq_OrnsteinUhlenBeck} as
\begin{equation}
    \frac{\partial p}{\partial t}=\gamma\frac{\partial}{\partial\bar{\omega}}\left(\bar{\omega}p\right)+\frac{1}{2}\frac{1}{M^2}\left[\sum_{i=1}^{N}\epsilon_{i}^2 \right]\frac{\partial^{2}p}{\partial\bar{\omega}^{2}}.
\end{equation}
The resulting probability density function $p\left(\bar{\omega}\right)$ is a normal distribution with mean 0 and standard deviation 
\begin{equation}
    \sigma=\sqrt{\frac{\sum_{i=1}^{N}\epsilon_{i}^{2}}{\gamma M^{2}}}.
\end{equation}
As an additional simplification, let us assume identical noise $\epsilon_i=\xi$ and identical inertia $M_i=m$ at all nodes, i.e., $M=Nm$. Then, the standard deviation is given as
\begin{equation}
    \sigma=\sqrt{\frac{N \xi^2}{\gamma m^{2}N^2}}=\sqrt{\frac{\xi^2}{\gamma m^{2}}}\frac{1}{\sqrt{N}}.
    \label{eq:scaling_law}
\end{equation}
This means the standard deviation decays approximately as $\sigma \sim 1/\sqrt{N}$. An important assumption when comparing areas with Supplementary Eq. \eqref{eq:scaling_law} is that we assume similar noise $\xi$, damping $\gamma$, and inertia $m$. While the inertia per machine will likely be similar in the different areas, the effective damping depends on the control applied and the noise on the mix of generators (nuclear, hydro, coal, wind, solar, ...) and the nature of the demand fluctuations.

To take these additional external factors into account when comparing the empirical data of the various synchronous areas, we do not compare the standard deviation  several areas but their aggregated noise amplitudes $\epsilon$. 
To retrieve the aggregated noise amplitude $\epsilon$ we employ a non-parametric Nadaraya--Watson estimator to extract the Kramers--Moyal moments of the underlying stochastic dynamics.
For a timeseries $x(t)$, the $n$th Kramers--Moyal moment can be extracted via
\begin{equation}
M_n(x,t) \!=\!\! \lim_{\Delta t\to 0} \!\frac{1}{\Delta t} \!\left \langle \left ( x'(t\!+\!\Delta t)\!-\!x'(t)\right )^n \!\vert x'(t)\!=\!x\right \rangle\!,
\end{equation}
where the averaging process is made more precise by implementing a kernel-density function $K_h(\cdot)=h^{-1}K(\cdot/h)$ with a bandwidth $h$.
The Nadaraya--Watson estimator $W_h(x)$, at point $i$, is given by \cite{Lamouroux09}
\begin{equation}
W_h(x)_i=\frac{K_h(x-x'_i)}{\sum_{j=1}^S K_h(x-x'_j)},
\end{equation}
where we take $K_h(x)$ to be an Epanechnikov function, compact in $\mathbb{R}_{[-1,1]}$, and $S$ is the number of data points.
The aggregated noise amplitude $\epsilon$ is retrieved studying the second Kramers--Moyal moment, and employing the aforementioned estimator, resulting in
\begin{equation}
M_2(x) = \frac{1}{S \Delta t} \sum_{i=1}^S W_h(x)_i [ x'(t\!+\!\Delta t)-x'(t) ]_i^2  = \epsilon^2,
\end{equation}
with $\Delta t = 1s$.
All Kramers--Moyal moments are easily retrievable, see Refs.~\citep{Gorjao2019, Rinn16}. The so extracted aggregated noise amplitude $\epsilon$ will closely follow the empirical Fokker--Planck equation and should therefore resemble a $\epsilon \sim 1/\sqrt{N}$ decay, as any deterministic effects are filtered out.

Finally, while we expect the noise to decay with the size, it is well-justified to add a constant to our previously derived expression \eqref{eq:scaling_law}, modifying it to 
\begin{equation}
    \epsilon \sim \sigma = \frac{a}{\sqrt{N}}+b.
    \label{eq:scaling_law_with_constant}
\end{equation}
We add the constant $b$ to take the effect of deadbands into account. All synchronous areas have deadbands \cite{Machowski2011}, i.e. frequency ranges for which there is no (primary) control active and the frequency dynamics evolves freely. This means for a certain range $|\omega|<\omega_\text{deadband}$ the damping to inertia ratio $\gamma$, which explicitly includes primary control, is much smaller and the frequency randomly evolves and contributes a minimum noise contribution $b$.

In addition to the standard deviation $\sigma$ or the aggregated noise amplitude $\epsilon$, we might also consider using a Gaussian kernel to detrend the data and only analyse the standard deviation of the detrended data. Finally, we could also consider using the standard deviation of the increments (at the smallest available time lag $\tau=1$) as a proxy of the actual noise.
We compare these different approaches in Supplementary Fig. \ref{fig:Scaling_fluctuations_measures}:
Moving from top to bottom: Taking the unfiltered standard deviation of the data (Supplementary Fig. \ref{fig:Scaling_fluctuations_measures} a), does give a rough trend but areas as South Africa (ZA) or Great Britain (GB) have a much larger standard deviation than we would expect from the scaling law. When we introduce the de-trending, the overall standard deviation drops (Supplementary Fig. \ref{fig:Scaling_fluctuations_measures} b) but the problems with GB and ZA persist. Likely this is caused by control regulations that allow larger deviations than they are allowed for example in Continental Europe. Next, we extract the aggregated noise amplitude $\epsilon$ (Supplementary Fig. \ref{fig:Scaling_fluctuations_measures} c) and observe a decay, well-approximating the conjectured scaling law. Finally, we note that using the standard deviation of the increments at lag $\tau=1$s is almost identical to the aggregated noise amplitude $\epsilon$ (Supplementary Fig. \ref{fig:Scaling_fluctuations_measures} d). Comparing all four plots, in particularly in terms of the standard deviation of the best fit (shaded area), leads to the conclusion that panels c and d are the best descriptors. Both the extracted noise $\epsilon$ as well as the increments on the one second scale $\Delta f_1$ are good proxies of the diffusive forces acting on the synchronous area.

\begin{figure}[ht]
\includegraphics[width=0.5\linewidth]{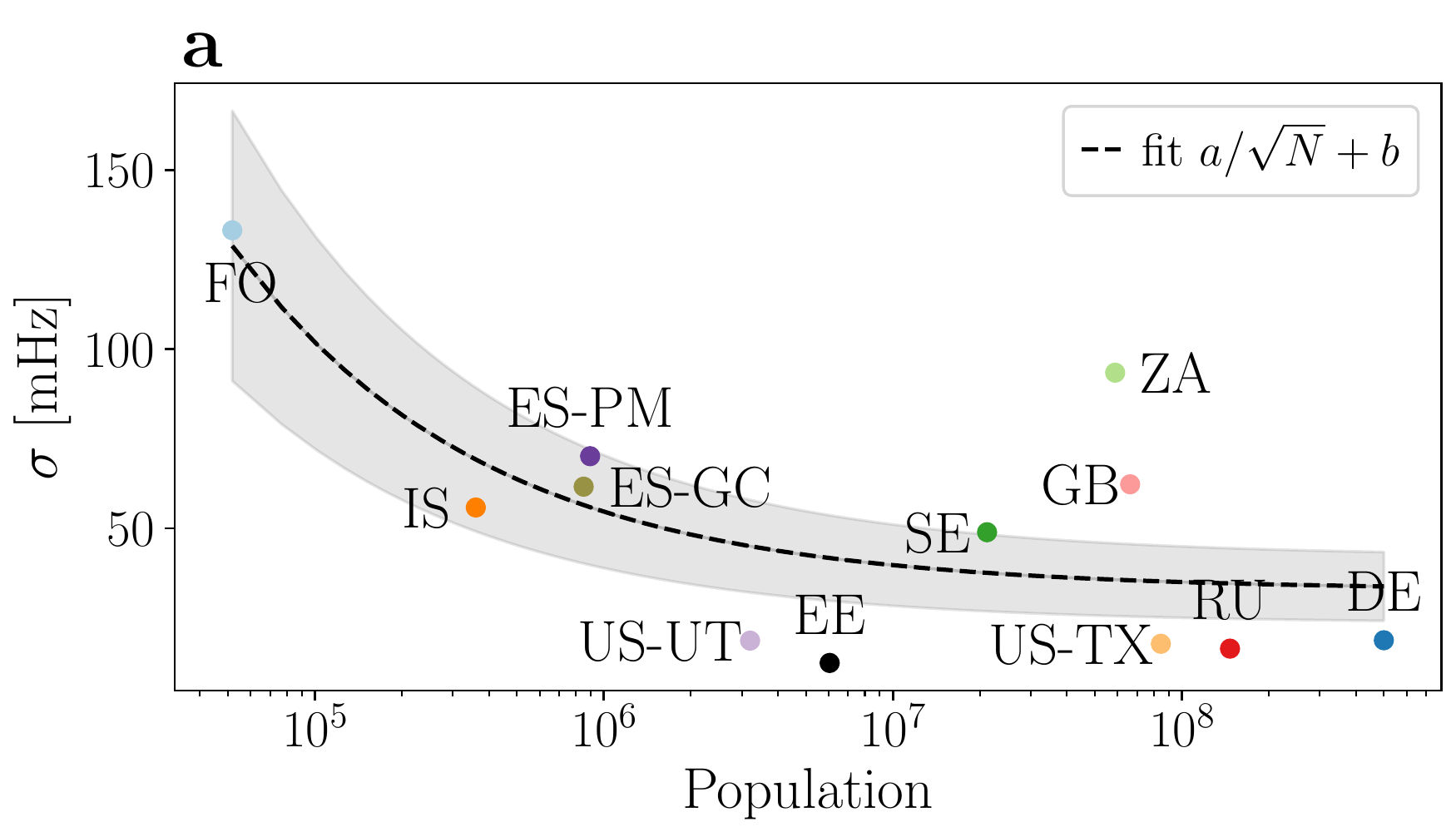}  
\includegraphics[width=0.5\linewidth]{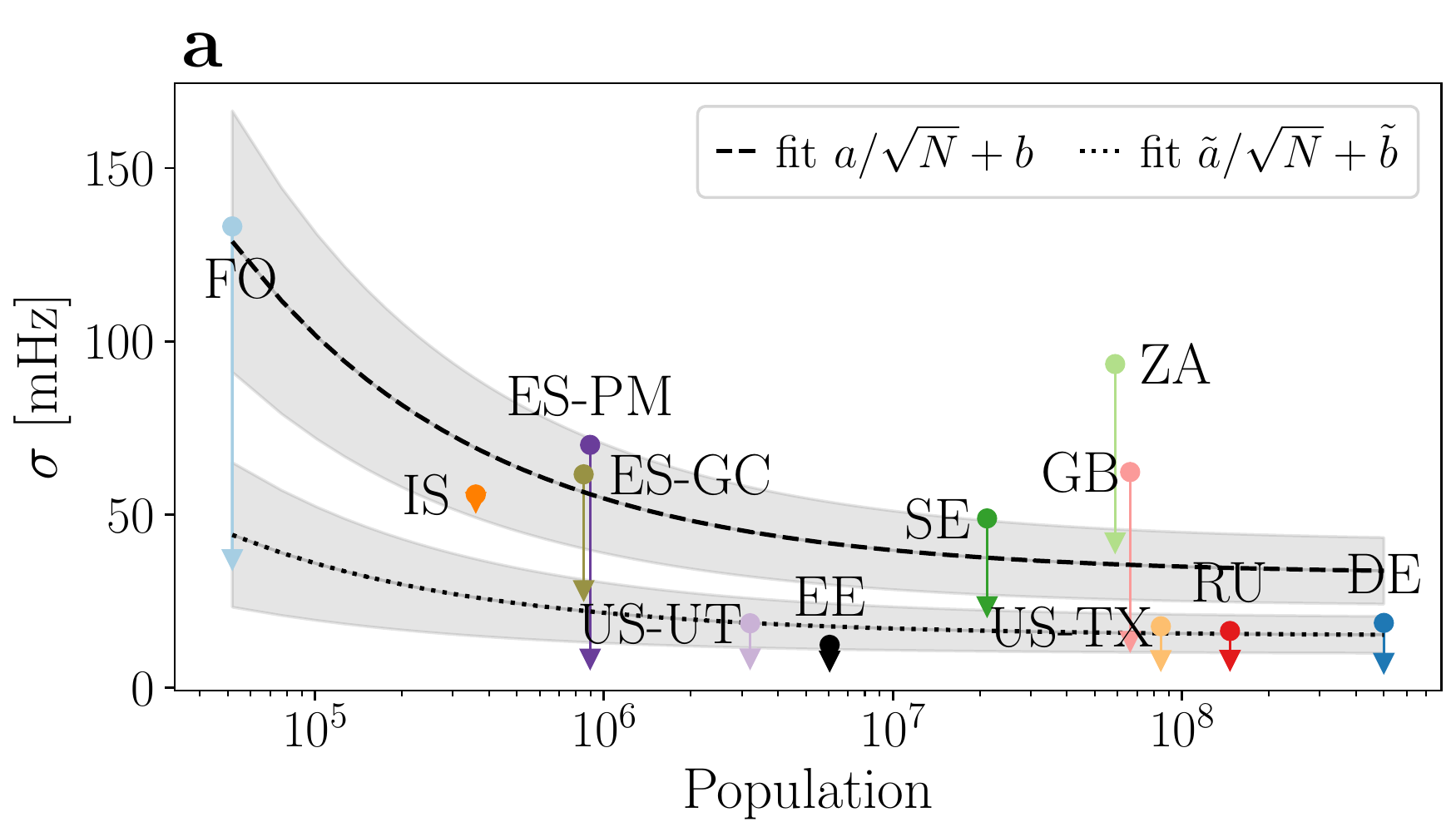}  
\includegraphics[width=0.5\linewidth]{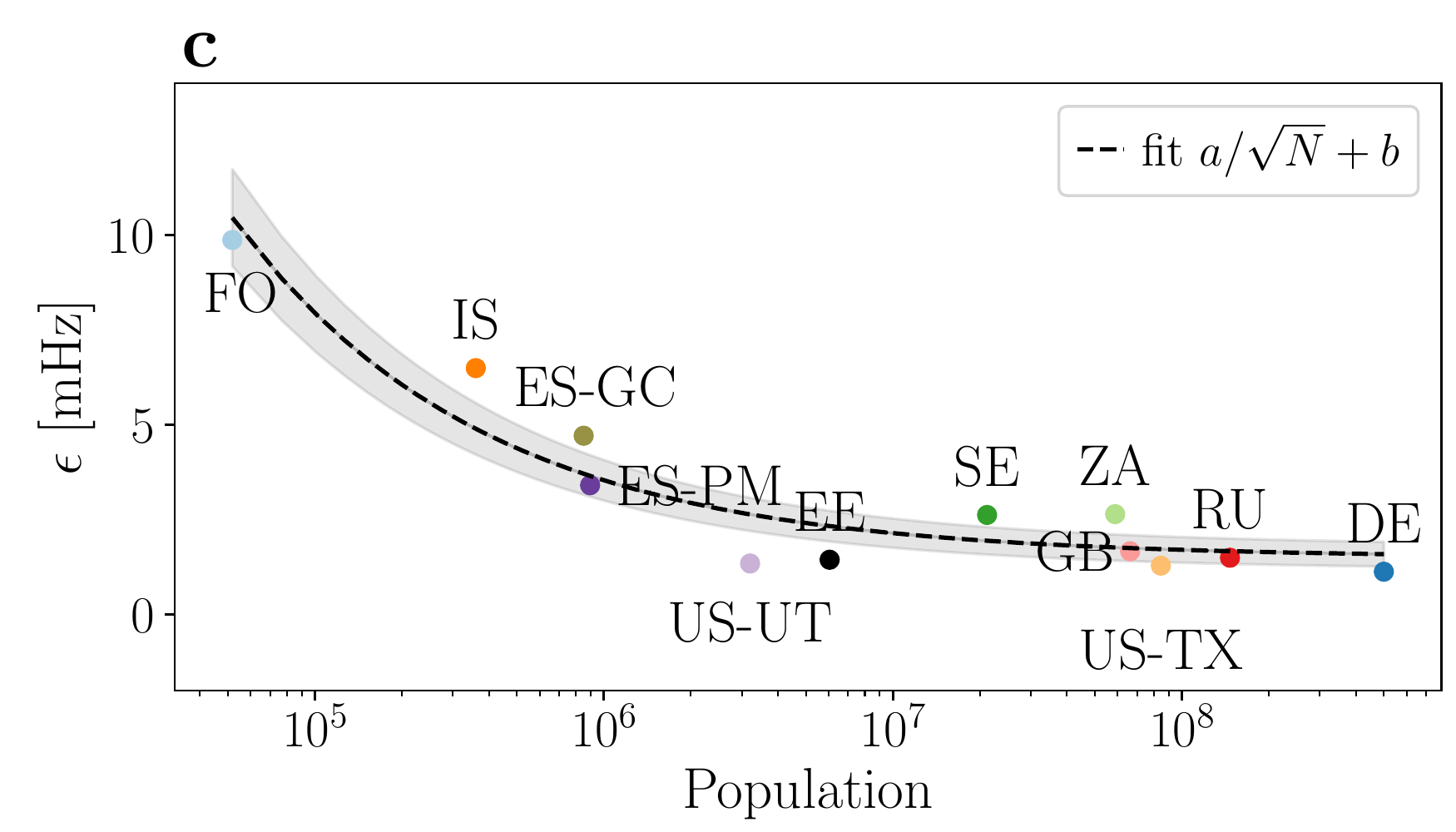}  
\includegraphics[width=0.5\linewidth]{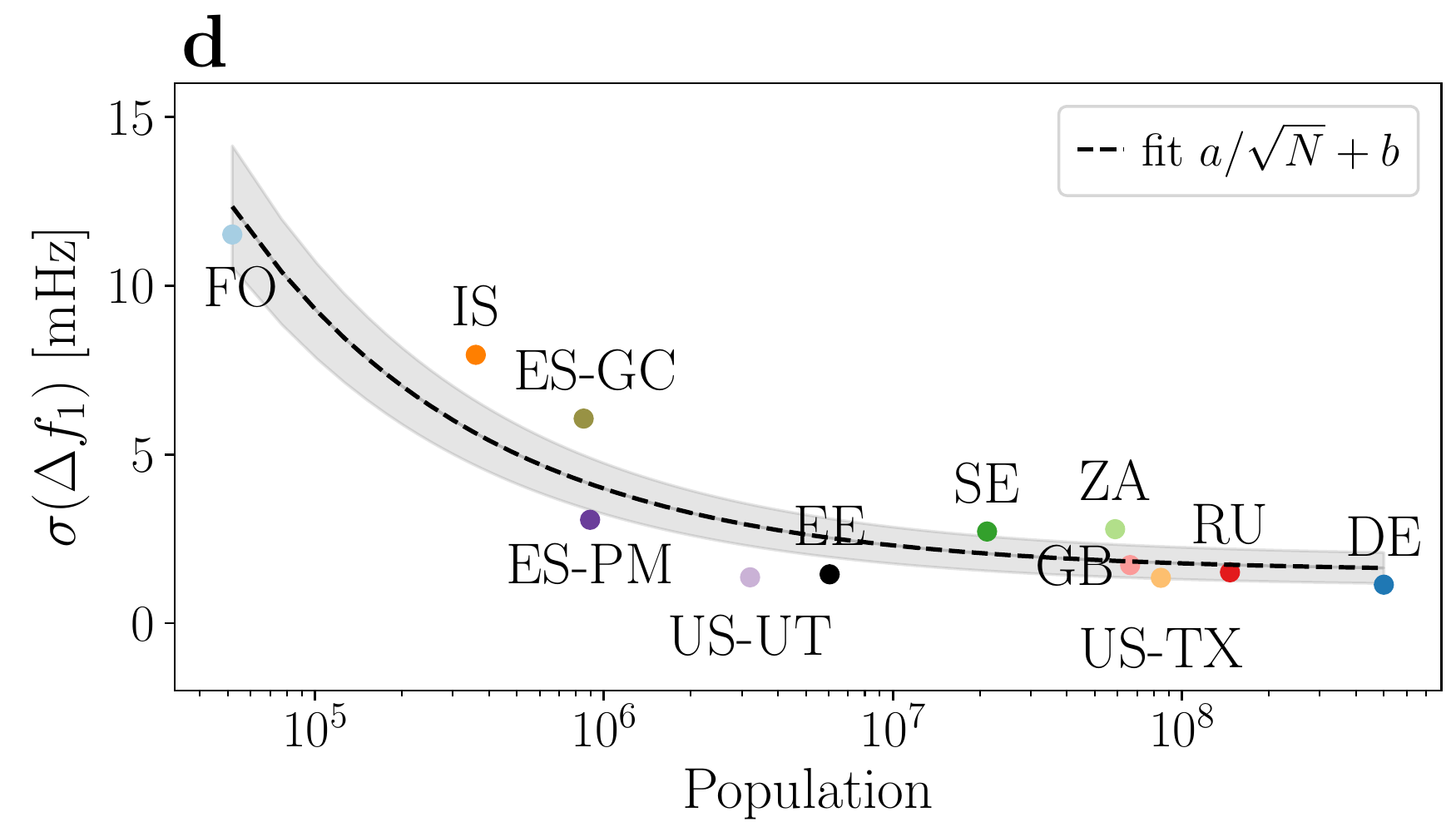}  
\caption{Average fluctuations decrease with increasing grid size with increments $\Delta f_1$ and aggregated noise amplitude $\epsilon$ as the best descriptors. We plot the standard deviation (a), filtered standard deviation (b), the aggregated noise amplitude (c) and the standard deviation of the increments at time lag $\tau=1$s (d). The shaded area gives the standard deviation of the fit.}
\label{fig:Scaling_fluctuations_measures}
\end{figure}

\clearpage
\section*{Supplementary Note 3}
\subsection*{Castaing's model for increments}

\begin{figure*}[ht]
\includegraphics[width=1\linewidth]{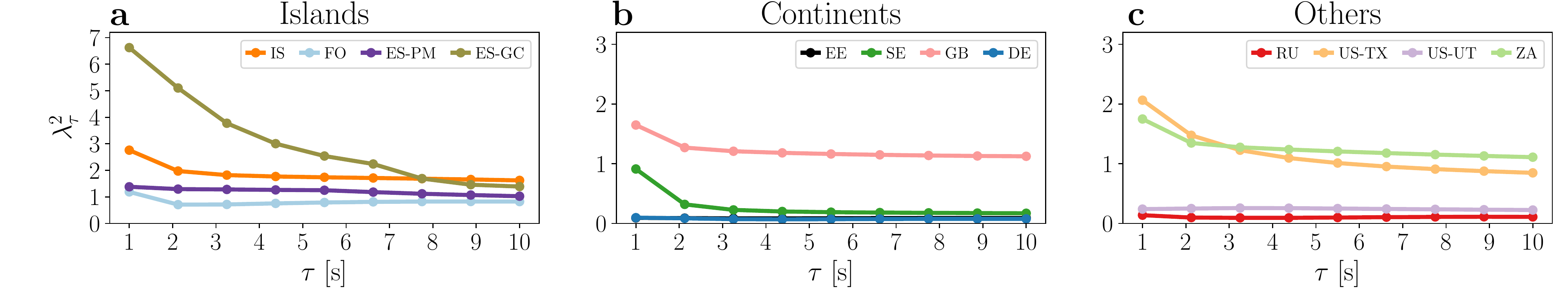}
\caption{Increment analysis reveals non-Gaussian characteristics, dominantly in islands. We plot the Castaing parameter $\lambda_\tau^2$, given by Supplementary Eq. \eqref{eq:Castaing}, for the different examined power grid frequency recordings.
We observe a non-vanishing intermittency in Gran Canaria (ES-GC), Iceland (IS), Faroe Islands (FO), Mallorca (ES-PM), Britain (GB), Texas (US-TX) , and South Africa (ZA).
In contrast, the increments' distribution of the Baltic (EE), Continental Europe (DE), Nordic (SE), Russia (RU) synchronous areas and the Western Interconnection (US-UT) approach a Gaussian distribution.}
\label{fig:Castaing}
\end{figure*}

In the main text we quantified deviations from Gaussianity of the increment distributions by the use of the excess kurtosis $\kappa-3$, which should decay to zero if the distributions under consideration are Gaussian. Here, we offer a more theoretical view by using Castaing's model.
This model describes the deviations of the increments' distribution from a Gaussian distribution \cite{Castaing1990,Castaing1994,Castaing1996} and has already been applied to frequency analysis \cite{Haehne2018,HaehnePropagation2019}.
The increments $\Delta f_\tau$ with the lag $\tau$ have a non-homogenous scaling, which leads to distributions with high kurtosis, and sometimes non-zero skewness \cite{Tabar2019}.
The rationale is that the process is a superposition of several subset processes with distinct scales, similar to superstatistics \cite{Beck2003,Beck2005}. Specifically, Castaing's model is a special case of log-normal superstatistics applied to increments.
The probability density function (PDF) of the increments $p(\Delta f_\tau, \sigma_\tau)$ is a function of the widths $\sigma_\tau$, given by
\begin{equation}\label{eq:P_deltaf}
    p(\Delta f_\tau, \tau) = \int_{-\infty}^\infty L_\lambda(\sigma_\tau) p_0(\Delta f_\tau, \sigma_\tau) d \ln{\sigma_\tau},
\end{equation}
where the underlying subset processes are assumed to have a Gaussian distribution $p_0(\Delta f_\tau, \sigma_\tau)\sim \mathcal{N}(0,\sigma_\tau)$ of some variance $\sigma$ and $L_\lambda(\sigma_\tau)$ accounts thus for the scales of each superposition.
This scale function $L_\lambda(\sigma_\tau)$ is conjectured to be log-normally distributed
\begin{equation}\label{eq:L_lambda}
    L_\lambda(\sigma_\tau) = \frac{1}{\lambda_\tau \sqrt{2\pi}}\mathrm{exp}\!\left[-\frac{\mathrm{ln}^2\!\left(\sigma_\tau/\sigma_0\right)}{2\lambda_\tau^2}\right],
\end{equation}
with $\lambda_\tau^2$ being the Castaing parameter.
For the case $\lambda_\tau^2 \to 0$, the distribution $L_\lambda(\sigma_\tau)$ approached a $\delta$-distribution, and the increments $\Delta f_\tau$ are purely Gaussian distributed.
As $\lambda_\tau^2$ increases, the convolution includes more scales and the tails of the PDF of the increments enlarge.
Inserting Supplementary Eq.~\eqref{eq:L_lambda} into Supplementary Eq.~\eqref{eq:P_deltaf} yields the explicit PDF of the increments as
\begin{equation}\label{eq:full_p_deltaf}
\begin{aligned}
    p( \Delta & f_\tau, \tau) = \\&\frac{1}{\lambda_\tau 2\pi}\int_0^\infty\! \frac{d \sigma_\tau}{\sigma_\tau^2} \mathrm{exp}\!\left[-\frac{\Delta f_\tau^2}{2\sigma_\tau^2}-\frac{\mathrm{ln}^2\!\left(\sigma_\tau/\sigma_0\right)}{2\lambda_\tau^2}\right].
\end{aligned}
\end{equation}
The increments intermittency behaviour is thus solely described by the Castaing parameter $\lambda_\tau^2$.
Multiplying both sides of Supplementary Eq. \eqref{eq:full_p_deltaf} by $\Delta f_\tau^2$ and integrating over $[-\infty,\infty]$, we find
\begin{equation}
    \sigma_0^2 = \langle \Delta f_\tau^2 \rangle~ \mathrm{exp}\!\left[-2\lambda_\tau^2\right].
\end{equation}
One can now recover the Castaing parameter $\lambda_\tau^2$ by extracting the fourth-order statistical moment, i.e., the kurtosis $\kappa_{\Delta f}(\tau)$, of the PDFs of $\Delta f_\tau$ as function of the lag $\tau$.
The Castaing parameter $\lambda_\tau^2$ results in
\begin{equation}\label{eq:Castaing}
    \lambda_\tau^2 = \ln{\left(\frac{\kappa_{\Delta f}(\tau)}{3}\right)},
\end{equation}
i.e., for a Gaussian distribution with $\kappa_{\Delta f}(\tau)=3$ the Castaing parameter decays to $0$. 

To extract the Castaing parameter, we compute the  increment statistics $\Delta f_\tau$ of a certain timeseries $f_\tau$ and calculate the kurtosis $\kappa_\tau$ of the PDF of the increments.
Then, we take the normalised logarithm as in Supplementary Eq.~\eqref{eq:Castaing} and do so over the desired range of increments time-lag $\tau$.

Computing the Castaing parameter $\lambda_\tau^2$ for our data, we observe very heterogeneous results between the various synchronous areas, see Supplementary Fig.~\ref{fig:Castaing}.
In some areas, the intermittent behaviour of the increments $\Delta f_\tau$ is subdued and the overall distribution approaches a Gaussian distribution (in EE, DE, SE, RUS, and US-TX), i.e., the Castaing parameter $\lambda_\tau^2$ approaches 0.
On the other hand, all islands display large and non-vanishing intermittent behaviour, as well as GB, US-TX, and ZA.
Iceland (IS) but particularly Gran Canaria (ES-GC) shows impressive deviations from Gaussianity that require detailed modelling in the future. 

With Castaing's model and parameter we used an alternative approach to quantify heavy tails, instead of purely using the kurtosis $\kappa$. 
A further advantage of Castaing's approach is that it allows us to model the increments as superimposed distributions, complementary to superstatistics of the aggregated frequency distributions  \cite{Schaefer2017a}.
An explicit increment model will be left for future work. 

\clearpage
\section*{Supplementary Note 4}
\subsection*{Further increment analysis}

\begin{figure*}[ht]
\includegraphics[width=.45\linewidth]{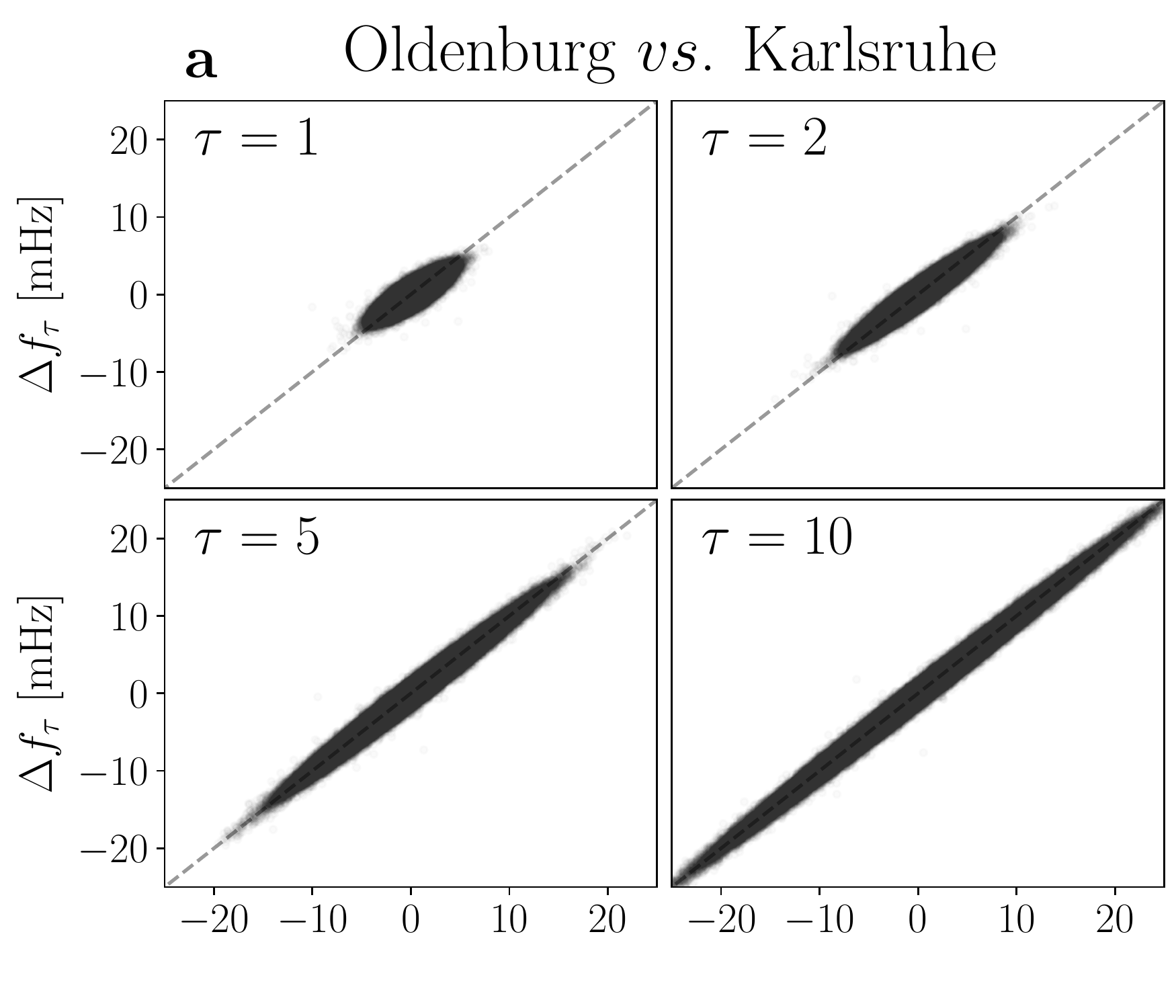}\includegraphics[width=.45\linewidth]{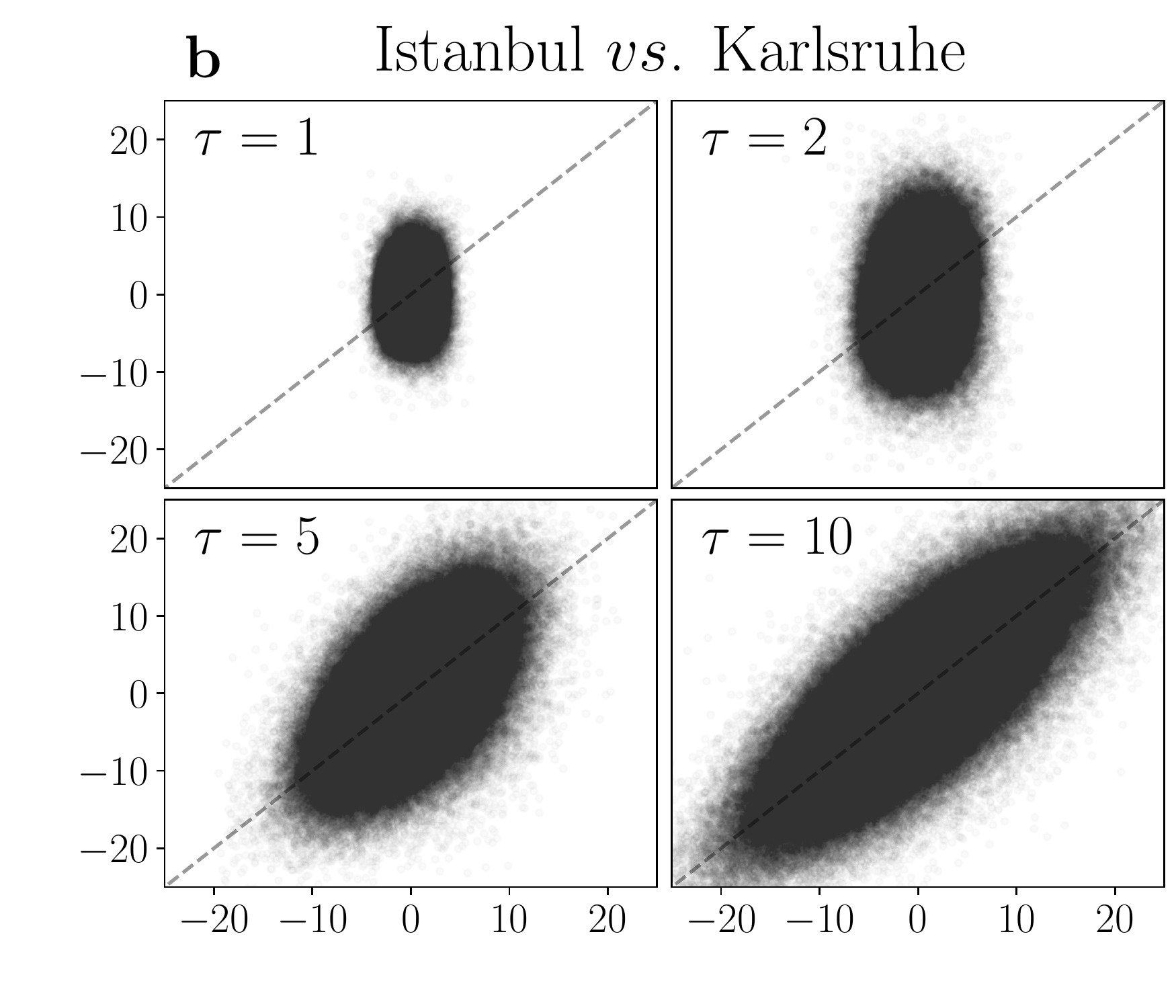}
\includegraphics[width=.45\linewidth]{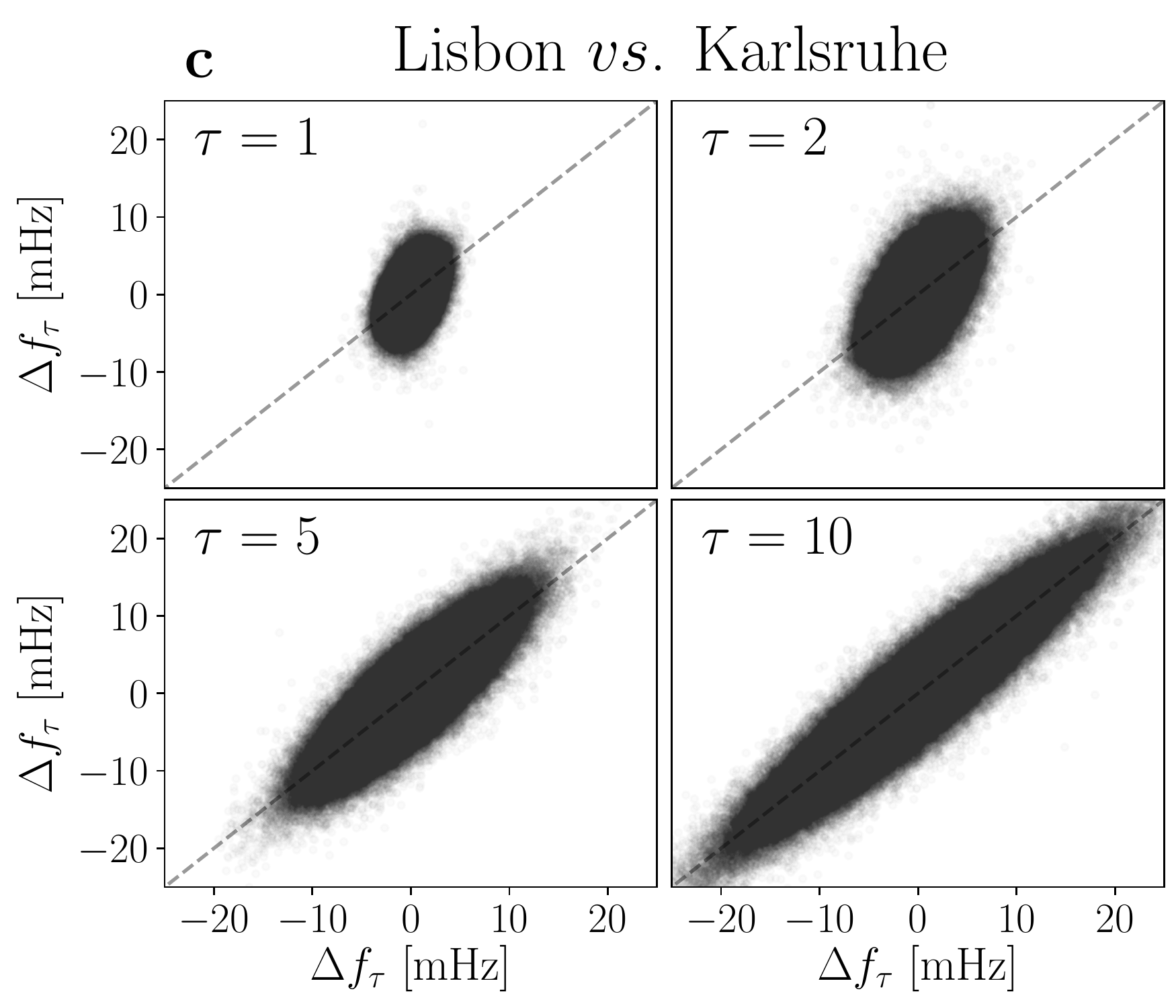}\includegraphics[width=.45\linewidth]{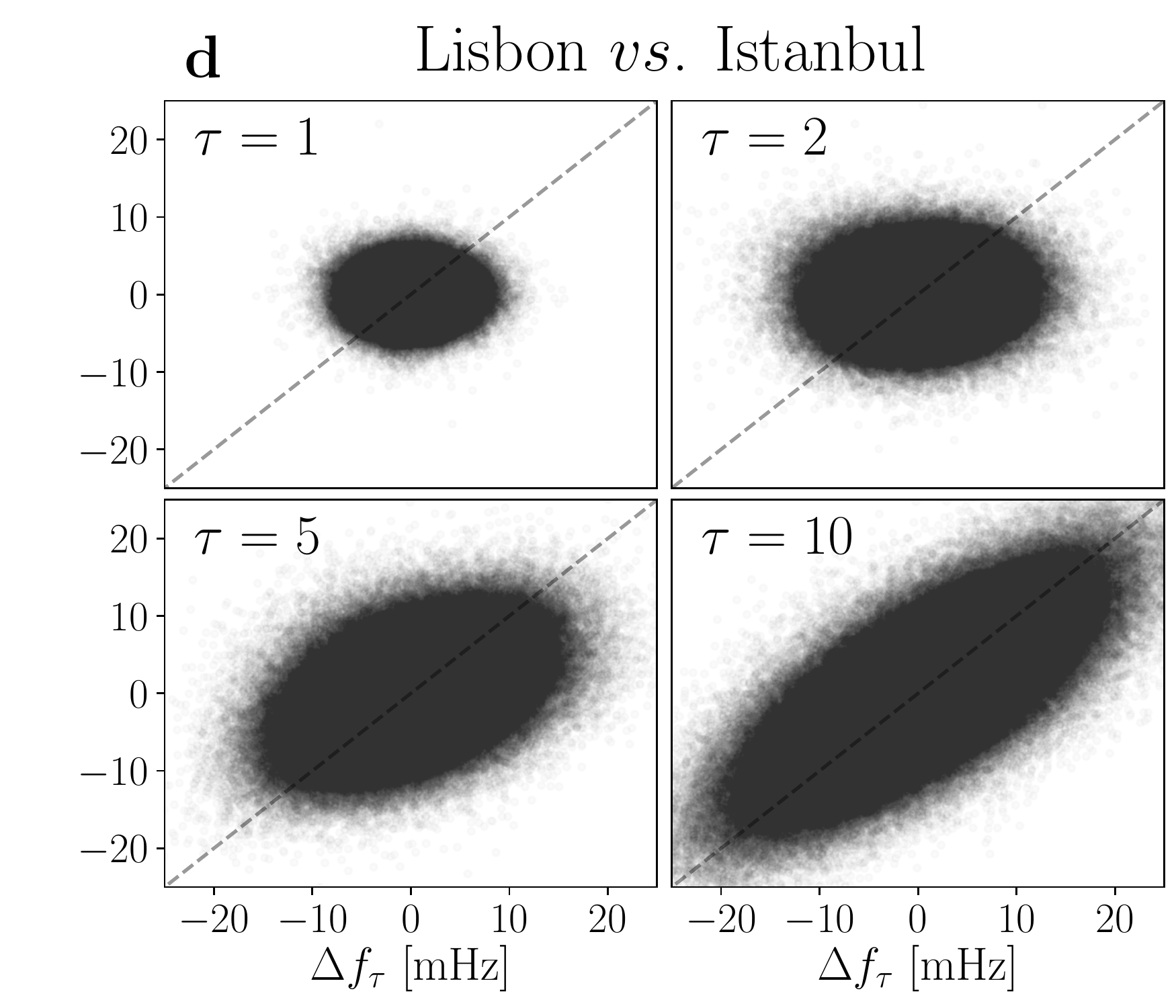}
\caption{Long increment lags lead to increased correlations.
We repeat the increment analysis from the main text but with larger lags $1$\!~s $\leq \tau \leq 10$\!~s.
Note that each of the 2$\times$2 subpanels is now using four different lags $\tau$ but the overall figure still follows the same arrangement as in the main text.}
\label{fig:DifferentIncrements}
\end{figure*}

We complement the increment analysis from the main text by considering larger time lags $\tau>1$\!~s.
In particular, we compute the increments \cite{Anvari2016,Haehne2018} $\Delta f_\tau =f(t+\tau)-f(t)$ for the four measurement sites in Continental Europe: Karlsruhe, Oldenburg, Istanbul and Lisbon. For the pairs investigated in the main text, we consider a lag $\tau = 1, 2, 5 , 10$ seconds in Supplementary Fig.~\ref{fig:DifferentIncrements}.
Since the scatter plots report increments at two locations at the same time $t$, points on the diagonal indicate large correlations, while circles or ellipses aligned with one axis indicate no correlations. 
The results for Oldenburg vs. Karlsruhe are almost independent of the specific time lag $\tau$. The magnitude of the increments increases for increasing time lag $\tau$ but almost all values follow the diagonal, indicating a high correlation on the full time scale for $1$ to $10$ seconds. 
In contrast, the increment plots involving Istanbul and Lisbon change much more with increasing lag $\tau$ as their increments on the time scale of 1 second are almost completely uncorrelated but at $5$ to $10$ seconds, an increasing number of points follow the diagonal, i.e., fluctuation events become correlated. Similar to the detrended fluctuation analysis (DFA) from the main text, we again observe that short time scales are independent, while we approximate a bulk description for longer time scales. This observation is also consistent with claims found e.g in \cite{zhang2019fluctuation} that high frequency fluctuations do not penetrate the grid over long distances but lower frequency fluctuations do.

\begin{figure}[ht]
\includegraphics[width=0.5\linewidth]{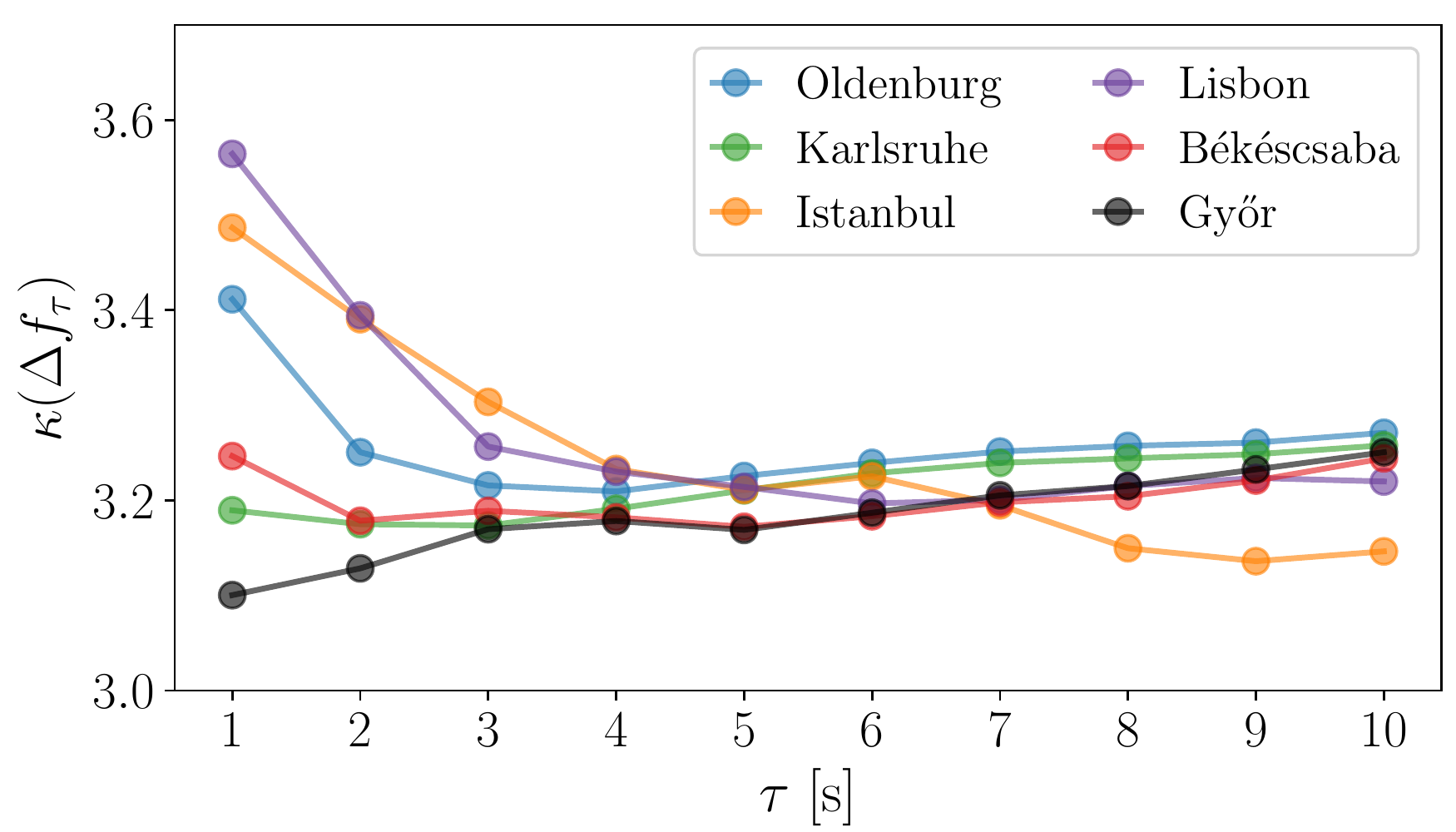}
\caption{The kurtosis $\kappa$ of the increment statistics $\Delta f_{\tau}$ decreases with increasing lag $\tau$.
We plot the kurtosis $\kappa$ of each recording site as a function of the increment lag $\tau$.}
\label{fig:KurtosisIncrements_vs_Delay}
\end{figure}

Finally, we may further investigate how the increment distributions look like, in particular with respect to their large deviations, i.e., their heavy tails measured by the kurtosis of the increment distributions. 
Computing the kurtosis $\kappa$ of the increment statistics $\Delta f_\tau$ at different lags $\tau$, shows that the deviations from the Gaussian ($\kappa^\text{Gaussian}=3$) decrease \emph{on average}. While the kurtosis shows a small increase in Győr with increasing time lag, the kurtosis at Istanbul and Lisbon is substantially reduced, see Supplementary Fig.~\ref{fig:KurtosisIncrements_vs_Delay}.
This finding is consistent with \cite{Haehne2018}, where the authors also found that long lags lead to approximately Gaussian increment statistics.
Here, we go further in that we observe spatial differences already at a time resolution of $1$ second, in particular for locations far away from Karlsruhe.

\clearpage
\section*{Supplementary Note 5}
\subsection*{Details on Detrended Fluctuation Analysis (DFA)}
Detrended Fluctuation Analysis (DFA) \cite{Peng1994,Kantelhardt2002} studies the fluctuation of a given process by considering increasing segments of the
timeseries.
Take a timeseries $X(t)$ with $N$ elements $X_i$, $i=1,2, \dots, N$. Obtain the detrended profile of the process by
defining $$ Y_i = \sum_{k=1}^i \left ( X_k - \langle X \rangle \right),~\text{for}~i=1,2, \dots, N,\nonumber $$ i.e., the cumulative sum
of $X$ subtracting the mean $\langle X \rangle$ of the data.
Section the data into smaller non-overlapping segments of length $s$,
obtaining therefore $N_s = \text{int}(N/s)$ segments. Given the total length of the data is not always a multiple of the segment's
length $s$, discard the last points of the data.
Consider the same data, apply the same procedure, but this time discard instead the first points of the data.
One has now $2N_s$ segments. 
To each of these segments fit a polynomial $y_v$ of order $m$ and calculate the variance of the difference of the data to the
polynomial fit 
\begin{equation}
F^2(v,s) = \frac{1}{s} \sum_{i=1}^s [Y_{(v-1)s,i} - y_{v,i}]^2, ~\text{for}~v=1,2, \dots, N_s,\nonumber
\end{equation}
where $y_{v,i}$ is
the polynomial fitting for the segment $i$ of length $v$.
One also has the freedom to choose the order of the polynomial fitting.
This gives rise to the denotes DFA1, DFA2, $\dots$, for the orders chosen. 
Notice $F^2(v,s)$ is a function of each variance of each $v$-segment of data and of the different $s$-length segments chosen.
One can define the fluctuation function $F^2(s)$ by averaging each row of segments of size $s$ 
\begin{equation}
F^2(s) = \frac{1}{N_s}\left\{ \sum_{v=1}^{N_s} F^2(v,s) \right\}^{1/2}. \nonumber
\end{equation}
The inherent scaling properties of the data, if the data displays power-law correlations, can now be studied in a log-log plot of $F^2(s)$ versus $s$. 
Herein the scaling of the data obeys a power-law with exponent $h$ as 
\begin{equation}
F^2(s) \sim s^{h},\nonumber
\end{equation}
where $h$ is the \textit{self-similarity} exponent (which may be multifractal) and relates directly to the Hurst index.
The self-similarity exponent $h$ is calculated by finding the slope of this curve in the log-log plots.
For a detailed explanation of DFA, see \cite{Ihlen2012}.
The analysis implemented here is based on \cite{MFDFA2020}.

\clearpage
\section*{Supplementary Note 6}
\subsection*{Principal Component Analysis}
\begin{figure*}[ht]
\includegraphics[width=1\linewidth]{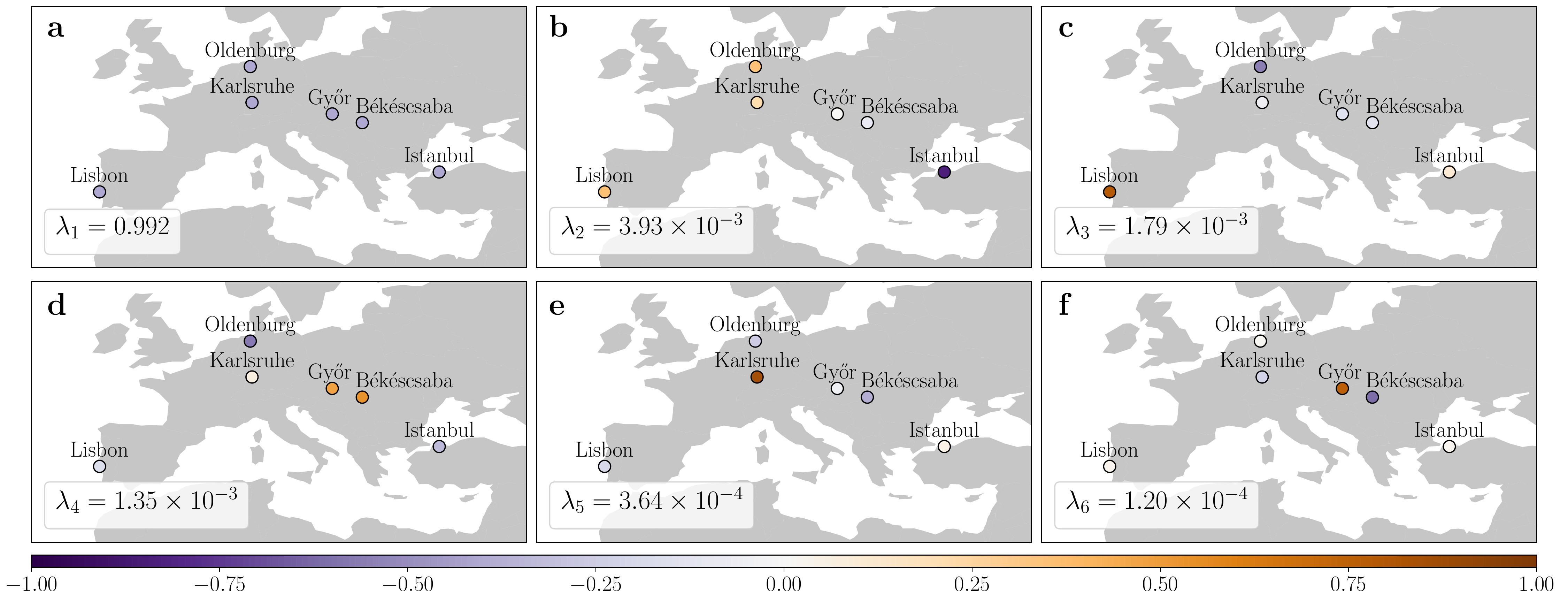}
\caption{The principal components reveal the spatial structure of inter-area modes.
The maps show the colour-coded entries of the normalised principal components $\mathbf{f}_m$.
The spatial structure corresponds to synchronous behaviour (a), East-West oscillations (b), North-South oscillations (c), Coastal-inland fluctuations (d-e), and intra-Hungarian oscillations (f).
The synchronous component describes the frequency bulk behaviour and thus explains already $\lambda_1=99.2$\% of the total variance.}
\label{fig:PCA_Maps}
\end{figure*}

\begin{figure}[ht]
\includegraphics[width=0.6\linewidth]{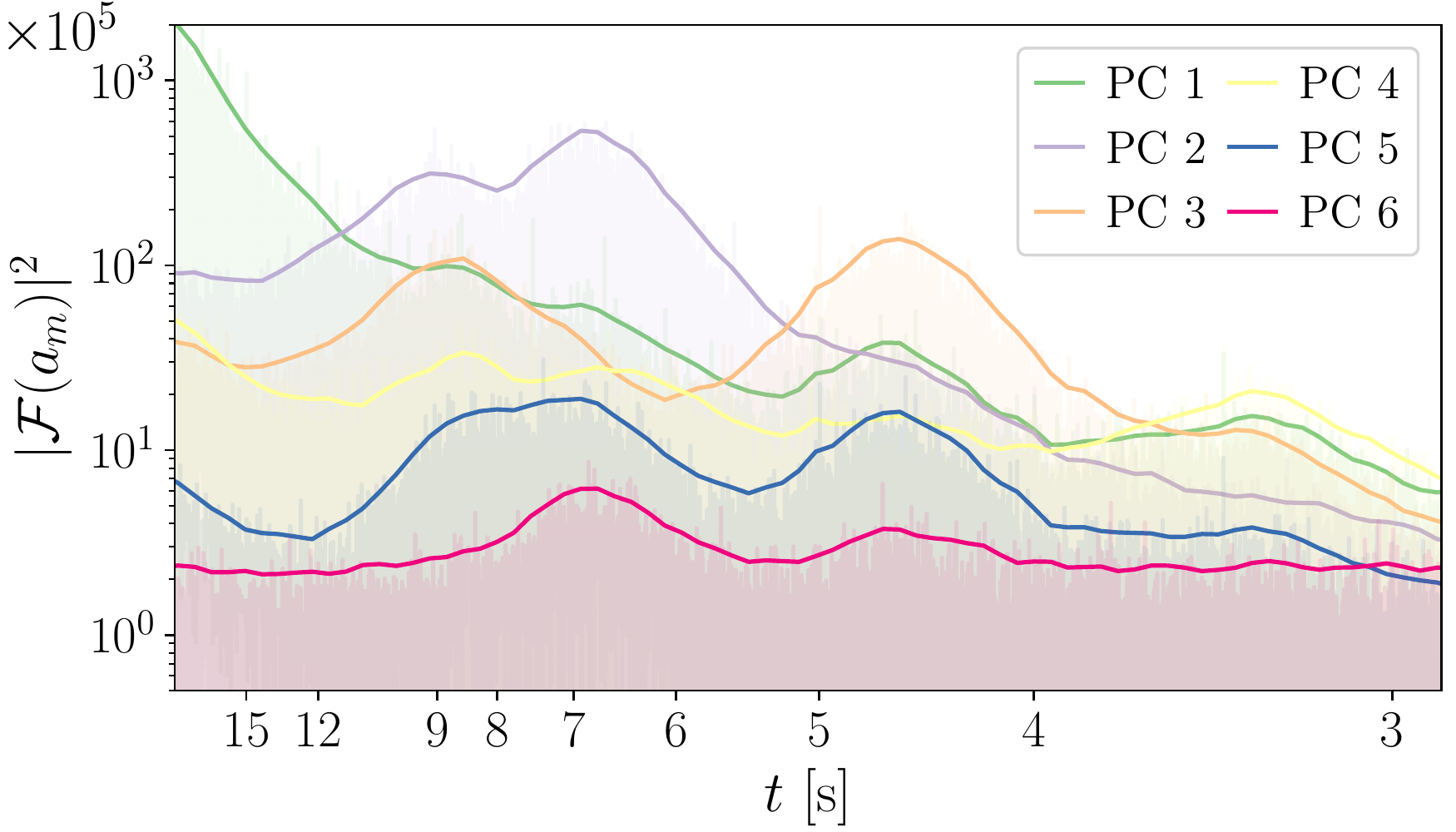}
\caption{The inter-area modes oscillate with a period between $3$ and $9$ seconds. The figure displays squared Fourier amplitudes $\left|\mathcal{F}\left(a_m(t)\right)\right|^2$ of the mode amplitudes $a_m(t)$.
The largest inter-area oscillations occur with a period of $t=7$~s and $t=4.5$~s, which corresponds well to the results in \cite{grebe_low_2010}.
In the main text, we only discussed the first three modes as the most dominant ones.}
\label{fig:PCA_all_modes}
\end{figure}

We apply a principal component analysis (PCA) to our multivariate frequency time series in order to extract the dominant modes of inter-area oscillations.
Generally, a PCA identifies the orthogonal linear subspaces (principal components) that maximise the projected variance of the data \cite{bishop_pattern_2007}.
Let $\mathbf{f}(t)$ be the vector containing the frequency recordings from all six locations within Europe.
The principal components are then given by the eigenvectors $\mathbf{f}_m(t)$ of the data covariance matrix.
Based on the principal components, we can decompose the centred frequency recordings into uncorrelated time series $a_m(t)$:
\begin{align*}  
    \mathbf{f}(t) = \langle \mathbf{f} \rangle + \sum_{m=1}^N  a_m(t) \mathbf{f}_m.
\end{align*}
The time series $a_m(t)$ describe the amplitude of the frequency recordings $\mathbf{f}(t)$ projected onto the m-th principal component.
Their variance, i.e., the projected variance, is given by the eigenvalue $\tilde \lambda_m$. By rescaling this variance, we obtain the share of total variance explained by the $m$-th component:
\begin{align*}
    \lambda_m = \frac{\tilde \lambda_m}{\sum_m \tilde \lambda_m}.
\end{align*}

We interpret the principal components (PCs) of $\mathbf{f}(t)$ as spatial inter-area modes and the time series $a_m(t)$ as their amplitudes.
Supplementary Fig.~\ref{fig:PCA_Maps} displays all six modes obtained from our synchronised measurements in Continental Europe.
The first mode (PC1) represents a synchronous frequency oscillation at all locations and thus corresponds to the bulk behaviour of the frequency.
The other modes (PC2-PC6) display asynchronous spatial patterns and thus represent inter-area oscillations.
Due to their small amplitude, these modes only explain a low portion of the variance ($\lambda_{2}, ..., \lambda_6 < 0.4\%$), while the bulk mode already covers $\lambda_1 \approx 99.2$\%.
However, PC2 to PC6 uncover the dominant spatial structure of inter-area modes across Continental Europe.
Let us analyse these modes based on their visualisation in Supplementary Fig.~\ref{fig:PCA_Maps} in more detail: PC2 contains one coherent area in Western Europe that oscillates in phase opposition to Istanbul.
In PC3 Lisbon and Istanbul swing in opposition to the Northern measurement locations.
PC2 thus resembles an East-West dipole, while PC3 is similar to a North-South dipole.
The other modes correspond to a Coast-Inland fluctuation (PC4 and PC5) and an intra-Hungarian oscillation (PC6). 

\begin{figure}[ht]
\includegraphics[width=0.6\linewidth]{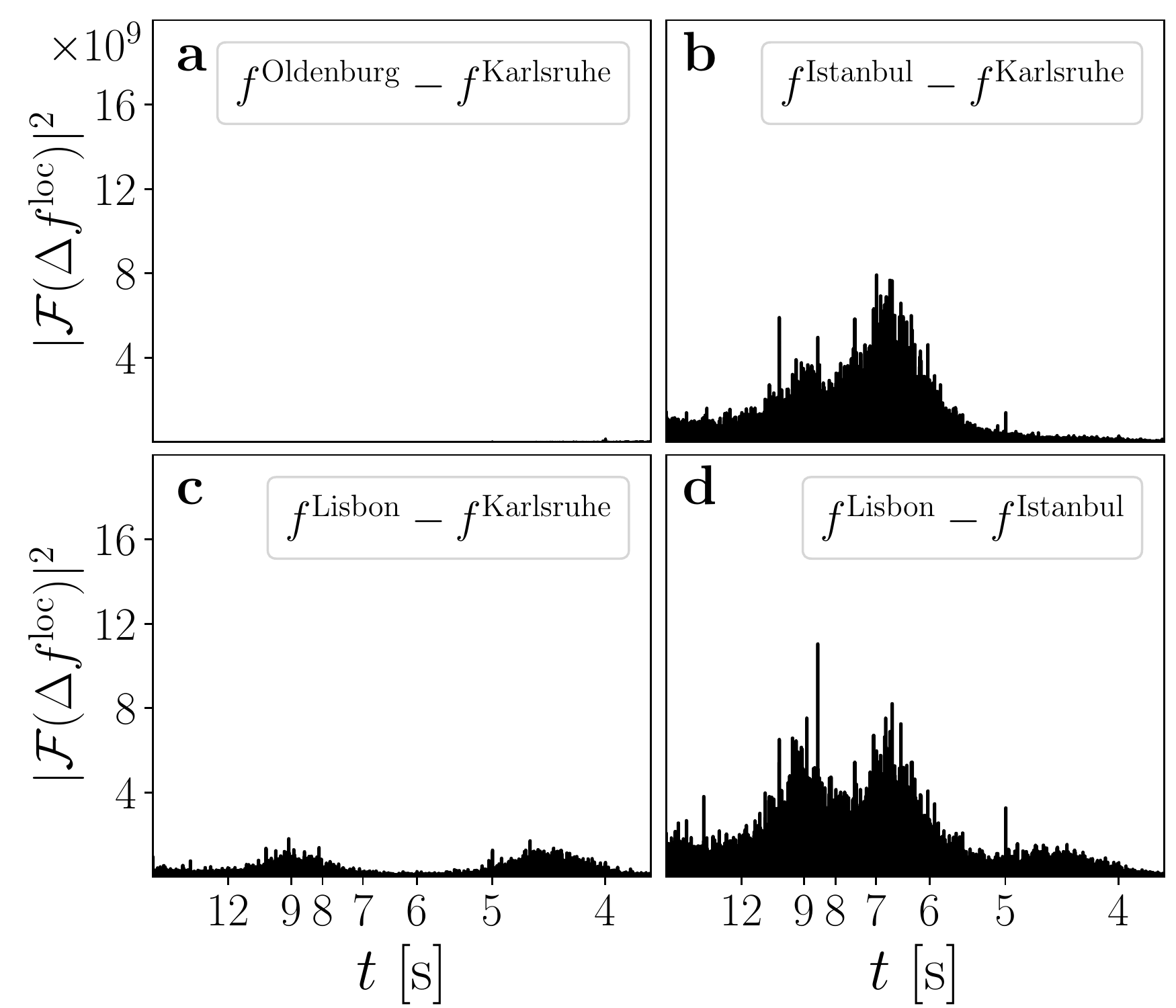}
\caption{Inter-area oscillations between different grid sites.
We perform a spectral analysis of the frequency differences by computing the spectrum of the frequency difference $\Delta f^\text{loc}$ between two sides, e.g. in (e): $\Delta f^\text{loc} = f^\mathrm{Oldenburg}-f^\mathrm{Karlsruhe}$.}
\label{fig:SynchronousMeasurements_Spectrum}
\end{figure}

To reveal the oscillation period of these modes, we analyse the power spectral density (PSD) of their amplitudes $a_m(t)$ (Supplementary Fig.~\ref{fig:PCA_all_modes}).
The East-West dipole (PC2) exhibits a main period length of $t\approx7$~s and a smaller contribution at $t\approx9$~s.
PC3 mainly oscillates with a period pf $t\approx4.5$~s, but there is also a distinct peak at $t\approx9$~s, which also appears in PC2 and PC4.
PC5 and PC6 mainly exhibit oscillations with a period of $t\approx7$~s and $t\approx4.5$~s.
The PSD of bilateral differences between single measurement locations reveals the same main period lengths (Supplementary Fig.~\ref{fig:SynchronousMeasurements_Spectrum}).
Interestingly, no inter-area oscillations can be observed between Karlsruhe and Oldenburg, which could indicate well-balanced power within Germany or could be caused by the limited spatial resolution of the available 6 modes.
Overall, we identify three main oscillations periods of inter-area modes with $t\approx9$~s, $t\approx7$~s, and $t\approx4.5$~s, which is consistent with the typical frequency of inter-area modes \cite{Klein1991}. 

The comparison to a more detailed study suggests that our principal components probably relate to overlaps of different global inter-area modes.
A first indicator for this conclusion is the occurrence of multiple substantial peaks in our PSD in Supplementary Fig.~\ref{fig:PCA_all_modes} (e.g. for PC3).
The detailed analysis in \cite{grebe_low_2010} revealed four dominant inter-area modes in Continental Europe with distinct period lengths.
The authors describe a mode G1, which resembles a North-South dipole and oscillates with $t=5$~s.
Furthermore, they present mode T1, which represents an East-West dipole oscillating at $t=6.7$s, and similar mode G2 with $t=3.3$s.
In mode G3, Eastern Europe, Spain, and Portugal swing in phase opposition to Central Europe with $t=2$s.
Comparing this to our results, we conclude that PC2 corresponds well to mode T1, while PC3 is similar to mode G1.
However, PC3 contains another substantial oscillation with $t=9s$.
This could be an effect of aliasing due to our low Nyquist-frequency of $0.5$ Hz.
Modes with period lengths below $2$s thus re-appear in our PSD at multiples of their oscillation period.
The mode G3 could thus correspond to the peak at $t\approx9$s in Supplementary Fig~\ref{fig:PCA_all_modes}.
Finally, PC4 contains the mode G2 among others, while PC5 and PC6 both contain the period lengths of G1 and T1.
The PCA modes are thus very similar to the results in \cite{grebe_low_2010}, but they do not isolate inter-area modes with single distinct periods. 

The overlap of different (linear) oscillation modes in our PCA results can have different reasons.
Our low spatial resolution could make it impossible to retain the spatial modes identified in an earlier study \cite{grebe_low_2010}.
On the other hand, a linear separation of the inter-area modes through a PCA and a Fourier decomposition could generally fail due to the non-linearity of power system dynamics.
Finally, our time resolution of one second could lead to aliasing and the incorrect reconstruction of inter-area modes.
In the future, a higher spatial and temporal resolution of frequency recordings would be necessary to further investigate these effects.  

\section*{Supplementary References}

\bibliographystyle{naturemag}
\addcontentsline{toc}{section}{\refname}
\bibliography{References}